\documentclass[twocolumn,tighten,times,trackchanges]{aastex62} %Times New Roman font slightly reduces page number
\received{12 July, 2019}
\revised{16 September, 2019}
\accepted{3 October, 2019}
\submitjournal{The Astrophysical Journal} %\apj

\usepackage{natbib}
\usepackage{subfigure,graphicx}
\usepackage{soul}
\usepackage{url}
\usepackage{color}
\usepackage{comment}
\usepackage{enumitem}
\definecolor{darkgray}{gray}{0.4}
\definecolor{patinared}{rgb}{.72,0,0}
\definecolor{patinablue}{rgb}{0,0,.65}
\definecolor{orange}{rgb}{1,0.5,0}
\hypersetup{
  colorlinks= true, %Colors links instead of ugly boxes
  urlcolor  = darkgray, %Color for external hyperlinks
  linkcolor = patinared, %Color of internal links
  citecolor = patinablue %Color of citations
}

\newcommand{\herschel}{\textit {Herschel}}
\newcommand{\hst}{\textit {HST}}
\newcommand{\halpha}{H$\alpha$}
\newcommand{\htwo}{H$_2$}
\newcommand{\oi}{[\ion{O}{1}]}
\newcommand{\um}{ $\mu$m}
\newcommand{\casa}{{\sc{Casa}}}
\newcommand{\kms}{km~s$^{-1}$}
\newcommand{\msun}{$M_\odot$}
\newcommand{\cotwoone}{\mbox{CO $J$=2~$\!\rightarrow\!$~1}}
\newcommand{\cosixfive}{\mbox{CO $J$=6~$\!\rightarrow\!$~5}}
\newcommand{\cothreetwo}{\mbox{CO $J$=3~$\!\rightarrow\!$~2}}
\newcommand{\siofivefour}{\mbox{SiO $J$=5~$\!\rightarrow\!$~4}}
\newcommand{\siosixfive}{\mbox{SiO $J$=6~$\!\rightarrow\!$~5}}
\newcommand{\siosevensix}{\mbox{SiO $J$=7~$\!\rightarrow\!$~6}}
\newcommand{\sioeightseven}{\mbox{SiO $J$=8~$\!\rightarrow\!$~7}}

\shorttitle{Dust and Molecules in the Ejecta of SN 1987A}   % Short form of title for running header in final publication
\shortauthors{Cigan et al.} % Short author list for running header in final publication

\begin{document}

\title{High angular resolution ALMA images of dust and molecules in the SN~1987A ejecta}

%###### aastex6.2 formatting #######

\author[0000-0002-8736-2463]{Phil Cigan}
\affiliation{School of Physics and Astronomy, Cardiff University, Cardiff CF24 3AA, UK}
\email{ciganp@cardiff.ac.uk}

\author[0000-0002-5529-5593]{Mikako Matsuura}
\affiliation{School of Physics and Astronomy, Cardiff University, Cardiff CF24 3AA, UK}

\author[0000-0003-3398-0052]{Haley L. Gomez}
\affiliation{School of Physics and Astronomy, Cardiff University, Cardiff CF24 3AA, UK}

\author[0000-0002-4663-6827]{Remy Indebetouw}
\affiliation{Department of Astronomy, University of Virginia, PO Box 400325, Charlottesville, VA 22904, USA}

\author[0000-0002-5724-1636]{Fran Abell{\'a}n}
\affiliation{Departamento de Astronom{\'i}a y Astrof{\'i}sica, Universidad de Valencia, C/Dr. Moliner 50, E-46100 Burjassot, Spain}

\author{Michael Gabler}
\affiliation{Max-Planck-Institut f{\"u}r Astrophysik, Karl-Schwarzchild-Stra{\ss}e 1, D-85748 Garching, Germany}

\author[0000-0002-3880-2450]{Anita Richards}
\affiliation{JBCA, School of Physics and Astronomy, University of Manchester, Manchester M13 9PL, UK}

\author[0000-0002-0427-5592]{Dennis Alp}
\affiliation{Department of Physics, KTH Royal Institute of Technology, The Oskar Klein Centre, AlbaNova, SE-106 91 Stockholm, Sweden}

\author[0000-0003-4932-9379]{Tim Davis}
\affiliation{School of Physics and Astronomy, Cardiff University, Cardiff CF24 3AA, UK}

\author[0000-0002-0831-3330]{Hans-Thomas Janka}
\affiliation{Max-Planck-Institut f{\"u}r Astrophysik, Karl-Schwarzchild-Stra{\ss}e 1, D-85748 Garching, Germany}

\author[0000-0001-6815-4055]{Jason Spyromilio}
\affiliation{European Southern Observatory, Karl-Schwarzschild-Stra{\ss}e 2, 85748 Garching, Germany}

\author[0000-0002-3875-1171]{M. J. Barlow}
\affiliation{Department of Physics and Astronomy, University College London, Gower Street, London, WC1E 6BT, UK}

\author[0000-0003-0729-1632]{David Burrows}
\affiliation{Department of Astronomy and Astrophysics, Pennsylvania State University, University Park, PA 16802, USA}

\author[0000-0001-8033-1181]{Eli Dwek}
\affiliation{Observational Cosmology Lab., Code 665, NASA Goddard Space Flight Center, Greenbelt, MD 20771, USA}

\author[0000-0001-8532-3594]{Claes Fransson}
\affiliation{Department of Astronomy, Stockholm University, The Oskar Klein Centre, AlbaNova, SE-106 91 Stockholm, Sweden}

\author[0000-0002-3382-9558]{Bryan Gaensler}
\affiliation{Dunlap Institute for Astronomy and Astrophysics, 50 St. George Street, Toronto, ON M5S 3H4, Canada}

\author[0000-0003-0065-2933]{Josefin Larsson}
\affiliation{Department of Physics, KTH Royal Institute of Technology, The Oskar Klein Centre, AlbaNova, SE-106 91 Stockholm, Sweden}

\author{P. Bouchet}
\affiliation{DRF/IRFU/DAp, CEA-Saclay, F-91191 Gif-sur-Yvette, France}
\affiliation{NRS/AIM, Universit{\'e} Paris Diderot, F-9119, Gif-sur-Yvette, France}

\author[0000-0002-3664-8082]{Peter Lundqvist}
\affiliation{Department of Astronomy, Stockholm University, The Oskar Klein Centre, AlbaNova, SE-106 91 Stockholm, Sweden}

\author{J. M. Marcaide}
\affiliation{Departamento de Astronom{\'i}a y Astrof{\'i}sica, Universidad de Valencia, C/Dr. Moliner 50, E-46100 Burjassot, Spain}

\author[0000-0002-5847-2612]{C.-Y.~Ng}
\affiliation{Department of Physics, The University of Hong Kong, Pokfulam Road, Hong Kong}

\author[0000-0003-3900-7739]{Sangwook Park}
\affiliation{Department of Physics, University of Texas at Arlington, Arlington, TX 76019, USA}

\author{Pat Roche}
\affiliation{Astrophysics, Department of Physics, University of Oxford, DWB, Keble Road, Oxford OX1 3RH, UK}

\author[0000-0002-1272-3017]{Jacco Th. van Loon}
\affiliation{Lennard-Jones Laboratories, Keele University, ST5 5BG, UK}

\author[0000-0003-1349-6538]{J. C. Wheeler}
\affiliation{Department of Astronomy, The University of Texas at Austin, 1 University Station C1400, Austin, TX 78712-0259, USA}

\author[0000-0003-2742-771X]{Giovanna Zanardo}
\affiliation{International Centre for Radio Astronomy Research, M468, University of Western Australia, Crawley, WA 6009, Australia}

\begin{abstract}
We present high angular resolution ($\sim$80\,mas) ALMA continuum images of the SN~1987A system, together with \cotwoone, $J$=6~$\!\rightarrow\!$~5, and \siofivefour\ to $J$=7~$\!\rightarrow\!$~6 images, which clearly resolve the ejecta (dust continuum and molecules) and ring (synchrotron continuum) components.
Dust in the ejecta is asymmetric and clumpy, and overall the dust fills the spatial void seen in H$\alpha$ images, filling that region with material from heavier elements.
The dust clumps generally fill the space where \cosixfive\ is fainter, tentatively indicating that these dust clumps and CO are locationally and chemically linked. 
In these regions, carbonaceous dust grains 
might have formed after dissociation of CO. 
The dust grains would have cooled by radiation, and subsequent collisions of grains with gas would also cool the gas, suppressing the \cosixfive\ intensity. 
The data show a dust peak spatially coincident with the molecular hole seen in previous ALMA \cotwoone\ and \siofivefour\ images. 
That dust peak, combined with CO and SiO line spectra, suggests that the dust and gas could be at higher temperatures than the surrounding material, though higher density cannot be totally excluded. 
One of the possibilities is that a compact source provides additional heat at that location. 
Fits to the far-infrared--millimeter spectral energy distribution give ejecta dust temperatures of 18--23K. We revise the ejecta dust mass to $\mathrm{M_{dust}} = 0.2-0.4$\msun\ for carbon or silicate grains, or a maximum of $<0.7$\msun\ for a mixture of grain species, using the predicted nucleosynthesis yields as an upper limit. 
\pagebreak
\end{abstract}

%%%%---------------- Introduction --------------------------

\section{Introduction}
\label{sec:intro}

Multiwavelength studies of Supernova 1987A (SN~1987A), located at a distance of $51.4 \pm 1.2$\,kpc in the Large Magellanic Cloud \citep{Panagia1999}, have provided unprecedented details of how supernova (SN) explosions trigger the dynamical distribution of gas in a supernova remnant (SNR), and how this SN/SNR system evolves over time.
The morphology of SN~1987A is well studied (see the recent review in \citealp{McCray2016}), with the system consisting of ejecta, and a bright and distinct equatorial ring (hereafter the ring), together with two fainter outer rings.
The ring is composed  of circumstellar material that radiates in UV, optical, X-rays, and radio over an extent of 1\farcs6 (0.3\,pc) \citep[e.g.][]{Burrows2000, Sonneborn1998, Ng2013}, as well as thermal dust emission due to shock heating of pre-existing dust formed during the red supergiant phase \citep{Bouchet2006, Dwek2010}.
The ejecta have a complex morphology.
The H$\alpha$ emission, originating from warm gas irradiated by X-rays from the ring \citep{Larsson2011,Fransson2013}, exhibits an elongated north-south structure and a `hole' in the center. 
Along with hydrogen lines from the ejecta, near-infrared (NIR) emission from warm ($\sim$2000 K) CO and mid-infrared (MIR) emission from SiO in the SN ejecta were detected early (as early as 112 days) after the explosion \citep[e.g., ][]{Spyromilio1988,Roche1991}.
After day 9,000, cold expanding CO, SiO and HCO$^+$ molecules were detected in the submillimeter (submm) part of the spectrum \citep{Kamenetzky2013, Matsuura2017}, highlighting that a significant part of the ejecta is cold (13--132\,K).
Interestingly, the inner ejecta of SN~1987A have not yet mixed with the circumstellar medium (CSM) or interstellar medium (ISM) and the majority has not yet passed through the reverse shock \citep{France2010,Frank2016}. 
Thus this young SNR is an ideal source for studying the footprints of the gas dynamics since the very early days of the SN, as the gas has been assumed to be free-expanding since its explosion  \citep{McCray1993}. 
Indeed, recent high angular resolution emission line images of SiO and CO \citep{Abellan2017} from the Atacama Large Millimeter/submillimeter Array (ALMA) have been used to compare the distribution of the molecular gas ejecta with the predictions from models of the gas dynamics after the SN \citep[][Gabler et al., \textit{in preparation}]{Wongwathanarat2015}.

The progenitor of SN~1987A, Sanduleak --69$^\circ$ 202, was a blue supergiant \citep{West1987,White1987,Gilmozzi1987,Kirshner1987}, thought to have had a zero-age main sequence mass of  $\sim 19\,\rm M_{\odot}$ \citep{Woosley1997,Hashimoto1989}, with a mass of $\sim14\,\rm M_{\odot}$ at the time of the explosion \citep{Woosley1988, Smartt2009,Sukhbold2016}.
From its mass, the expectation is that a neutron star should have formed at the time of explosion.
Despite prompt neutrino emission observed at the burst \citep{Hirata1987} indicating the formation of a neutron star \citep{Burrows1988, Sukhbold2016},
the search for a compact object associated with SN~1987A has been difficult: observational searches have proven unfruitful  \citep[e.g.][]{Manchester2007, Alp2018a, Esposito2018, Zhang2018}.
The possible detection of radio polarization towards the ejecta \citep{Zanardo2018} hints at the presence of magnetized shocks, potentially due to a compact object.
\citet{Alp2018a} proposed that a thermally-emitting neutron star could be dust-obscured, and that this may be detectable as a point source in far-infrared (FIR) or submm images of the remnant, though this has not yet been detected.

It is still largely debated whether or not SNe are net dust producers or destroyers in galaxies \citep[e.g., ][]{Morgan2003a,Matsuura2009,Gall2011,Gomez2013,Dwek2014,Rowlands2014b,Michalowski2015, Watson2015,Lakicevic2015,Temim2015}.
Due to its youth and proximity, SN~1987A is an excellent laboratory for studying  SN dust.
It is also rare, since any dust emission seen in the inner region of the remnant can be attributed unambiguously to dust formed in the supernova ejecta and not from the swept-up CSM/ISM or unrelated foreground/background material \cite[a common issue with Galactic SNRs, e.g.,][]{Morgan2003b,Gomez2012a,DeLooze2017,Chawner2018}.
SN~1987A also provides insight into dust formation at an early stage compared to previously studied Galactic supernova remnants -- here we can probe timescales on the order of decades rather than centuries, filling in the gap between very young SNe \citep[e.g.,][]{Gall2014} and historical remnants.

Thermal emission from small quantities ($10^{-4}\,\rm M_{\odot}$) of dust was detected in the early days after the SN explosion (day$\sim$300--600) using MIR observations \citep{Danziger1989, Lucy1989, Bouchet1991, Roche1993, Wooden1993}.
More surprisingly, the \textit{Herschel Space Observatory} (\citealp{Pilbratt2010}, hereafter \textit{Herschel})  revealed a large amount of cold dust ($\sim0.5 \rm M_{\odot}$) at the location of the remnant (\citealp{Matsuura2011,Matsuura2015}). 
ALMA resolved the emission from dust in SN~1987A on scales of $0.3^{\prime \prime}$ and confirmed that the $\sim \rm 0.5\,M_{\odot}$ of cold (20\,K) dust discovered with \textit{Herschel} originates from the inner ejecta region \citep{Indebetouw2014,Matsuura2015}.
\cite{Dwek2015} and \cite{Wesson2015} re-visited the dust emission at early times ($<$1200 days). 
\cite{Dwek2015} find that a large mass of dust can be present early on (0.4\msun\ at $\sim$615 days) with a model of silicates and amorphous carbon.  
\cite{Wesson2015} conclude instead, from comparing radiative transfer models to the optical--IR SED limits, that the dust mass increased more slowly over the first 10 years.
This substantial mass of dust observed in the inner debris of SN~1987A demonstrates that a large fraction of the heavy elements ejected in a SN may be locked up in a dust reservoir.

Whether dust grains formed in the ejecta of a supernova are carbon or silicate-rich remains an unanswered question: the models of \cite{Cherchneff2009,Cherchneff2010} and \cite{Sarangi2013, Sarangi2015} predict that for abundance ratios $\rm C/O < 1$, carbon atoms will mostly be locked up in CO molecules in the first 1000 days, preventing the formation of a large mass of amorphous carbon dust.
Though depending on the gas density, CO may be dissociated by electrons produced by radioactive decay \citep{Clayton2011}, and/or (to a lesser extent) X-rays from the ring, however, the models of \cite{Sarangi2013} and \cite{Sarangi2015} indicate the dissociation of CO is insignificant.
In contrast, in order to explain the FIR dust emission, a substantial fraction of the dust grains must be composed of amorphous carbon (amC, \citealt{Matsuura2015}), as the emissivity of amC grains is higher than that of silicates in general, thus leaving an unresolved tension between observations and theory.  
We note that a model which explains the FIR emission and requires only a small amount of amC grains, with the majority of mass in silicates, was proposed by \cite{Dwek2015}. 
There they fit the FIR SED with amC-silicate composite grains assuming a ``continuous distribution of ellipsoids'' (CDE) model and found a reduced dust mass, though the majority of the reduction in mass in this case arises from the inclusion of dust grains with long axial ratios (so-called needles), which allows the CDE model to surpass the FIR emissivity of amorphous carbon.
No evidence of the silicate signature was found in the warm dust emission in the first two years after the explosion, suggesting that small silicate grains were not the first condensates \citep{Roche1993}.

In this work, we present high angular resolution ALMA (Cycle 2) dust images for SN~1987A, where we resolve dust clumps on scales of $\sim$80 mas. 
Here, we revisit the dust mass and grain composition using the ALMA photometry. We discuss the implications of our results for the gas phase chemistry leading to dust formation, and find evidence for warmer gas at the center of the inner ejecta hinting at the possible indirect detection of a compact source.

%%%%---------------- Data and Analysis --------------------------
\section{Data}
\label{sec:data}

\subsection{Observations and Reduction}
\label{sec:observations}

\begin{deluxetable*}{ llccccccccc }[t]
\def\arraystretch{0.8}
\tablecaption{ Observations \label{table:Obs}}
\tablehead{ \colhead{} & \colhead{Sub-} & \colhead{Frequency} & \colhead{Baselines} & \colhead{Angular Scales} & \colhead{Observation} & \colhead{SN} & \colhead{Bandpass} & \colhead{Phase} & \colhead{Check} & \colhead{Time}  \\
			 & band & Range (GHz) & (m) & (arcsec) & Date & Day & Calibrator & Calibrator & Source & (min) }

\startdata
B7 & A1 & 299.88--315.87 & 45.4--1574.4 & 0.13--4.31 & 2015-06-28 & 10352 & J0538-4405 & J0635-7516 & J0601-7036 & 18.4 \\
   & A2 & 299.88--315.87 & 43.3--2269.9 & 0.09--4.52 & 2015-09-22 & 10438 & J0538-4405 & J0635-7516 & J0601-7036 & 18.3 \\
   & B  & 342.48--358.34 & 15.1--1574.4 & 0.11--11.46 & 2015-07-25 & 10379 & J0538-4405 & J0635-7516 & J0601-7036 & 20.9 \\
   & C  & 346.23--362.09 & 15.1--1574.4 & 0.11--11.34 & 2015-07-25 & 10379 & J0538-4405 & J0635-7516 & J0601-7036 & 19.9 \\
   & D1 & 303.62--319.48 & 45.4--1574.4 & 0.13--4.26 & 2015-06-28 & 10352 & J0538-4405 & J0635-7516 & J0601-7036 & 18.8 \\
   & D2 & 303.62--319.48 & 43.3--2269.9 & 0.09--4.47 & 2015-09-22 & 10438 & J0538-4405 & J0635-7516 & J0601-7036 & 18.8 \\
B9 & A  & 673.44--681.06 & 43.3--2269.9 & 0.04--2.10 & 2015-09-25 & 10441 & J0522-3627 & J0601-7036 & J0700-6610 & 12.5 \\
   & B  & 680.94--688.56 & 43.3--2269.9 & 0.04--2.07 & 2015-09-25 & 10441 & J0522-3627 & J0601-7036 & J0700-6610 & 12.5 \\
   & C  & 688.44--696.06 & 43.3--2269.9 & 0.04--2.05 & 2015-09-25 & 10441 & J0522-3627 & J0700-6610 & J0450-8101 & 12.5 \\
\enddata

\tablecomments{
Observations for proposal ID 2013.1.00063.
Each sub-band is comprised of four 2 GHz blocks of 128 channels (15.625 MHz each).
The same flux calibrator, J0519-454, was used for all observation blocks.
} 

\end{deluxetable*}

\begin{figure*}[t!]
\centering
\includegraphics[trim=0mm 0mm 0mm
  0mm,clip=true,width=1.0\textwidth]{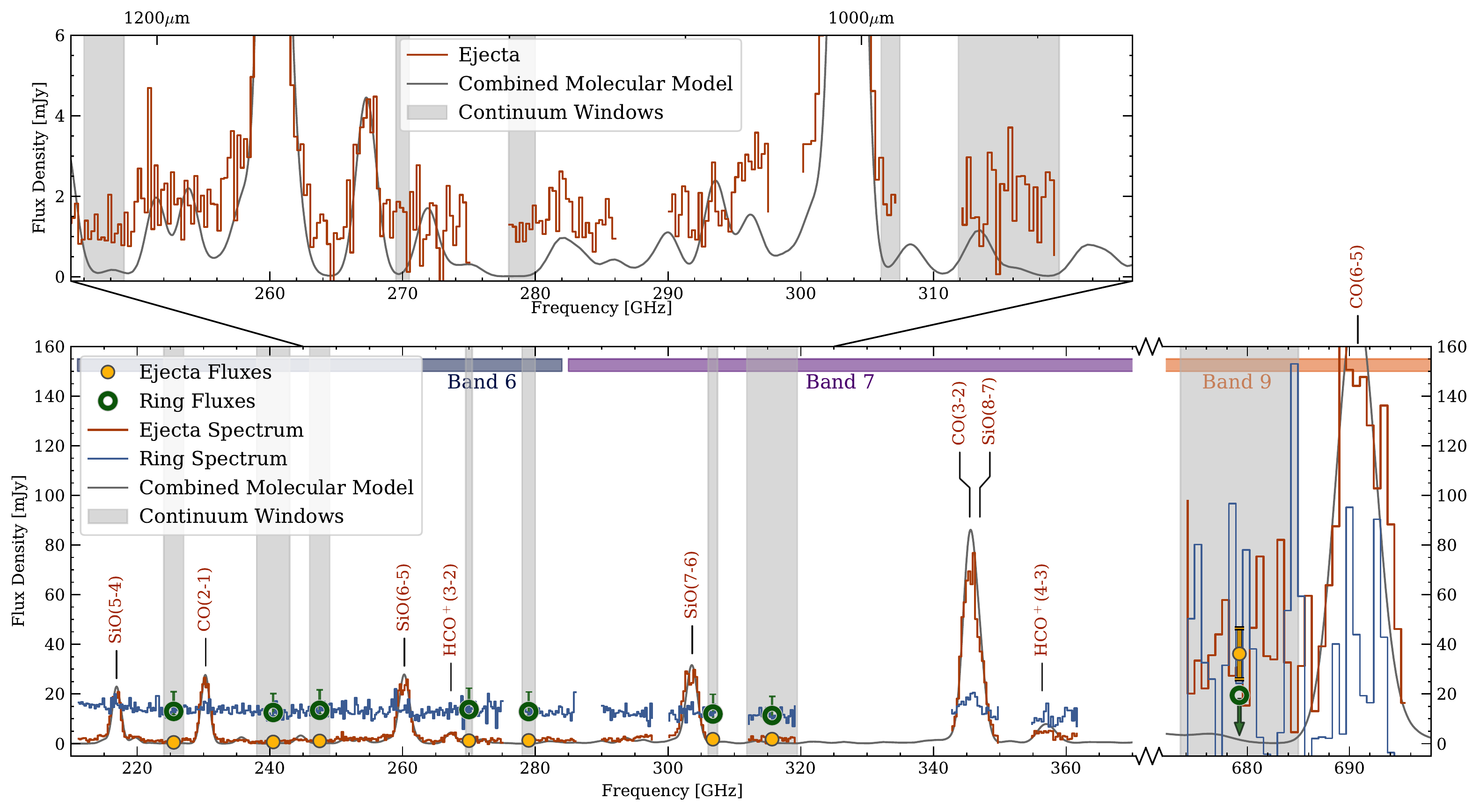}
\caption{
Spectra and integrated continuum flux densities of the ejecta and ring for Bands 6, 7, and 9.
The molecular line emission model is taken from \citet{Matsuura2017}.
Vertical grey bands indicate the portions of the spectrum deemed to be line-free and therefore used in creating the continuum images at 307, 315, and 679 GHz (the 315\,GHZ image is shown in Fig.~\ref{fig:hstcomp}).
Data points indicate the flux densities in that band for the ring (green) and the ejecta (yellow).
\vspace{0.6cm}
}
\label{fig:ejectavsring_b679}
\end{figure*}

\begin{deluxetable}{ ccccc }
\def\arraystretch{0.7}
\tablecaption{ Imaging Properties\label{table:Imaging}}
\tablehead{ \colhead{Frequency Range} & \colhead{$\nu_{c}$} & \colhead{Beam FWHM} & \colhead{Beam PA} & \colhead{RMS Noise} \\
			(GHz) & (GHz) & (arcsec$^2$) & (deg) & (mJy bm$^{-1}$) }
\startdata
\multicolumn{1}{l}{Continuum} & & & & \\
224.00 -- 227.00   & 225.50 &   0.30 $\times$ 0.30   &   0.00   &   0.12  \\
238.00 -- 243.00   & 240.50 &   0.30 $\times$ 0.30   &   0.00   &   0.09  \\
246.00 -- 249.00   & 247.50 &   0.30 $\times$ 0.30   &   0.00   &   0.10  \\
269.50 -- 270.50   & 270.00 &   0.30 $\times$ 0.30   &   0.00   &   0.18  \\
278.00 -- 280.00   & 279.00 &   0.30 $\times$ 0.30   &   0.00   &   0.10  \\ \\
306.06 -- 307.47   & 306.76 &   0.20 $\times$ 0.15   &   124.32   &   0.07  \\
311.88 -- 319.48   & 315.68 &   0.19 $\times$ 0.14   &   119.11   &   0.04  \\ \\
673.45 -- 685.00   & 679.22 &   0.08 $\times$ 0.06   &   74.37   &   0.71  \\
\hline
\multicolumn{1}{l}{Spectral Lines} & & & & \\
\cotwoone    & 230.54 & 0.06 $\times$ 0.04 & 27.43  & 0.04 \\
\cosixfive   & 691.47 & 0.09 $\times$ 0.07 & 185.35 & 2.82 \\
\siofivefour & 217.10 & 0.06 $\times$ 0.04 & 19.74  & 0.05 \\
\siosixfive  & 260.52 & 0.04 $\times$ 0.03 & 173.66 & 0.06 \\
\siosevensix & 303.93 & 0.13 $\times$ 0.10 & 35.47  & 0.47
\enddata

\tablecomments{
The position angles are counter-clockwise from north.   \cotwoone, \siofivefour, and \siosixfive\ parameters are for the data cubes from \cite{Abellan2017}.
The $\nu_{c}$ values listed for the CO and SiO lines are rest frequencies.
RMS values for the spectral lines are per velocity channel.
For observation dates and epochs, see Table~\ref{table:Obs}.
} 

\end{deluxetable}

Our observations of SN~1987A were obtained with the Atacama Large Millimeter/Submillimeter Array (ALMA), as part of the Cycle 2 observing program 2013.1.00063.S.
The data were taken over several days in the latter half of 2015, between 10352 and 10441 days after the initial explosion.
The Band--7 (870$\mu$m) and 9 (450$\mu$m) integrations utilized between 34 and 36 antennae with baselines spanning 15m to 2.3km.
See Table~\ref{table:Obs} for a summary of the observations.

Each data set was reduced separately with Common Astronomy Software Applications package \citep[\casa\footnote{\url{http://casa.nrao.edu/}},][]{CASA}, version 4.5.1.
Once calibrated, the \texttt{tclean} algorithm was used to deconvolve and image the data.

The check source (reference quasar with precisely known position) and phase calibrator coordinates, determined with {\texttt{imfit}} in \casa, were offset by no more than 0.4 mas from the catalog values. 
Other measures of astrometric quality for our observing configurations include the ALMA baseline measurement accuracy (2 mas), noise-limited signal error $\sim$(beam size)/(S/N) (3--4 mas), and the phase transfer error from the measured phase RMS ($<$ 12 mas), where S/N is the signal-to-noise ratio and RMS denotes root-mean-square.
Combining these, the overall astrometric accuracy we assume for the data presented in this work is 10 mas in Band--6, 12 mas in Band--7, and 15 mas in Band--9. 

Decorrelation due to factors such as weather was investigated using the flux calibrator, phase calibrator, and check source by phase-averaging over several intervals and integrating the resulting flux densities -- a large variation in flux density for different phase averaging intervals would suggest that decorrelation is pronounced enough to decrease the recovered flux.
The variations of the calibrator flux densities in all Band--7 and Band--9 windows were within the systematic uncertainties except for Band--7C (346--362\,GHz), which had significantly worse weather than the other segments, with an estimated decorrelation of $\sim$35\%.
Bands 7A and 7D also suffered from poor weather in the original June 2015 observations, with $\sim$1.45 mm precipitable water vapor (PWV), and were therefore repeated in September 2015.
These are denoted as 7A2 and 7D2.
Despite the poorer quality of the June data, combining them with the September data results in higher S/N images.

Self-calibration, a common technique for high S/N data where calibrating the data against itself in successive deconvolution cycles can often result in improved dynamic range, was determined to have a negligible impact on the images.
Final images were cleaned with natural weighting applied to the baselines to optimize sensitivity per beam.
The imaging parameters, including resolution and sensitivity, are given in Table~\ref{table:Imaging}.

\subsection{Defining Continuum Wavelength Ranges}
\label{sec:continuumdata}

The wavelengths covered by these observations include many spectral lines from molecular species -- primarily CO and SiO, with contributions from various SO lines and potentially others.
The $\pm\sim$1000 km s$^{-1}$ expansion velocity of the ejecta means that the linewidths span a substantial fraction of the observed bands.
The continuum bands selected relative to the modelled molecular line emission are shown in Fig.~\ref{fig:ejectavsring_b679}, using the ALMA spectra and the emission line model of CO, SiO, SO, and SO$_2$ from \citet{Matsuura2017}.
Only windows that were free from molecular line emission (shown by the grey vertical bands) were used to make continuum images, centered at roughly 307, 315, and 679 GHz.
The 315~GHz continuum image is shown in Fig.~\ref{fig:hstcomp}.
We also utilize here the Cycle 2 Band--6 imaging data presented by \citet{Matsuura2017}, to provide continuum information below 300~GHz.
The Band--6 images were restored to a common circular beam with full-width at half-maximum (FWHM) of 0\farcs30.

\subsection{Molecular Line Data}
\label{sec:moleculardata}

In this section we present the molecular line data observed in the same blocks as the continuum discussed above: \cosixfive\ with rest frequency 691.47 GHz and \siosevensix\ at $\nu_\mathrm{rest}$ = 303.93 GHz.
The 345.80 GHz \cothreetwo\ and 347.33 GHz \sioeightseven\ lines were also covered in these observing blocks, but as they are heavily blended we do not consider them in the present work.

The \siosevensix\ and \cosixfive\ cubes were created with \texttt{tclean} in \casa, with a spectral resolution of 300 km s$^{-1}$, which gives a reasonable balance between velocity resolution and sensitivity per channel.
\cosixfive\ was imaged with natural weighting to maximize sensitivity per beam.
\siosevensix\ was imaged with \texttt{robust}$=-1$ in order to better spatially resolve the central features of interest.
A comparison of the integrated (Moment-0) maps of the CO and SiO lines is given in Fig.~\ref{fig:multiwav}.

In addition to the molecular line data described above, we also utilize the \cotwoone, \siofivefour, and  \siosixfive\footnote{The images for the \siosixfive \, transition were described but not shown in \citet{Abellan2017}.} data as described in \citet{Abellan2017}.
Although both sets were taken in Cycle 2, the molecular line data presented by \cite{Abellan2017} have higher signal to noise as they were combined with Cycle 3 data, and due to \cosixfive\ being in a band with poorer atmospheric transmission than \cotwoone.
The additional Cycle 3 data also give their \cotwoone, \siofivefour, and \siosixfive\ maps finer spatial resolution than the observations presented in the current work, with FWHM 0\farcs04--0\farcs06 (see Table~\ref{table:Imaging}).
For full details of their data reduction technique, we refer the reader to their Section~2.
The channel maps are shown in Figs.~\ref{fig:chanmaps_co} and \ref{fig:chanmaps_sio}.
The given velocities are the observed values, not shifted to the reference frame of SN~1987A.
The systemic velocity (Kinematic Local Standard of Rest frame; LSRK) of SN~1987A is 287 \kms\ receding from Earth \citep{Groningsson2008a}.

\section{Description of Images}
\label{sec:analysis}

\begin{figure*}[t]
\centering
\includegraphics[width=1.0\textwidth]{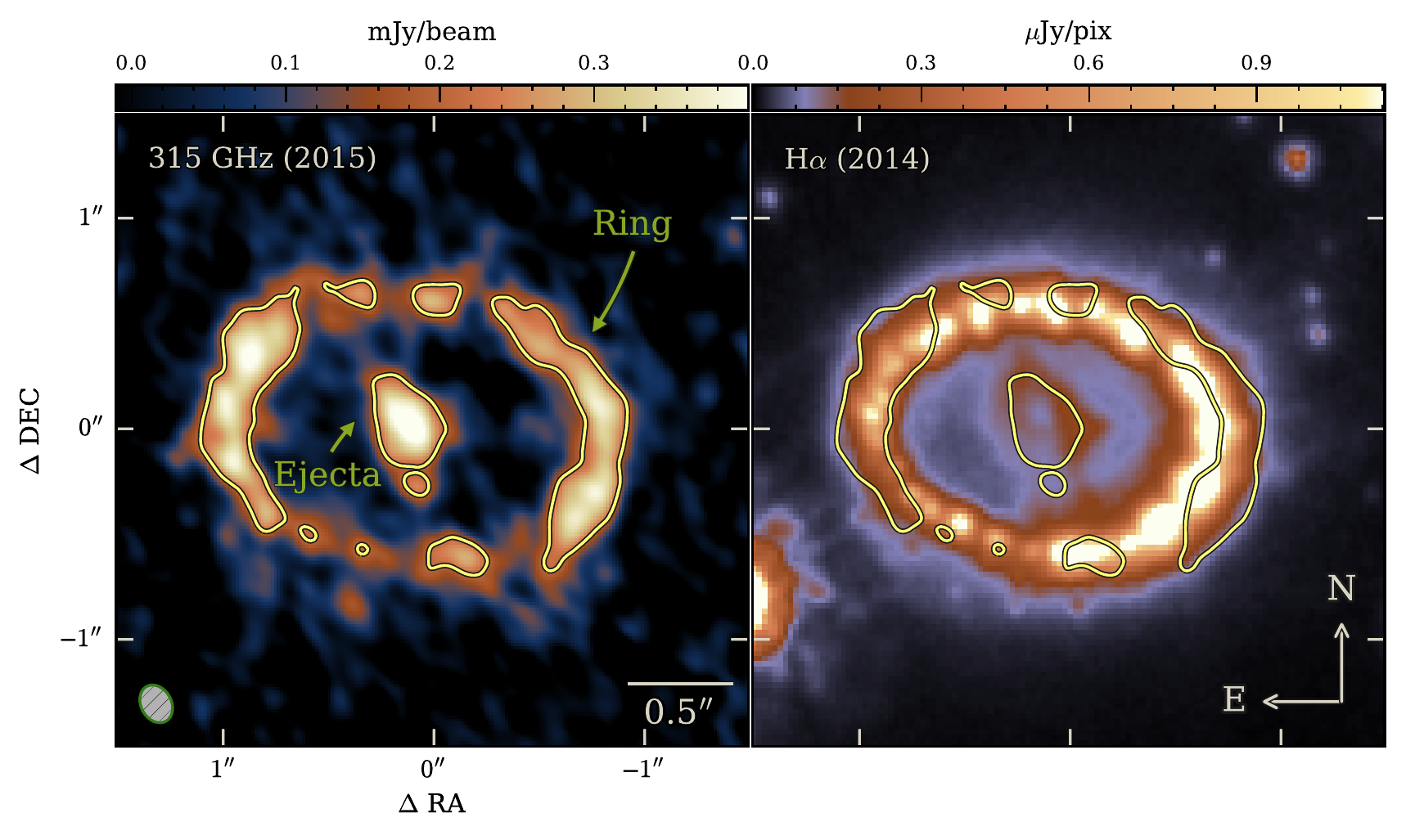}
\caption{
ALMA 315 GHz (with beam) and 2014 \hst\ 625W band image \citep{Fransson2015}, which includes \halpha.
The yellow contours display 315\,GHz emission at 0.2 mJy/beam.
The 315~GHz continuum in the inner ejecta originates from thermal dust emission, while in the ring it is due to synchrotron emission.
The 18 mas uncertainty on the relative alignment due to Band--7 astrometric error (12 mas) and \hst\ image registration based on fitting the ring (6 mas) is of order 1 pixel in these images. 
}
\label{fig:hstcomp}
\end{figure*}

\begin{figure*}[t!]
\centering
\includegraphics[width=1.0\textwidth]{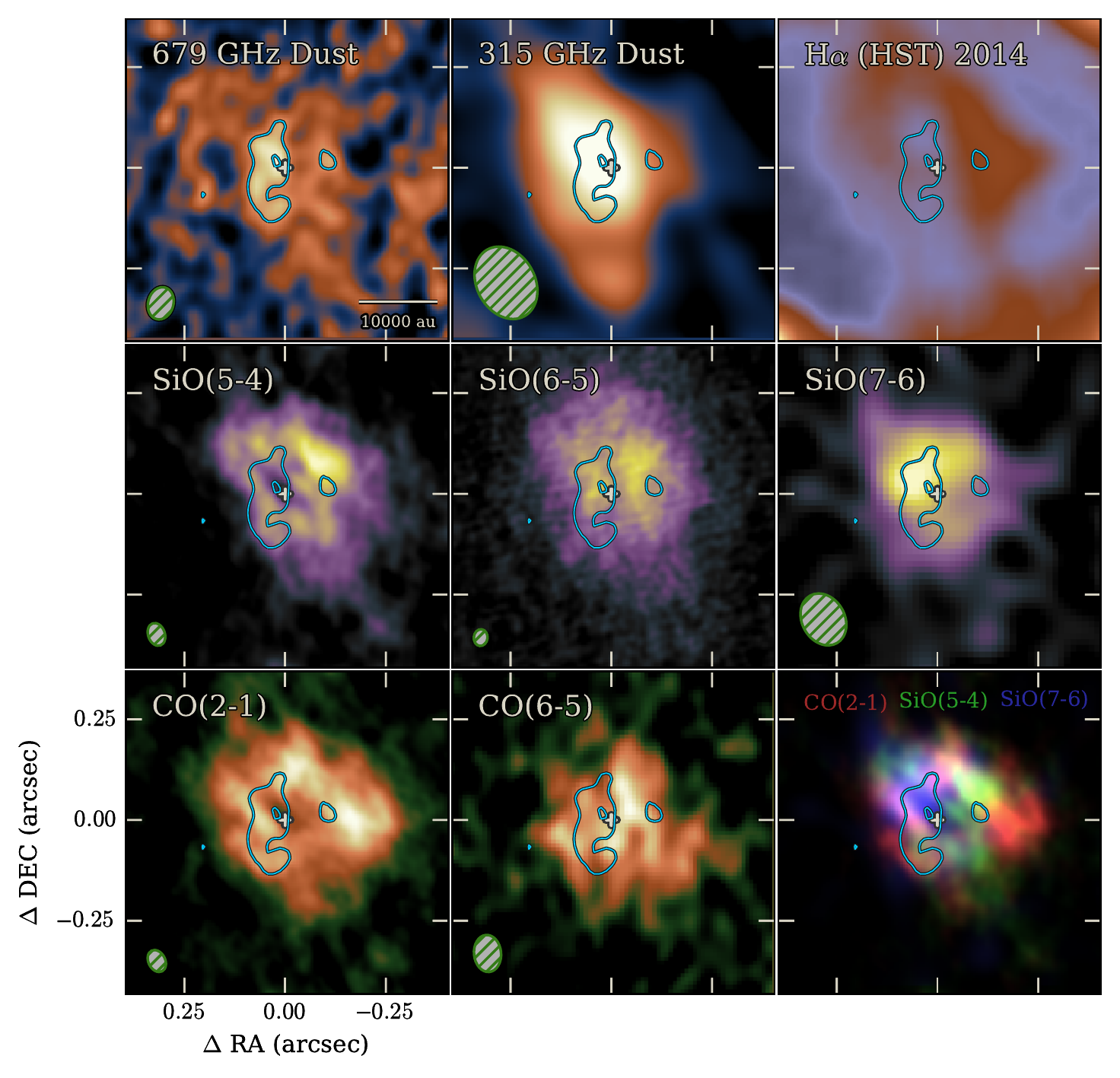}
\caption{ Multi-wavelength view of the ejecta in SN~1987A.  
The cyan contours represent the 679 GHz dust emission at 3$\sigma$ and 5$\sigma$. 
Beam sizes for individual maps are denoted by the green ellipses.
The small cross denotes the system center as defined in Appendix~\ref{app:center}.
The bottom right panel is a 3-color image of \cotwoone\ in red, \siofivefour\ in green, and \siosevensix\ in blue, and highlights how varied the spectrally-integrated emission is between the various line transitions.
The brightest areas are generally distinct patches of primary color instead of blended, demonstrating that the CO and SiO peaks are not cospatial, and none matches the distribution of the 679 GHz dust.
Some of the CO falls in the H$\alpha$ hole (the lower left), but the majority of the CO peaks on the little H$\alpha$ `wing' to the right of the hole.
The small $5\sigma$ cyan contour just northeast of the center of the remnant is the so-called ``blob''.
}
\label{fig:multiwav}
\end{figure*}

\begin{figure}[h!]
\centering
\includegraphics[width=1.0\linewidth]{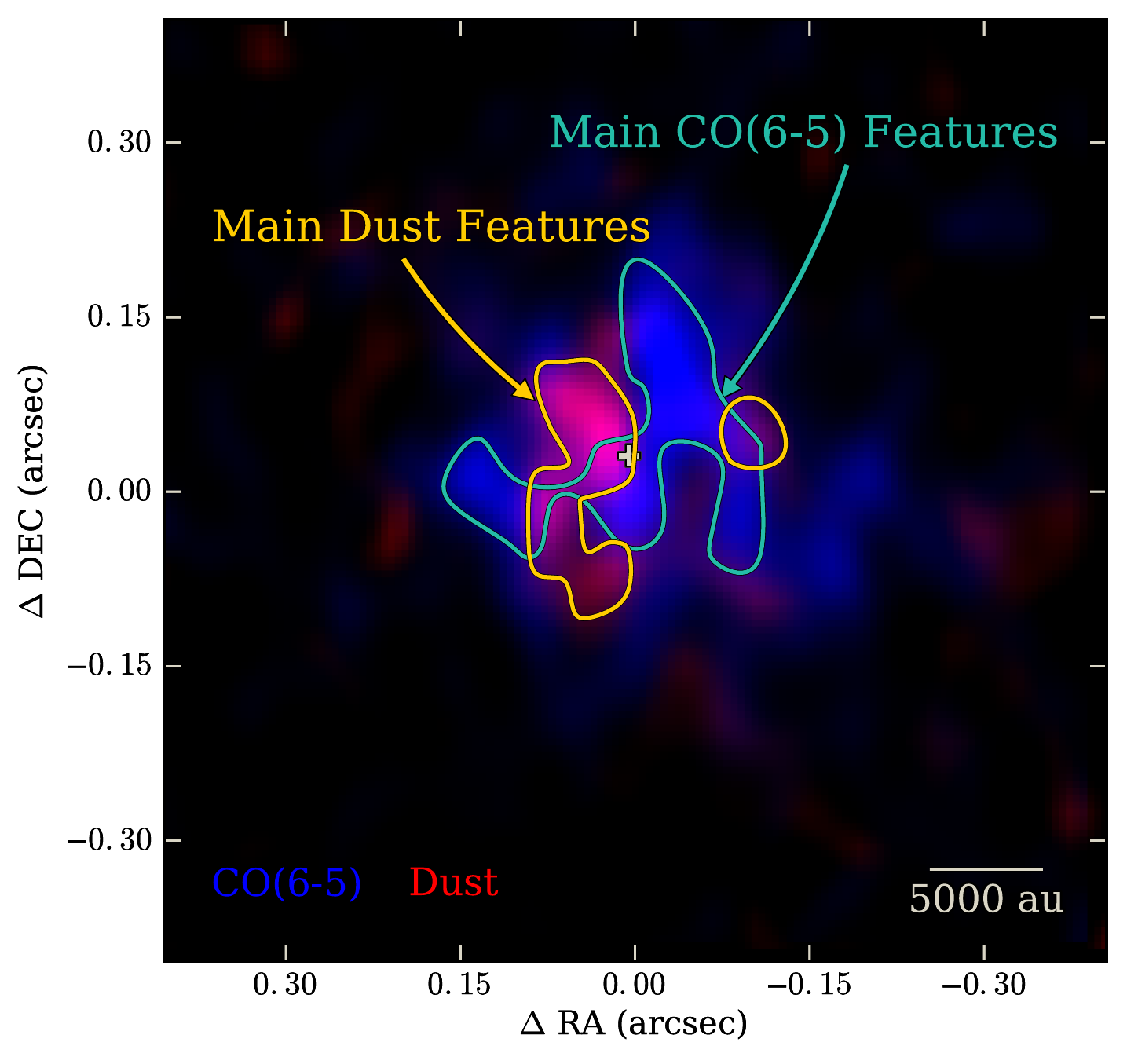}
\caption{
The spatial anti-correlation of dust and \cosixfive.
In this overlay, \cosixfive\ is in blue and 679\,GHz is in red, showing their relative spatial distributions.
The off-white cross denotes the system center position as defined in Appendix~\ref{app:center}. 
The gold line highlights the extent of the major features in the dust, while the teal line demarcates the major \cosixfive\ features.
The contrast of each component image was set independently to emphasize its major features -- visible blue and red features roughly correspond to S/N$>$3 for the dust and S/N$>$5 for \cosixfive, respectively.
The dust and \cosixfive\ emission exhibit a notable anti-correlation.
}
\label{fig:3color}
\end{figure}

The SN~1987A ring and ejecta continuum image at 315\,GHz is shown in comparison to \halpha\ in Fig.~\ref{fig:hstcomp}. 
These images have been aligned following a technique in \citet{Alp2018a} where the ring emission is used to derive a reference center, though here we take a simpler approach (see Appendix~\ref{app:center} for details).  
Our derived ring+ejecta system center used in this work is $\alpha$=5$^{\rm h}$35$^{\rm m}$27\fs998, $\delta$=--69$^{\circ}$16$^{\prime}$11\farcs107 (ICRS), $\pm$18 mas (Fig.~\ref{fig:centercoordsmethod}).  
At 315\,GHz, the ring is clumpy and the brightness contrast in the east and west components of the ring is different to that observed in the \halpha\ ring emission.  
The brighter emission observed in the NE and SW regions of the ring in the radio is similar to that seen in hard X-rays \citep{Helder2013,Frank2016}. 
The ejecta are located at the center of the image inside of the ring structure. 
The ring emission at 315\,GHz is attributed to synchrotron (see \S~\ref{sec:ringemission}), and the inner region is thermal dust emission from the SN ejecta \citep{Indebetouw2014}.

Fig.~\ref{fig:multiwav} shows an enlarged view of the ejecta images of dust continuum and lines. The majority of the submm ejecta continuum is distributed in a roughly symmetrical ellipsoid, with fainter asymmetrical emission protruding west and south-west.
At 315\,GHz, the ejecta are moderately resolved, and show a conspicuously separate clump of emission south of the main body of the ejecta.
This clump persists in images produced with lower \texttt{robust} settings in \texttt{tclean}, where sensitivity is lower and spatial resolution is higher.
Both the primary ejecta material and the smaller clump as observed in the 315\,GHz image appear to fill in the gaps seen in the \halpha\ image, like a `lock in the keyhole'.
This is shown in Fig.~\ref{fig:hstcomp}, where the 3$\sigma$ contours highlighting the major continuum features are overlaid onto the continuum and \halpha\ images. 
The alignment accuracy is $\sim$1 pixel in the images, given the astrometric uncertainties discussed in \S~\ref{sec:observations} (12 mas for Band--7 continuum) and Appendix~\ref{app:center} (6 mas for registration of the \textit{Hubble} Space Telescope (\hst) image to the 315~GHz ALMA image).

The 679 GHz image provides the highest resolution view of the continuum (top left panel of Fig.~\ref{fig:multiwav}).
This figure shows that the dust is asymmetrically distributed and is composed of several clumps, with the brightest feature (hereafter the ``blob'') just northeast of the center of the remnant.
The beam resolution provides a limit on the clump size -- assuming a distance of 51.4 $\pm$ 1.2 kpc \citep{Panagia1999}, the Band--9 beam FWHM of 0\farcs08$\times$0\farcs06 corresponds to a physical scale of 0.020$\times$0.016 pc, or 4125$\times$3230 au.
Nevertheless, the resolved 679 GHz image indicates that dust is not smoothly distributed across the ejecta, and the locations of dust clumps are not identical to clumps in the CO or SiO.
The S/N in the 679\,GHz image is moderate -- the outer cyan contours in Fig.~\ref{fig:multiwav} and the dust emission (in red) in Fig.~\ref{fig:3color} have pixel S/N$>$3, and the surrounding ejecta area has pixel S/N values of $\sim$2 in the 679\,GHz image.
The area between the ejecta and the ring -- outside of the outermost ejecta contour in Fig.~\ref{fig:hstcomp} -- is consistent with noise.

The molecular images provide a probe of different conditions in the ejecta, where lower transitions probe lower temperature gas (if optically thin, see Section~\ref{sec:radex}). 
One prominent feature is the central hole seen in the \cotwoone\ and \siofivefour\ images (middle left and lower left panels of Fig.~\ref{fig:multiwav}).
This was first reported by \citet{Abellan2017} and was seen both in the integrated 2D spatial maps and the 3D data cubes.
Although the integrated \siosixfive\ map (middle panel of Fig.~\ref{fig:multiwav}) does not show the hole clearly in the same manner as \siofivefour\ and \cotwoone, the hole is also visible in the central channels ($v = 0 -300$ \kms) of the velocity map (Fig.~\ref{fig:chanmaps_sio}).
Because of the additional $-$600--0~\kms\ components located within the same line of sight as the hole (Fig.~\ref{fig:chanmaps_sio}) in the integrated maps, the hole is not clear in the \siosixfive\ map.
The CO and SiO molecular hole is just to the south of the `keyhole' that is seen in \halpha\ (Fig. 8 of \citet{Fransson2015}; top right panel of our Fig.~\ref{fig:multiwav}), though the molecular hole appears to be slightly smaller in scale and located on the southern edge of the hole in \halpha\ emission.
The centers of the holes are offset by $\sim$50 mas, or $\sim$4$\times$ the astrometric and alignment errors. 

\cotwoone\ and \siofivefour\ have similar structures in the integrated images, however the spatial distributions of the higher transitions of each species have some differences. 
\siosixfive\ is more evenly distributed in a shell pattern while the lower S/N image of \cosixfive\ appears clumpy (Fig.~\ref{fig:multiwav}), though this is likely affected by the noise. 

\cosixfive\ has emission coincident with the \cotwoone\ hole, in that its channel maps (Fig.~\ref{fig:chanmaps_co}) show emission around the hole location, albeit at low S/N.
However, the integrated spatial distribution appears different from \cotwoone.
The brightness peaks are distributed differently, and the hole is not visible in the integrated \cosixfive\ map due to some emission at those coordinates in the 600--900 \kms\ channels (the far side).
The presence of a molecular hole in \siosevensix\ cannot be confirmed in these data, as the systemic line center ($v_\mathrm{LSRK} \sim$300 \kms) falls at the edges of two sidebands observed separately, which were concatenated during reduction, and suffers from roll-off at the edge of the spectral window; the resulting S/N is poor in that channel.
The other molecular lines do not share this limitation as they fell well within the sideband spectral windows.
We do note a peak of \siosevensix\ emission, however -- the brightest source of emission in the entire cube -- overlapping with the spatial location of the hole and the dust blob but offset from the systemic velocity by \mbox{$\sim -400$ \kms} (this corresponds to the 0 \kms\ channel of Fig.~\ref{fig:chanmaps_sio}). 

The resolved dust peak (small 5$\sigma$ contour in Fig.~\ref{fig:multiwav}) is co-located with the molecular hole in the low transitions of CO and SiO, and slightly extends to the north and east into the relative depression visible in the \siofivefour\ channels near the systemic velocity.
The brightest points of dust emission tend to coincide with relative depressions in the \cosixfive\ brightness, giving the appearance of an anti-correlation between the main dust and \cosixfive\ features.
This is more clearly demonstrated in Fig.~\ref{fig:3color}, where the dust (red) and \cosixfive\ (blue) images are overlaid.
The individual images were normalized independently to emphasize the main features of each, with the visible colors shown roughly corresponding to areas of S/N$>$3.
The gold and teal lines are guides highlighting the highest-S/N features of the dust and CO, respectively, in order to compare peaks in the emission.
While there is some overlap in the faint features of the dust and CO, and the southern extent of the dust peak starts to fade to a combined magenta, the dust and \cosixfive\ peaks do not generally overlap.
Rather, the brightest dust features are located in areas of relatively faint \cosixfive\ emission and vice-versa.

To test whether the apparent dust-CO anti-correlation is an artifact or result of the data reduction or continuum subtraction, we performed several checks.
First, the Band--9 dust continuum was reconstructed in different ways, by imaging (\textsc{Casa} \texttt{mfs}-mode) in a variety of spectral windows and also by making a data cube across the entirety of Band--9 (including the CO line), and fitting the continuum emission.
These techniques gave consistent results. 
Initially we used \casa\ to subtract a (zeroth order) continuum in the visibility plane.
We compared this with an order-0 subtraction in the image plane, and found no significant differences. 
This means that the structures seen in our final \cosixfive\ map are robust to variations in how the continuum is determined and subtracted.
The anti-correlation in \cosixfive\ is visible, even before continuum subtraction, in the \cosixfive\ dirty map (i.e., with no cleaning to deconvolve the interferometer sidelobes). 
Thus the apparent anti-correlation seen in the dust and CO distributions is robust to changes in the data processing. 
Lastly, we test whether the anti-correlation is statistically robust by calculating the weighted version of the normalized cross-correlation function $\sum \mathrm{cov}_{XY}/\sqrt{\mathrm{cov}_{XX} \cdot \mathrm{cov}_{YY}}$, which returns a standard correlation measure $r$ between the range --1 and +1.  
The pixel weights used were the map (S/N)$^2$.  
The resulting correlation measure is $r$=--0.30 $\pm$ 0.08 -- a moderate anti-correlation -- using the accuracy estimate from, e.g., \cite{Frick2001}.  
Due to the relatively low S/N in these images and therefore scatter in pixel-by-pixel comparisons, $r$ will always be pulled closer to 0 and will not approach $\pm$1. 

We tested the robustness of the correlation measure by investigating different angular scales using the wavelet analysis described in \cite{Frick2001} and \cite{Arshakian2016}.  
On scales of 1--2 beam widths (i.e., convolutions with kernels of those scales), the images start to become increasingly positively correlated, which is expected due to the peaks being separated by that amount ($\sim 5 \times$ the astrometry error).  
Below these scales, $r$ remains negative, so the anti-correlation is not sensitive to small changes in image resolution. 
Using the same wavelet analysis on the other images, the dust correlates more positively with the other CO and SiO lines than with \cosixfive. 
The standard correlation measures agree, with $r$=+0.04 for \cotwoone\ and $r$=+0.36 for \siosixfive.  

The brightest dust feature is located one beam width northeast of the secondary \cosixfive\ peak, and the brightest \cosixfive\ features curve around the main dust peak.
The CO peaks are obvious because the line emission is brighter than the continuum (c.f. the integrated profile in Fig.~\ref{fig:ejectavsring_b679}), but there is some fainter (S/N$<$5) \cosixfive\ emission overlapping with some of the dust emission.
There is also low-level (S/N$<2$) dust emission that roughly spans the full extent of the ejecta.
The other molecular species further complicate the picture, as noted earlier -- the peak of \siosevensix\ is coincident with the dust peak and the \siofivefour\ hole (as seen in the bottom right panel of Fig.~\ref{fig:multiwav}).
At the southern edge of the hole, some faint \halpha\ emission appears to be aligned with \cotwoone\ in projection, but their velocity ranges differ. 
The 3-D view of \halpha\ \cite[][their Figure 6]{Larsson2016} indicates that the \halpha\ emission in this region peaks at velocities around --1500 \kms\ while the peak \cotwoone\ is between 0 and +200 \kms.  
While no velocity information is available for the dust continuum emission, it is spatially offset from the nearby \halpha\ and \cotwoone\ by $\sim$1 dust resolution element.  
That is, the \cotwoone\ and \halpha\ in this region are offset in velocity, and the dust peak is spatially offset from both.  

To summarize, we find that the dust emission is clumpy.
The Band--9 image enables us to resolve the dust in the ejecta to angular scales of 62$\times$81 mas.  
The peaks of the ejecta CO and SiO emission are not cospatial with the peaks of the ejecta dust emission (with anti-correlated \cosixfive\ and dust structures).
The small peak/clump in the dust emission revealed in the Band 9 image (the blob) overlaps with holes previously observed in the lower line transitions of SiO and the CO molecular ejecta, and is coincident with some emission observed in the \siosevensix\ line.

\section{The Spectral Energy Distribution of SN~1987A}
\label{sec:analysis_SED}

The three physical mechanisms primarily responsible for emission in the FIR--radio portion of the continuum are thermal IR greybody emission from dust (from the ejecta region - \citealp{Matsuura2011,Indebetouw2014}), non-thermal radio/mm synchrotron emission (from the ring - \citealp{Manchester2007,Potter2009,Zanardo2010,Lakicevic2012b,Zanardo2013,Indebetouw2014}) and a lesser contribution from free-free mm/sub-mm bremsstrahlung emission from hot ionized material (see \citealt{Zanardo2014} for a full review of the different components).
In this Section we measure the photometry, analyze the emission from dust in the ejecta using the ALMA data, investigate the properties derived using a variety of dust models from the literature, and investigate the synchrotron emission in the ring.

\subsection{Photometry}
\label{sec:photometry}

%%%%%%%%%% Photometry data table %%%%%%%%%%%
\begin{deluxetable*}{ cccccccc }[t!]
\tablecaption{ Continuum Photometry   \label{table:Photometry}} %Brief caption (title)
\tablehead{  \colhead{$\nu_{c}$} & \colhead{$\Delta \nu$} & \colhead{S$_\nu$} & \multicolumn{2}{c}{Aperture Center} & \colhead{R$_\mathrm{MAJ}$\tablenotemark{$\dag$}} & \colhead{R$_\mathrm{MIN}$\tablenotemark{$\dag$}} & \colhead{P.A.} \\ (GHz) & (GHz) & (mJy) & RA (deg) & DEC (deg) & (arcsec) & (arcsec) & (deg)  }
\startdata
\\[-0.2cm]\multicolumn{8}{c}{Ejecta} \\ \hline 
225.50 & 3.0 & 0.5 $\pm$ 0.2 $^{+0.3}_{-0.0}$ & 83.866586 & -69.269733 & 0.32 & 0.29 & 125 \\ 
240.50 & 5.0 & 0.7 $\pm$ 0.2 $^{+0.4}_{-0.1}$ & 83.866600 & -69.269720 & 0.37 & 0.36 &  75 \\ 
247.50 & 3.0 & 1.1 $\pm$ 0.4 $^{+0.6}_{-0.1}$ & 83.866639 & -69.269739 & 0.41 & 0.35 & 125 \\ 
270.00 & 1.0 & 1.2 $\pm$ 0.6 $^{+0.7}_{-0.1}$ & 83.866652 & -69.269709 & 0.40 & 0.31 &  60 \\ 
279.00 & 2.0 & 1.3 $\pm$ 0.4 $^{+0.8}_{-0.1}$ & 83.866662 & -69.269747 & 0.41 & 0.40 & 125 \\ 
306.76 & 1.4 & 1.8 $\pm$ 0.6 $^{+1.1}_{-0.2}$ & 83.866667 & -69.269753 & 0.42 & 0.36 & 125 \\ 
315.68 & 7.6 & 1.8 $\pm$ 0.6 $^{+1.1}_{-0.2}$ & 83.866667 & -69.269753 & 0.42 & 0.36 & 125 \\ 
679.22 & 11.5 & 36.2 $\pm$ 7.2 $^{+7.2}_{-7.2}$ & 83.866728 & -69.269740 & 0.42 & 0.36 & 125 \\[0.2cm] 
\multicolumn{8}{c}{Ring} \\ \hline 
225.50 & 3.0 & 13.0 $\pm$ 0.5 $^{+7.8}_{-1.3}$ & 83.866585 & -69.269731 & 1.45, 0.35 & 1.35, 0.33 & 175 \\ 
240.50 & 5.0 & 12.5 $\pm$ 0.5 $^{+7.5}_{-1.3}$ & 83.866600 & -69.269722 & 1.45, 0.42 & 1.35, 0.39 & 175 \\ 
247.50 & 3.0 & 13.4 $\pm$ 0.8 $^{+8.1}_{-1.3}$ & 83.866630 & -69.269736 & 1.45, 0.43 & 1.35, 0.40 & 175 \\ 
270.00 & 1.0 & 13.7 $\pm$ 1.1 $^{+8.2}_{-1.4}$ & 83.866592 & -69.269720 & 1.45, 0.45 & 1.35, 0.42 & 175 \\ 
279.00 & 2.0 & 12.9 $\pm$ 0.8 $^{+7.7}_{-1.3}$ & 83.866661 & -69.269747 & 1.45, 0.44 & 1.35, 0.41 & 175 \\ 
306.76 & 1.4 & 12.0 $\pm$ 1.5 $^{+7.2}_{-1.2}$ & 83.866666 & -69.269753 & 1.45, 0.45 & 1.35, 0.42 & 175 \\ 
315.68 & 7.6 & 11.3 $\pm$ 1.7 $^{+6.8}_{-1.1}$ & 83.866667 & -69.269753 & 1.45, 0.45 & 1.35, 0.42 & 175 \\ 
679.22 & 11.5 & 19.4 $\pm$ 19.3 $^{+3.9}_{-3.9}$ & 83.866720 & -69.269737 & 1.45, 0.45 & 1.35, 0.42 & 175 \\[0.2cm] 
\multicolumn{8}{c}{Total System} \\ \hline 
225.50 & 3.0 & 13.5 $\pm$ 0.5 $^{+8.1}_{-1.3}$ & 83.866585 & -69.269731 & 1.45 & 1.35 & 175 \\ 
240.50 & 5.0 & 13.2 $\pm$ 0.5 $^{+7.9}_{-1.3}$ & 83.866600 & -69.269722 & 1.45 & 1.35 & 175 \\ 
247.50 & 3.0 & 14.6 $\pm$ 0.7 $^{+8.7}_{-1.5}$ & 83.866630 & -69.269736 & 1.45 & 1.35 & 175 \\ 
270.00 & 1.0 & 15.4 $\pm$ 0.8 $^{+9.2}_{-1.5}$ & 83.866592 & -69.269720 & 1.45 & 1.35 & 175 \\ 
279.00 & 2.0 & 14.2 $\pm$ 0.8 $^{+8.5}_{-1.4}$ & 83.866661 & -69.269747 & 1.45 & 1.35 & 175 \\ 
306.76 & 1.4 & 13.9 $\pm$ 1.7 $^{+8.3}_{-1.4}$ & 83.866666 & -69.269753 & 1.45 & 1.35 & 175 \\ 
315.68 & 7.6 & 13.3 $\pm$ 1.8 $^{+8.0}_{-1.3}$ & 83.866667 & -69.269753 & 1.45 & 1.35 & 175 \\ 
679.22 & 11.5 & 55.5 $\pm$ 20.9 $^{+11.1}_{-11.1}$ & 83.866720 & -69.269737 & 1.45 & 1.35 & 175 
\enddata

\tablecomments{Integrated flux densities of the ejecta, ring, and total system for each continuum band with central frequency $\nu_c$ and bandwidth $\Delta \nu_c$.
Integrated flux densities are quoted as \textit{value} $\pm$ \textit{measurement uncertainty} $\pm$ \textit{systematic uncertainty}.
The systematic uncertainty includes calibration uncertainties (10\% in Bands 6\&7 or 20\% in Band 9), and an additional 50\% on the positive side in Bands 6\&7 due to the systematic offset from Cycle 0 and Cycle 6 levels. 
Apertures for the ejecta and total system are ellipses, apertures for the ring are annuli.  
The center coordinates are in ICRS.
\tablenotetext{$\dag$}{The ring annulus radii are given as (R$_\mathrm{outer}$, R$_\mathrm{inner}$).}
}
\end{deluxetable*}

\begin{figure*}
\centering
\includegraphics[trim=2mm 0mm 0mm
  0mm,clip=true,width=1.0\textwidth]{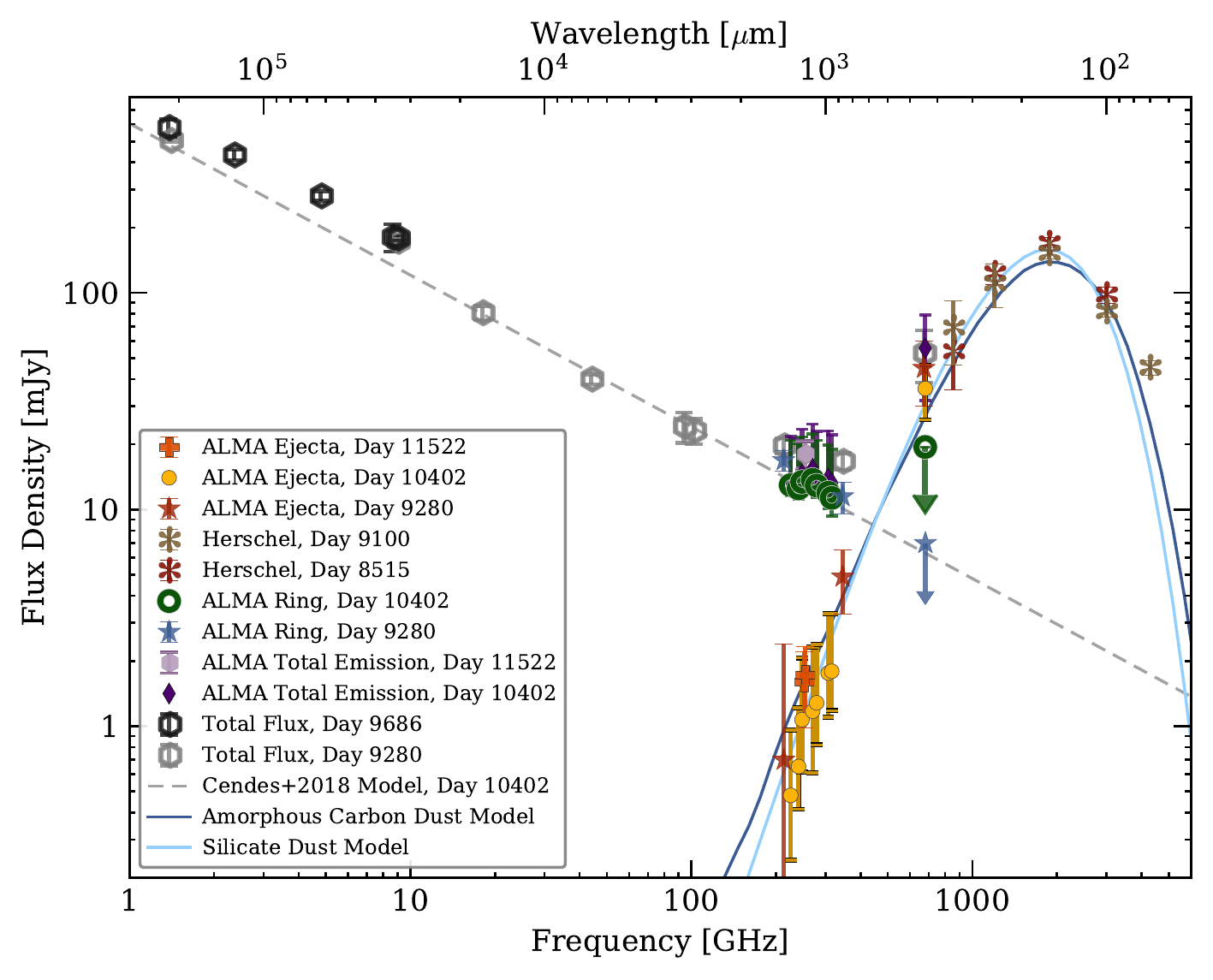}
\caption{
Continuum values from 225--679~GHz for the integrated ejecta (yellow), ring (green), and total system (purple) from this work, for observations taken an average of 10,402 days after the SN explosion.
For reference, ALMA Cycle 0 corresponds to day 9280, Cycle 2 is day 10402, and Cycle 6 is day 11522.
The higher angular resolution ALMA data confirm the ejecta emission follows a thermal dust profile down to $\sim$200~GHz.
The ring emission, on the other hand, shows no evidence of submm dust emission but instead is consistent with a synchrotron emission profile -- ATCA observations from Days 9280--9686 \citep{Zanardo2014,Callingham2016} combined with the Cycle 2 ALMA ring flux densities give a power-law index of $\alpha = -0.70 \pm 0.06$. 
The dark blue and cyan lines show SED fits for amorphous carbon \citep[ACAR sample,][]{Zubko1996} and silicate \citep[forsterite,][]{Jaeger2003} dust emission models, respectively, demonstrating that disparate models give reasonable fits to the data.
The previous ALMA ejecta measurements from \cite{Zanardo2014} are shown as red stars, and previous ALMA ring measurements from \cite{Zanardo2014} are denoted as blue stars.
Preliminary Cycle 6 ALMA flux densities (Matsuura et al., \textit{in preparation}) are shown as the orange crosses and light purple hexagons for the ejecta and total emission, respectively.  
The positive error bars for the Cycle 2 Band 6\&7 data include an additional 50\% of the flux density values to reflect the observed systematic offset from the Cycle 0 and Cycle 6 levels.
The empty grey hexagons are measures of the total system flux density in various parts of the spectrum at day 9280: 1.4\,GHz, 18\,GHz, 44\,GHz \citep{Zanardo2013}; 9\,GHz \citep{Ng2013}; 94\,GHz \citep{Lakicevic2012b}; 102, 213, 345, \& 672\,GHz \citep{Zanardo2014}. 
The empty black hexagons represent the ATCA total system flux densities between 1--9~GHz at day 9686 \citep{Callingham2016}. 
The ring flux density at day 9280 (blue stars) is larger than ALMA total emission (purple diamonds) at day 10402 due to the ALMA band 7 systematic error discussed in Section~\ref{sec:photometry}.
The brown asterisks are the unresolved \herschel\ 70--500\um\ flux densities \citep{Matsuura2015}, and the crimson asterisks are the 2010 HERITAGE flux densities \citep{Meixner2013, Matsuura2011}. 
The ring emission model S($\nu$,t) from \cite{Cendes2018} for day 10402 and $\alpha=-0.70$ is shown by the dashed grey line.
}
\label{fig:ejectavsring_continuumpoints}
\end{figure*}

The continuum bands are defined as those frequencies that are molecular line free, as demonstrated in Fig.~\ref{fig:ejectavsring_b679} and \S\ref{sec:continuumdata}, with the chosen frequency windows summarized in Table~\ref{table:Photometry}; these bands are different from the default ALMA wide band continuum. 
The centers of the apertures used for deriving photometry are the same as described in Section~\ref{sec:analysis} (Appendix~\ref{app:center}), with elliptical apertures selected to encompass the ejecta and the ring annulus with varying sizes in each ALMA band in order to include only the relevant signal (Table~\ref{table:Photometry}). 
For the 315 GHz ejecta, this results in an elliptical aperture with semi-major and semi-minor axes of  $r_\mathrm{MAJ} \times r_\mathrm{MIN}$ = 0\farcs42$\times$0\farcs36 and major axis P.A. of 25$^\circ$ (N through E).
For the 315 GHz circumstellar ring, an elliptical annulus with inner $r_\mathrm{MAJ}$ and $r_\mathrm{MIN}$ of 0\farcs45 $\times$ 0\farcs42, outer $r_\mathrm{MAJ}$ and $r_\mathrm{MIN}$ of 1\farcs45 $\times$ 1\farcs35, and 85$^{\circ}$ P.A. from N was chosen.
The extents of the regions were selected independently across the different bands to best match the features in each image, but only vary slightly.
For comparison with previous lower spatial resolution observations, the total system emission is also calculated by summing emission within an ellipse defined by the outer ring extent above.
This total system integration includes the contribution from the gap between the ring and the ejecta.

The Cycle 2 images, aside from Band 9, exhibit a slight decrease in integrated flux density for radii just beyond the outer edge of the ring, which could be due to undersampled flux or errors in calibration or deconvolution. 
We take the RMS of flux densities in background pixels from a large annulus beyond the ring to estimate the level of these effects, and the resulting uncertainty contribution is typically of order a few percent of the flux density.

As the reconstruction of the images may propagate systematic as well as random noise in the background, we have used our images to make a series of measurements using the same aperture as for the source.
The distribution of these measurements is roughly gaussian, and thus we adopt the RMS of this distribution as our aperture error, $\sigma_\mathrm{AP}$.

An additional empirical uncertainty component was included to account for the potential smearing of ring emission into the ejecta aperture.
This was estimated by taking the average deviation in the flux density after expanding and shrinking the semimajor and semiminor axes by 0\farcs1, a size which covers reasonable large differences in aperture choice yet avoids significant overlap between the ejecta and ring.

All of these uncertainties were added in quadrature to estimate the overall uncertainty in a given band (typically $\sim$15\%, dominated by the random-position aperture uncertainty).
We include an additional 10\% uncertainty for systematic (calibration) error in Bands 6 \& 7, and 20\% in Band--9\footnote{ALMA Cycle 2 Technical Handbook, \\ \url{https://almascience.eso.org/documents-and-tools/cycle-2/alma-technical-handbook/}}.
Finally, we considered the possibility that we may be missing diffuse emission from cold dust within the SN structure due to over-resolving an extended source.
To address this issue, we simulated observations for multiple synthetic sources resembling SN~1987A but with varying extended ellipse components, and found that this effect is at a level below the ALMA systematic uncertainties.
The reader is referred to Appendix~\ref{app:casasimulations} for more details.

The ejecta, ring and total system flux densities and uncertainties are shown in Fig.~\ref{fig:ejectavsring_continuumpoints} as gold circles, green rings, and purple diamonds, respectively, and their values are listed in Table~\ref{table:Photometry}.
Previous measurements are also shown in Fig.~\ref{fig:ejectavsring_continuumpoints} for reference.
Preliminary Cycle 6 flux densities (Matsuura et al., \textit{in preparation}) from 11,522 days after the explosion, are included here.
The ejecta flux density is 1.6\,mJy at 252.4\,GHz, and 1.7\,mJy at 254.3\,GHz.
The total system flux densities at these frequencies are 17.9\,mJy and 18.1\,mJy, respectively.
The uncertainty on each of these flux density measurements is estimated as 0.4\,mJy.

We note that our Cycle 2 total flux densities in Bands 6 and 7 are systematically lower than the ALMA Cycle 0 and 6 flux densities, though they agree within the error bars.
They are typically around 50\% lower than the equivalent levels from Cycles 0 and 6, therefore we include an additional 50\% to their positive systematic uncertainties.
Potential causes, as noted in \S~\ref{sec:observations}, include decorrelation from poor weather or a mis-scaling of the flux calibrator. 
The integrated 350 and 360 GHz ejecta flux densities are particularly low, either due to inherently low flux at this epoch, or due to data quality.
As mentioned in \S~\ref{sec:observations}, weather affected the phase stability of observations between 346--362 GHz, which can result in decorrelation and therefore reduced flux recovery.
As these measurements are less reliable, they have been omitted from the remainder of this study. 
We note that the systematic offset will not have affected the analysis of the resolved dust distribution discussed in Section~\ref{sec:analysis}, since the offset is not seen in the Band 9 data where the dust peak (the blob) was identified. 

The literature values for the total SN~1987A flux densities at various wavelengths are shown as grey hexagons, and represent the overall spectral energy distribution (SED) of the system.
The total emission at 1.4\,GHz, 18\,GHz, 44\,GHz \citep{Zanardo2013}, 9\,GHz \citep{Ng2013}, 94\,GHz \citep{Lakicevic2012b}, and 102\,GHz \citep{Zanardo2014} is dominated by synchrotron emission from the ring.
The total emission at 213, 345, \& 672\,GHz \citep{Zanardo2014}, on the other hand, gradually consists of a higher and higher fraction of thermal emission until that is dominant in the submm and the FIR.
As the synchrotron brightness increases in time \citep{StaveleySmith2014,Cendes2018}, these literature flux densities were scaled to their levels at day 9280 by \citet{Zanardo2014} to match the average epoch of the ALMA cycle 0 observations.
The details of the ejecta and ring portions of the SED will be discussed in turn in the following two sections.

\subsection{Modified Blackbody Fits to the Ejecta Dust Emission}
\label{sec:modBBfits}

\subsubsection{Description of the modified blackbody fits}

\begin{deluxetable*}{lllcccc}[h!]
\tablecaption{ Modified Blackbody Fits \label{table:modBBfits_kappacurves}} 

\tablehead{\colhead{Dust} & \colhead{Reference} & \colhead{Grain Density} & \colhead{M$_\mathrm{dust}$} & \colhead{T} & \colhead{$\kappa_{850}$ } & \colhead{Good Fit} \\
  &   &  (g cm$^{-3}$)  & (M$_\mathrm{dust}$ ) & (K) & (m$^2$ kg$^{-1}$) }

\startdata
amC (ACH2 sample), Mie 0.1$\mu$m & \cite{Zubko1996} & 1.81 & 1.46 $^{+ 0.09 }_{- 0.08 }$ & 17.5 $^{+ 0.1 }_{- 0.1 }$ & 0.087 & Y \\
amC (ACAR sample), Mie 0.1$\mu$m & \cite{Zubko1996} & 1.81 & 0.38 $^{+ 0.02 }_{- 0.02 }$ & 22.0 $^{+ 0.2 }_{- 0.2 }$ & 0.254 & N \\
amC (BE sample), Mie 0.1$\mu$m & \cite{Zubko1996} & 1.81 & 0.77 $^{+ 0.05 }_{- 0.04 }$ & 20.7 $^{+ 0.2 }_{- 0.2 }$ & 0.141 & N \\
amC (AC1 sample), Mie 0.1$\mu$m & \cite{Rouleau1991} & 1.85 & 0.43 $^{+ 0.03 }_{- 0.03 }$ & 21.6 $^{+ 0.3 }_{- 0.3 }$ & 0.203 & Y \\
Cellulose (800K sample), Mie 0.1$\mu$m & \cite{Jaeger1998} & 1.81 & 0.46 $^{+ 0.03 }_{- 0.03 }$ & 18.8 $^{+ 0.2 }_{- 0.2 }$ & 0.178 & Y \\
Graphite, Mie 0.1$\mu$m & \cite{Draine1984} & 2.26 & 1.62 $^{+ 0.11 }_{- 0.10 }$ & 17.8 $^{+ 0.2 }_{- 0.2 }$ & 0.069 & Y \\
PAH (neutral), Mie 0.01$\mu$m & \cite{Laor1993} & 2.24 & 1.69 $^{+ 0.11 }_{- 0.10 }$ & 18.0 $^{+ 0.2 }_{- 0.2 }$ & 0.071 & Y \\
Silicate -- Enstatite, Mie 0.1$\mu$m & \cite{Jaeger2003} & 2.71 & 4.10 $^{+ 0.23 }_{- 0.24 }$ & 18.0 $^{+ 0.2 }_{- 0.2 }$ & 0.029 & Y \\
Silicate -- Forsterite, Mie 0.1$\mu$m & \cite{Jaeger2003} & 3.2 & 4.03 $^{+ 0.26 }_{- 0.25 }$ & 17.9 $^{+ 0.2 }_{- 0.2 }$ & 0.029 & Y \\
Silicate -- ``Cosmic'', Mie 0.1$\mu$m & \cite{Jaeger1994} & 3.2 & 3.46 $^{+ 0.24 }_{- 0.22 }$ & 17.7 $^{+ 0.2 }_{- 0.2 }$ & 0.034 & Y \\
Silicate/Carbon Composite CDE -- f$_\mathrm{C}$=0.18 & \cite{Dwek2015} & 2.95 & 0.38 $^{+ 0.02 }_{- 0.02 }$ & 21.1 $^{+ 0.2 }_{- 0.2 }$ & 0.270 & N  \\ 
Silicate -- LMC Average & \cite{Weingartner2001} & \nodata & 2.49 $^{+ 0.15 }_{- 0.14 }$ & 18.0 $^{+ 0.2 }_{- 0.2 }$ & 0.047 & Y \\
Silicate -- 30K Average & \cite{Demyk2017} & \nodata & 0.27 $^{+ 0.02 }_{- 0.02 }$ & 18.5 $^{+ 0.2 }_{- 0.2 }$ & 0.265 & Y \\
Silicate -- Composite Aggregate & \cite{Semenov2003} & \nodata & 35.06 $^{+ 2.18 }_{- 2.18 }$ & 21.4 $^{+ 0.2 }_{- 0.2 }$ & 0.003 & Y \\
Silicate -- Porous Multilayer Spheres & \cite{Semenov2003} & \nodata & 6.13 $^{+ 0.36 }_{- 0.37 }$ & 29.4 $^{+ 0.4 }_{- 0.4 }$ & 0.010 & N \\
Silicate -- Bare Grains, 0.03 Myr & \cite{Ormel2011} & \nodata & 0.50 $^{+ 0.03 }_{- 0.03 }$ & 20.7 $^{+ 0.3 }_{- 0.2 }$ & 0.180 & Y \\
Silicate -- Icy Grains, 0.03 Myr & \cite{Ormel2011} & \nodata & 0.60 $^{+ 0.04 }_{- 0.04 }$ & 18.3 $^{+ 0.2 }_{- 0.2 }$ & 0.184 & Y \\
Silicate -- Naked Grains, n$_H = 10^5$ & \cite{Ossenkopf1994} & \nodata & 0.64 $^{+ 0.04 }_{- 0.04 }$ & 20.9 $^{+ 0.2 }_{- 0.2 }$ & 0.163 & N \\
Silicate -- Thin Ice Mantles, n$_H = 10^5$ & \cite{Ossenkopf1994} & \nodata & 0.74 $^{+ 0.04 }_{- 0.04 }$ & 19.1 $^{+ 0.2 }_{- 0.2 }$ & 0.142 & Y \\
SiC, Mie 0.1$\mu$m & \cite{Pegourie1988} & 3.22 & 1.57 $^{+ 0.08 }_{- 0.09 }$ & 23.3 $^{+ 0.2 }_{- 0.2 }$ & 0.058 & N \\
FeS, Mie 0.1$\mu$m & \cite{Henning1996} & 4.83 & 0.68 $^{+ 0.04 }_{- 0.04 }$ & 34.7 $^{+ 0.6 }_{- 0.6 }$ & 0.086 & N \\
FeO, Mie 0.1$\mu$m & \cite{Henning1995} & 5.7 & 0.28 $^{+ 0.01 }_{- 0.02 }$ & 28.2 $^{+ 0.4 }_{- 0.3 }$ & 0.259 & N \\
SiO$_2$, Mie 0.1$\mu$m & \cite{Henning1997} & 2.196 & 4.41 $^{+ 0.39 }_{- 0.33 }$ & 17.5 $^{+ 0.2 }_{- 0.2 }$ & 0.022 & N \\
TiO$_2$, Mie 0.1$\mu$m & \cite{Posch2003} & 3.78 & 81.77 $^{+ 5.10 }_{- 4.49 }$ & 17.7 $^{+ 0.2 }_{- 0.2 }$ & 0.001 & Y \\
Al$_2$O$_3$ ``Compact'' sample, Mie 0.1$\mu$m & \cite{Begemann1997} & 3.2 & 0.90 $^{+ 0.06 }_{- 0.06 }$ & 19.0 $^{+ 0.2 }_{- 0.2 }$ & 0.112 & Y \\
NaAlSi$_2$O$_6$, Mie 0.1$\mu$m & \cite{Mutschke1998} & 2.4 & 0.10 $^{+ 0.01 }_{- 0.01 }$ & 23.1 $^{+ 0.3 }_{- 0.3 }$ & 0.982 & Y \\
MgAl$_2$O$_4$, Mie 0.1$\mu$m & \cite{Fabian2001} & 3.64 & 121.74 $^{+ 7.54 }_{- 7.48 }$ & 17.7 $^{+ 0.2 }_{- 0.2 }$ & 0.001 & Y \\
Pure Iron, Mie 0.1$\mu$m & \cite{Henning1996} & 7.87 & 3.97 $^{+ 0.28 }_{- 0.23 }$ & 19.9 $^{+ 0.2 }_{- 0.2 }$ & 0.025 & Y
\enddata
\tablecomments{
Mass and temperature fits for greybodies for a selection of 28 of the dust models discussed in the text.
amC: amorphous carbon.
Fit values and uncertainties are from bootstrap resampling of the data within their error bars, and are determined from the 50\textsuperscript{th}, 84\textsuperscript{th}, and 16\textsuperscript{th} percentiles of the distributions of fits from 1000 samplings of the observed flux densities.
For $\kappa(\lambda)$ models derived from Mie theory, grains of radius $a=0.1 \mu$m were assumed except for the case of PAHs, where $a=0.01\mu$m was used.
Grain densities are given for the Mie and CDE cases: amorphous carbon from \cite{Zubko2004} (and \cite{Rouleau1991} for their AC1 sample), graphite and SiC from \cite{Laor1993}, PAHs from \citep{Li2001}, stoichiometric varieties of olivines and pyroxenes from \cite{Henning1996}, silicate/carbon composite CDE with carbon volume filling factor f$_\mathrm{C}$=18\% from \cite{Dwek2015}, jadeite from \cite{Mutschke1998}, and in general from the Jena optical constants database\footnote{\url{https://www.astro.uni-jena.de/Laboratory/OCDB/index.html}} \citep{Henning1999}.
$\kappa_{850}$, the mass absorption constant at 850\um, is listed for each model, extrapolated as power-laws for models where wavelength coverage falls short.
The quality of fit is denoted in the rightmost column where a good fit is defined as $\chi_\nu^2<2$.}
\end{deluxetable*}

Fig.~\ref{fig:ejectavsring_continuumpoints} displays the mm to FIR SED. In this figure, the brown asterisks show the FIR flux densities measured by \herschel\ for the total SN~1987A (unresolved) system, and the gold circles show the mm flux densities from this work measured with ALMA for the resolved ejecta. The shape of the SED shows that the ejecta emission arises from thermal (dust) radiation, all the way into the mm, confirming the results of \citet{Zanardo2014} and \citet{Matsuura2015}.

The next step is to fit the thermal dust emission using dust models.  
In order to cover the peak of the thermal emission, we use the \herschel\ flux densities from \citet{Matsuura2015} in our model fits since they are measuring the emission from the ejecta dust, albeit unresolved.
Two \herschel\ flux densities are treated as upper limits: 70\um, as it is possibly contaminated by warm ring dust \citep{Matsuura2019} and/or \oi\ 63\um\ emission; and 500\um, as it was a non-detection.
One potential issue with using the \herschel\ flux densities is that the \herschel\ measurements were obtained at an average of $\sim$1300 days before the ALMA cycle 2 data, and the FIR emission could potentially vary over time.
The heating source of the ejecta was suggested to be primarily from $^{44}$Ti decay, which has an estimated lifetime of 85 years \citep[][also see later \citet{Matsuura2011}]{Ahmad2006,Jerkstrand2011}.
The predicted decrease in this decay energy between the 2012 \herschel\ and 2015 ALMA observations is 4.2\%.  
Assuming the FIR luminosity decreased by this amount between the 2012 and 2015 epochs, the reduction in the temperature of a 20K blackbody would be $\sim$0.2K, translating into individual \herschel\ flux density decrements of 2--5\%.
This is several times smaller than the uncertainties on the PACS and SPIRE flux densities.
Therefore, if the ejecta heating is dominated by $^{44}$Ti decay, the use of the \herschel\ flux densities with the latest ALMA measurements is valid.
An alternative additional heating source will be discussed in Section 6.3.

We tested the robustness of the SED results using three common parameter estimation techniques: using a maximum likelihood estimation (MLE) with uncertainties determined by bootstrap resampling; bayesian estimation with Markov Chain Monte Carlo (MCMC) posterior distributions, using the \texttt{emcee} package \citep{emcee}; and finally by checking with ordinary least squares (OLS) regression.
The OLS, MLE, and MCMC routines all yield consistent fits for a given dust emission profile.  
In order to take into account the systematic offset in our Band 6\&7 flux densities from other cycles (see \S~\ref{sec:photometry}), we determine our best fits and uncertainties from resampling of the flux densities within their error bars.
The best fit and uncertainties are taken to be the 50$^\mathrm{th}$, 16$^\mathrm{th}$, and 84$^\mathrm{th}$ percentiles of the distributions from 1000 samplings.

The modified blackbody (modBB) function we use follows the form \mbox{$S_\nu(\lambda) = M_{dust} \kappa_{\rm {abs}} (\lambda) B_\nu(\lambda) /d^2$}, where $S_\nu(\lambda)$ is flux density, $\kappa_{\rm {abs}}(\lambda)$ is the mass absorption coefficient of dust grains, $B_\nu(\lambda)$ is the Planck function and $d$ is the distance to SN~1987A.
We assume that the emission is optically thin across this wavelength range.
$\kappa_{\rm {abs}} (\lambda)$ can be directly obtained from the literature for some cases, but for the majority of cases, assuming spherical dust grains of radius $a$ and density $\rho$, $\kappa_{\rm {abs}}(\lambda)$ is defined as $\kappa=3/4 \,\, Q \rho a$, where $Q$ is the absorption efficiency of the dust, which can be calculated from optical constants with Mie theory.  
$\kappa_{\rm {abs}}$ is often assumed to be a power law defined as \mbox{$\kappa_0 \, (\lambda/\lambda_0)^{-\beta}$} (where $\kappa_0$ is the reference value of $\kappa_{\rm {abs}}(\lambda)$ at wavelength $\lambda_0$ and $\beta$ is the power law emissivity index of $\kappa_{\rm {abs}}(\lambda)$).

Fitting a dust greybody to the FIR-mm SED with the power law approximation to $\kappa_{\rm {abs}}$ gives fit parameters of $\beta=2.05^{+0.11}_{-0.10}$, ${M_{\mathrm dust}}$ $=1.53^{+0.13}_{-0.13}$ \msun, and ${T_{\mathrm d}}$ $=17.83^{+0.60}_{-0.57}$\,K  if $\kappa_0 (850\mu \rm{m})$=0.07 m$^2$\,kg$^{-1}$ (using the empirical measurement of $\kappa_{\rm {abs}}$ assuming the fraction of metals locked in dust is constant across the local ISM, \citealp{James2002}).    
As this is a significant amount of dust, here we also fit the SED using a wide variety of compositions and the full characterization of $\kappa_{\rm {abs}}(\lambda)$, following \cite{Indebetouw2014} and \cite{Matsuura2015}.

We perform dust greybody fits to the FIR-mm SED using a wide variety of $\kappa_{\rm {abs}}(\lambda)$ profiles directly obtained from the literature or calculated using Mie theory; in total we used 134 $\kappa_{\rm {abs}}(\lambda)$ profiles.
These include: \cite{Weingartner2001} LMC average (as an approximation to the conditions near SN~1987A in the LMC), \cite{Demyk2017} amorphous silicate samples at 30K, \cite{Ormel2011} calculations of bare and icy silicate+graphite grains, \cite{Ossenkopf1994} bare and icy mantle grains (protostellar core coagulation models), and using Mie scattering calculations for amorphous carbon (amC, \citealp{Zubko1996,Rouleau1991}), graphite \citep{Draine1984}, `cosmic silicates' \citep[amorphous FeMgSiO$_4$,][]{Jaeger1994}, other silicates including enstatite and forsterite from \cite{Henning1996}, pure iron \citep{Henning1996}, and 68 profiles from the Jena database \citep{Henning1999} of minerals including silicates, amorphous carbons, carbides, oxides, and sulfides.
Many of the models in the Jena database only extend to 500\um\ or 1000\um.
In these cases we extrapolate to our longest ALMA wavelength of 1329\um\ in log space (as lines, as most models follow power laws in the submm--mm), from the last 300--500\um\ of each curve to ensure a smooth continuation of the general trend in each.
We also consider a ``continuous distribution of ellipsoids'' (CDE) model for a composite of carbons and silicates \citep{Dwek2015}, with a carbon volume filling factor f$_\mathrm{C}$ of 18\%. 

For Mie theory calculations, we adopt a grain radius of $a=0.01$\um\ for PAHs, and $a=0.1$\um\ for all other models.
For most dust models, Mie-derived FIR $\kappa_{\rm {abs}}$ curves are nearly identical for grain radii of $a < 5$\um.
A representative sample of 28 dust models was selected from this list to span a wide variety of dust types. Selected grain types include \cite{Zubko1996} amorphous carbons (amCs), \cite{Jaeger2003} amorphous silicates, and iron from \cite{Henning1996}, among others.
The fitted masses and temperatures for this sample are listed in Table~\ref{table:modBBfits_kappacurves}.

To determine whether the SED fit is `good', we set a quality threshold whereby $\chi_\nu^2 < 2$.
This on its own can be an insufficient indicator of fit quality, however -- for example, a fit that closely matches the majority of the ALMA flux densities can still satisfy $\chi_\nu^2 < 2$ even if it falls below all of the \herschel\ points due to the nature of the $\chi^2$ metric used in our three fitting methods.
We note that these fit criteria are purely formal and do not consider availability of mass from nucleosynthesis yields; those physical limits are discussed in \S~\ref{sec:SEDdiscussion}.

\subsubsection{Results of the dust fits}

\begin{figure}
\centering
\includegraphics[width=0.5\textwidth]{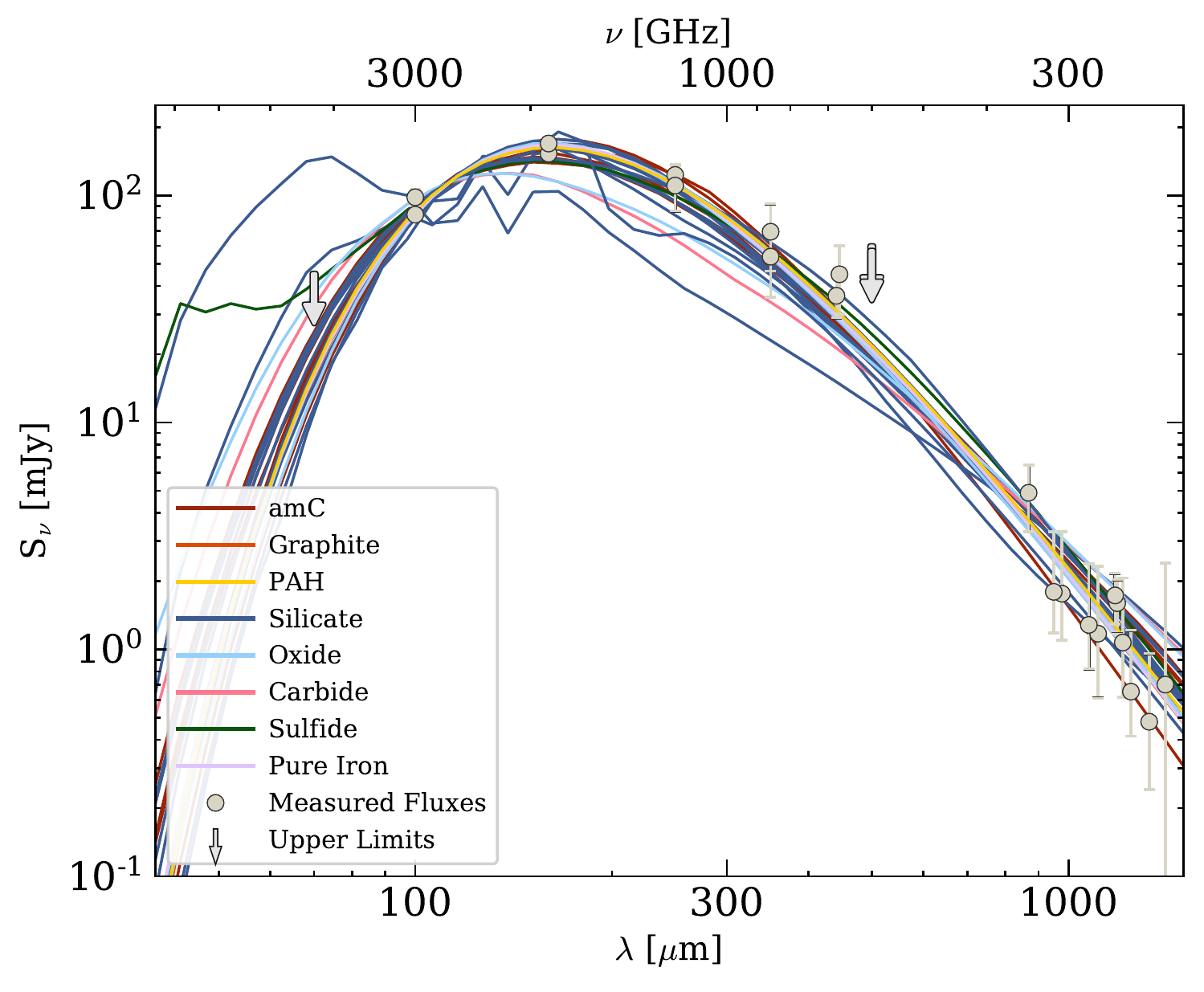}
\includegraphics[width=0.5\textwidth]{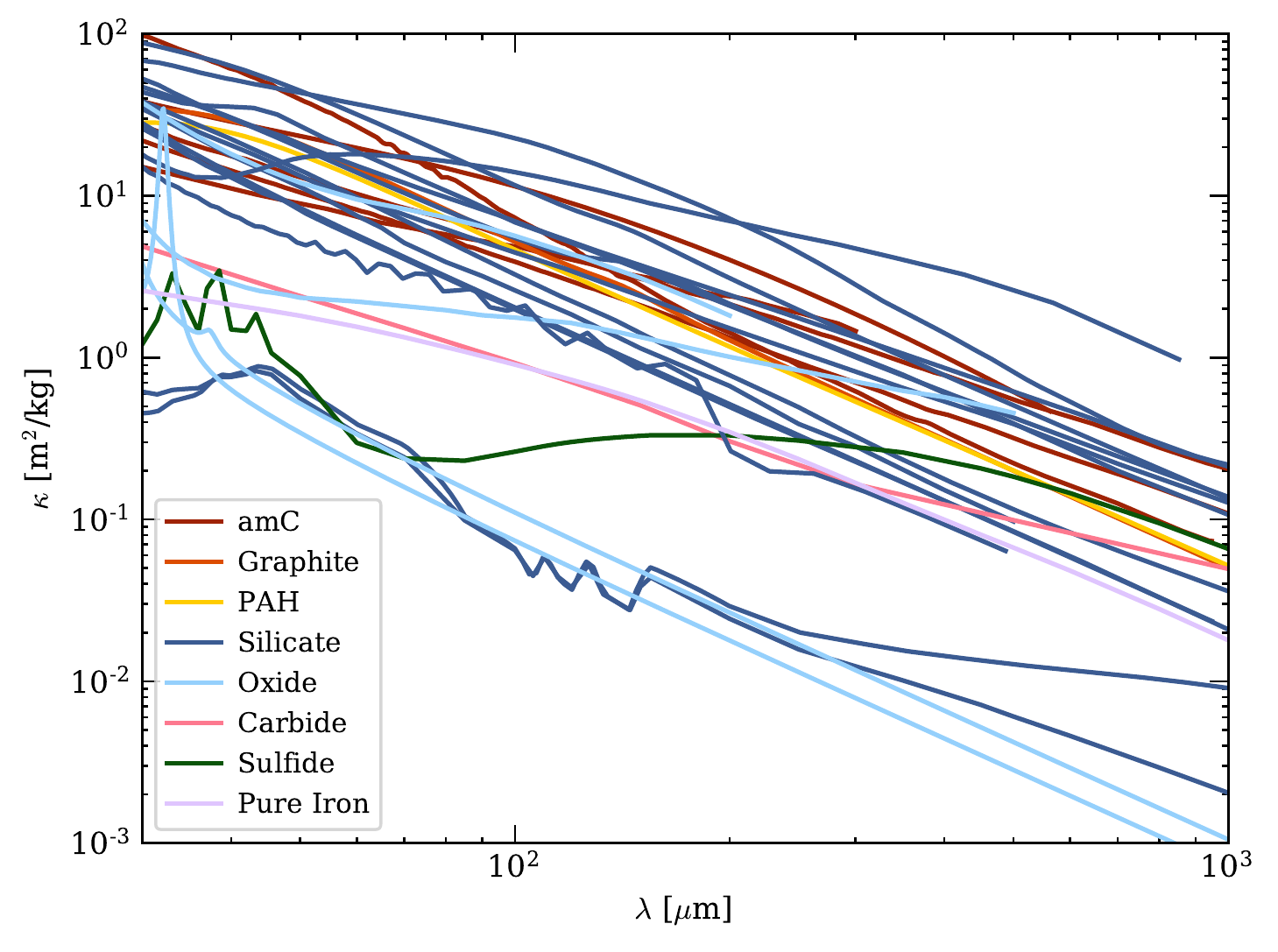}
\caption{
\textit{Top}: Modified blackbody fits to the \herschel\ and ALMA observations of the ejecta continuum in SN~1987A, using the 28 dust varieties and parameters listed in Table~\ref{table:modBBfits_kappacurves}.
The \herschel\ 70\micron\ and 500\micron\ flux densities are used as upper limits.
Most dust models give reasonable fits to the observed flux densities.
The colors denote the mineralogy of the dust models: amorphous carbon models are shown in red, graphite in orange, PAH in yellow, silicates and silicate composites in blue, oxides (including FeO, TiO$_2$, Al$_2$O$_3$, NaAlSi$_2$O$_6$, and MgAl$_2$O$_4$)
in light blue, carbide (SiC) in pink, sulfide (FeS) in green, and pure iron in light purple.
\textit{Bottom}: Dust $\kappa_{\rm {abs}}(\lambda)$ curves used for the modified blackbody fits, showing the large variation in emission profiles. 
}
\label{fig:sampleSEDfits}
\end{figure}

Fig.~\ref{fig:sampleSEDfits} shows 28 of the resulting dust fits to the FIR-submm SED (top panel of the figure) using the various dust emissivity curves (bottom panel of the figure) listed in Table~\ref{table:modBBfits_kappacurves}. 
These have been grouped into similar dust compositions: amC, graphite, PAHs, silicates, oxides, carbides, sulphides, and pure iron to show the qualitative differences for fits using the various composition types. 
(Individual models are not labeled; for quantitative measures of specific models, we refer the reader to Table~\ref{table:modBBfits_kappacurves}.) 
Fig.~\ref{fig:sampleSEDfits} illustrates that vastly different dust properties can still translate to relatively similar fits to the observed flux densities. 
All models that result in good fits to the SED have temperatures between $\sim$18--23K (Table~\ref{table:modBBfits_kappacurves}); nine of the 28 dust varieties listed here fail our $\chi^2$ criteria for a `good' fit.

Models that represent amorphous carbons and graphites tend to have higher $\kappa_{\rm {abs}}$ values and thus return lower inferred dust masses, $\mathrm{M_{dust}} \sim$0.4--0.8\msun (Table~\ref{table:modBBfits_kappacurves}).
Amorphous silicates from \cite{Demyk2017}, which have more than a factor of 10 larger $\kappa_{\rm 850}$ compared to those from \cite{Semenov2003}, also yield a relatively moderate mass of $\sim$0.27\msun.
The silicate models of \cite{Ormel2011} and \cite{Ossenkopf1994} also give moderate dust masses, and represent grains that originate from coagulation processes as proposed in dense molecular clouds of the ISM.
These aggregated grains, some of which are coated with ice, can be as large as 100\,$\mu$m in radius, and the mass absorption coefficients increase in the FIR and mm.
As a consequence, the inferred dust masses for these grain types are $\sim$0.5--0.74\,\msun\ for \cite{Ormel2011} and \cite{Ossenkopf1994}.
Because the dust grains required for ISM coagulation are calculated to form over timescales of tens of thousands to millions of years, it is unclear whether such icy grains can be formed in the SN ejecta in such a short timescale as $<$30 years, however we present the results of their fits here for comparison.
Models of larger composite silicate grains, such as those of \cite{Jaeger1994} or \cite{Jaeger2003}, require even larger masses of $\sim$1--4\,\msun\ to fit the observed SED.
The CDE composite model of \cite{Dwek2015} results in a moderate mass of 0.38\msun, due to the increased emissivity of the carbon inclusion. 

Dust varieties that generally satisfy our criteria for a good fit include several amorphous carbon and silicate models (amorphous pyroxene and olivine varieties), graphite, PAHs, and alumina. Varieties that tend to fit the data poorly in the optically thin limit include FeO, FeS, SiC, SiO$_2$, organics, and water ice.

The largest source of uncertainty in the observed dust mass is the choice of dust emission profile\footnote{We did not attempt to fit a two-temperature component modBB as the SED shape is narrow, which suggests only one component is necessary \citep{Matsuura2011,Matsuura2015,Mattsson2015}.}, specifically the value of the mass absorption coefficient ($\kappa_{\rm {abs}}$) in the submm.
We attempted to investigate spatial variations in the fitted dust parameters across the ALMA maps, however the limiting beam size translates to only 2--3 independent elements across the ejecta, and we found that the differences in these flux ratios are smaller than their uncertainties.

\begin{figure}
\centering
\includegraphics[width=0.5\textwidth]{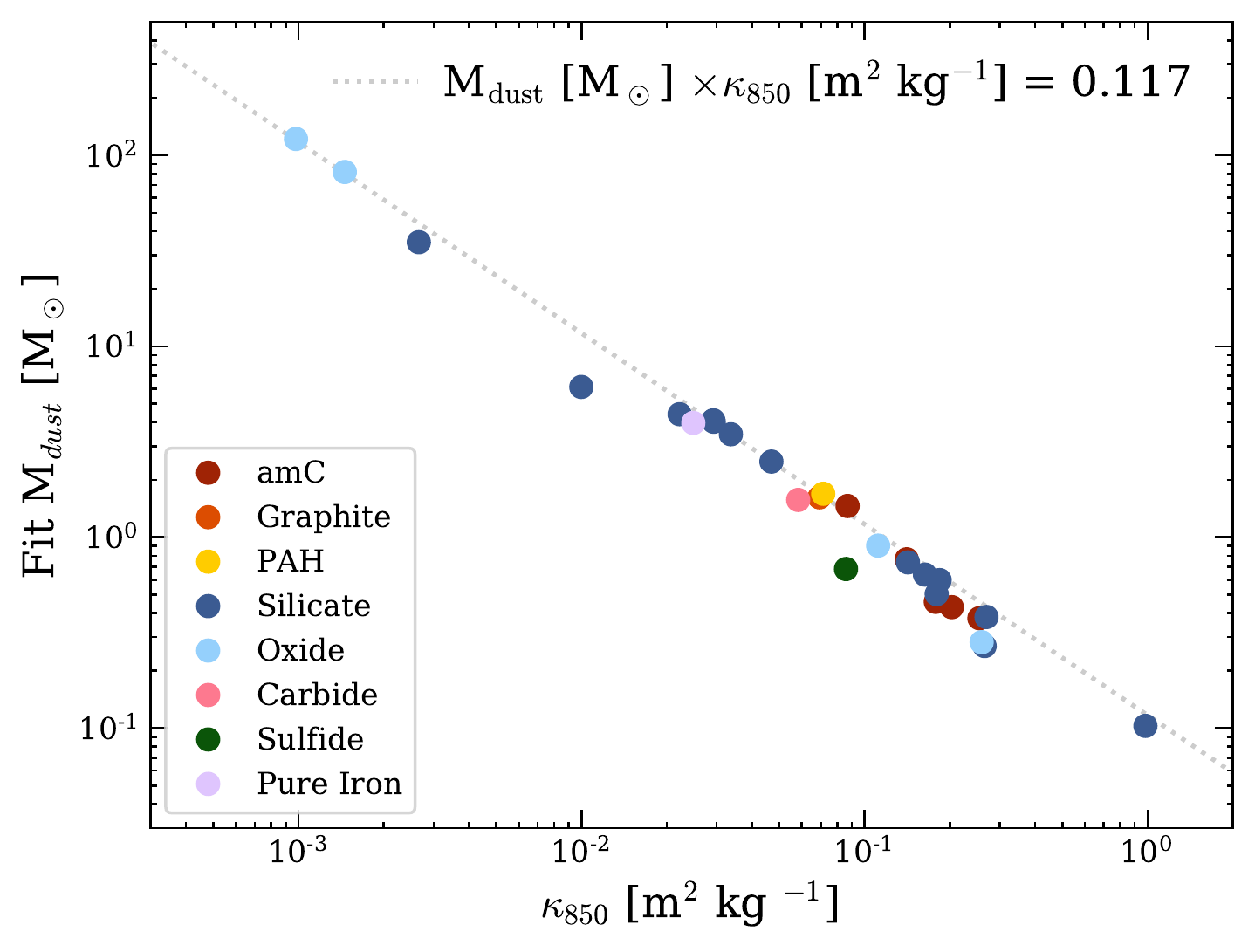}
\caption{ Fitted dust mass as a function of $\kappa_{850}$ from various dust models. The colors for the different dust varieties are the same as in Fig.~\ref{fig:sampleSEDfits}. The fitted dust mass closely follows a linear inverse trend over 3 orders of magnitude: M$_\mathrm{dust} \propto {\kappa_{850}}^{-1}$.   }
\label{fig:massvskappa}
\end{figure}

We compare the inferred dust masses from different dust absorption coefficients (Fig.~\ref{fig:massvskappa}).
The inferred masses very clearly follow an inverse linear relation based on the submm $\kappa_{\rm {abs}}$ value where \mbox{${M_{\mathrm dust} [M_\odot]} = 0.117 \times \kappa_{850}\, ^{-1}$}.
The significance of this is that for the current SN~1987A SED, the total ejecta dust mass can be estimated based on a single representative value of the desired dust mass absorption coefficients.

\subsection{The Ring}
\label{sec:ringemission}

The ring flux densities are shown in Fig.~\ref{fig:ejectavsring_continuumpoints}, along with previous measurements of the ring from ALMA and the Australia Telescope Compact Array (ATCA):
1.4\,GHz, 18\,GHz, 44\,GHz \citep{Zanardo2013}; 9\,GHz \citep{Ng2013}; 94\,GHz \citep{Lakicevic2012b}.
As the frequency increases towards the submm and FIR, the contribution of the thermal ejecta emission to the total emission becomes more significant, so the ring emission at day 9280 was estimated from the total flux densities at 102, 213, 345, \& 672\,GHz by scaling and subtracting an ejecta model component in Fourier space based on the Band--9 ejecta flux density at each frequency \citep{Zanardo2014}.

The Cycle 2 ring flux densities exhibit more scatter and are lower than the ATCA values by $\sim 30$\%.
The integrated ring emission follows a non-thermal power law profile of the form $S_{\nu} \propto \nu^{\alpha}$.
The spectral index $\alpha$ was previously estimated from ATCA radio \citep{Zanardo2014} %data at 1408 \um\ ??
and ALMA Cycle 0 data \citep{Indebetouw2014} to be $-0.73 \pm 0.02$ at day 9280 \citep{Zanardo2014}.  
An updated fit, using the ALMA Cycle 2 ring flux densities and the more recent 1--9 GHz measurements from \cite{Callingham2016} from day 9686, results in a slightly lower $\alpha = -0.70 \pm 0.06$.

The radio ring emission has steadily increased due to the synchrotron-producing electrons being accelerated by expanding shockwaves \citep{Zanardo2010,StaveleySmith2014}. 
Recently, \cite{Cendes2018} have fit the radio emission of the (2D) ring and (3D) torus emission models across many epochs as a power law of the form S($\nu,t$)=K$ \nu^\alpha (t-t_0)^\beta$, where $\alpha$ is the spectral index of the emission across the spectrum, $\beta$ is the power law slope, and K is the offset constant for the given model. 
The ALMA ring flux densities are generally in excellent agreement (within 5\%) with the \cite{Cendes2018} model prediction for day 10402 where K=1.5$\pm$0.1, $\alpha$=0.70 and $\beta$=0.59$\pm$0.02. 
We find no evidence of a contribution from dust in the ring to the mm wavelength flux densities (see, e.g., \citealt{Bouchet2006,Matsuura2019}).

\section{Analysis and results of molecular lines}
\label{sec:radex}

\begin{figure*}[]
\centering
\includegraphics[width=0.49\textwidth]{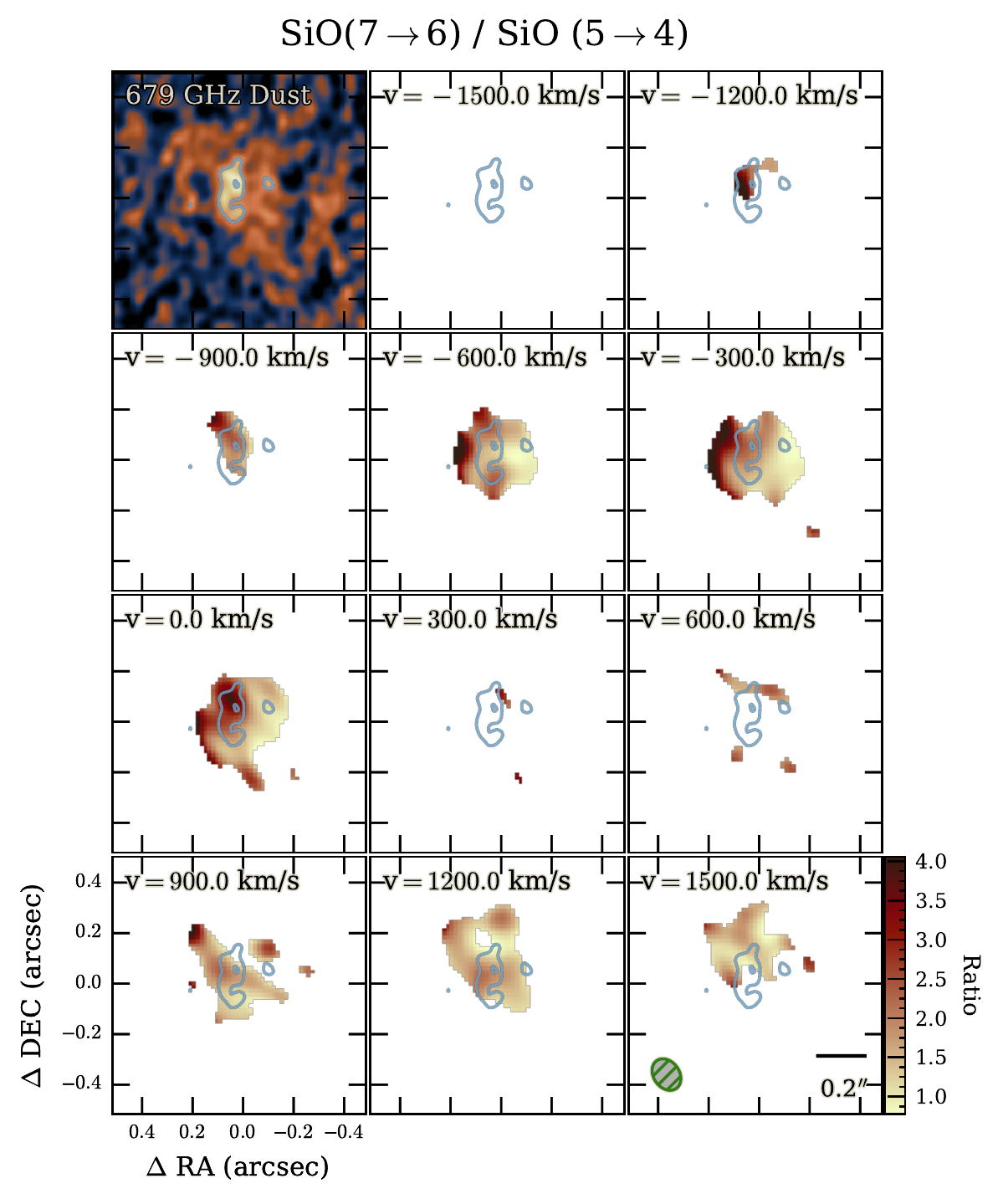}
\includegraphics[width=0.49\textwidth]{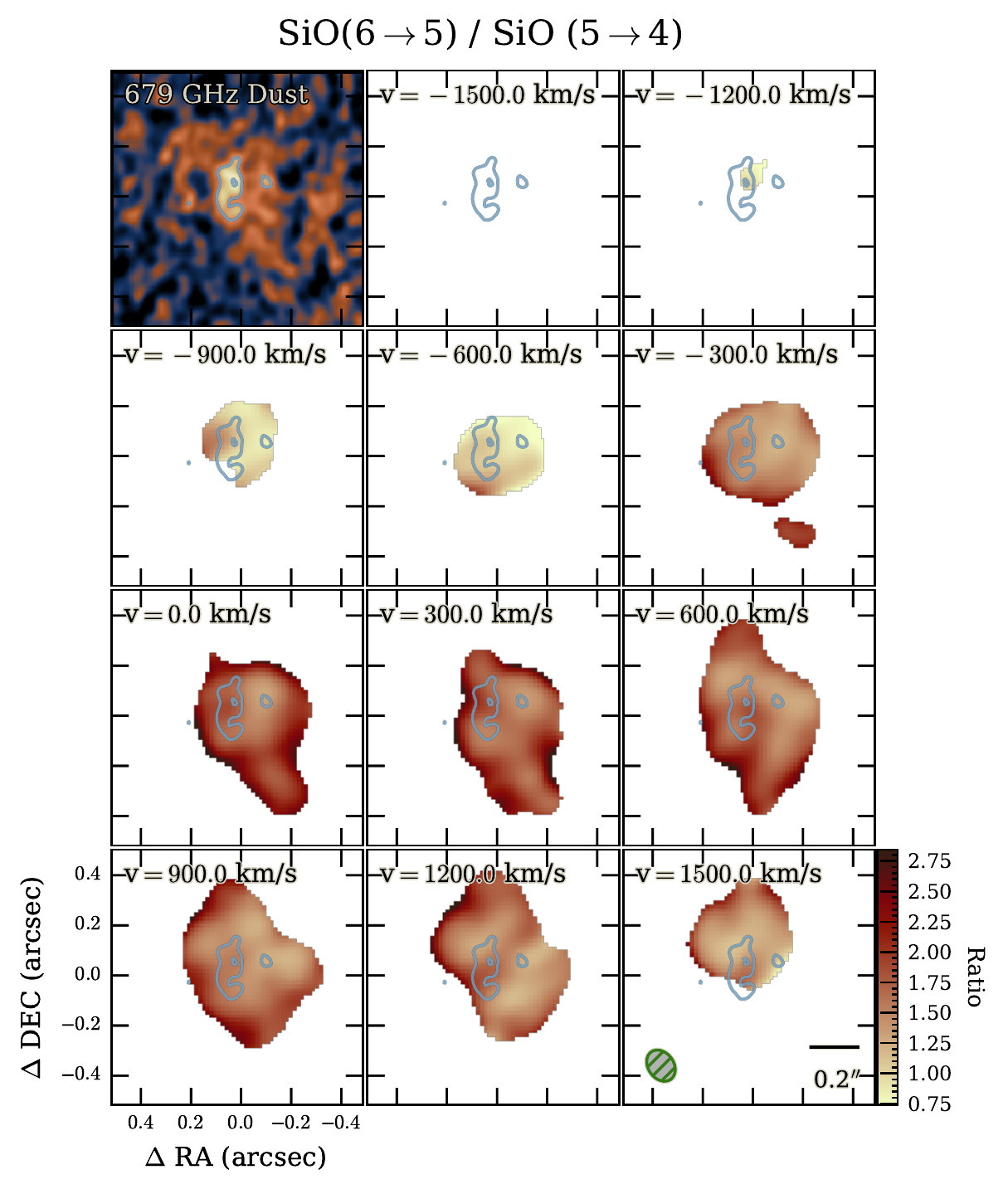}
\caption{
Molecular line ratios of SiO.
The line emission in each channel was converted to flux [W m$^{-2}$] integrated over the 300 \kms\ bin before division.
Channel velocity centers are LSRK; for reference, the SN~1987A systematic velocity is 287 \kms.
\textit{Left}: \mbox{\siosevensix\ / \siofivefour}.
Note that while the 300 \kms\ window covers the line centers, the \siosevensix\ line has poor coverage there due to that frequency falling at the edges of the two observed tunings.
\textit{Right}: \mbox{\siosixfive\ / \siofivefour}.
Intensity maps were convolved to the \siosevensix\ beam size before dividing.
Cyan contours are 679 GHz dust 3$\sigma$ and 5$\sigma$ levels.
The apparently high ratios at the edges of the map features are primarily due to reduced S/N in those outer regions. 
}
\label{fig:molecule_ratios}

\end{figure*}

\subsection{Analysis of Molecular Lines Using RADEX}
\label{sec:radexdesription}

In the previous Section, we modelled the integrated SED of the dust emission under the assumption of uniform temperature and density within the ejecta.
However, in \S\ref{sec:analysis}, we saw that the \siofivefour\ and \cotwoone\ images exhibit a hole (Fig.~\ref{fig:multiwav}) where the dust emission peaks, and the SiO line ratio indicates lower \siofivefour\ brightness with respect to \siosixfive\ and \siosevensix\ (seen in the SiO line ratio maps in Fig.~\ref{fig:molecule_ratios}).
In order to understand the spatial distribution of the line ratios and intensities qualitatively, we use the non-LTE (Local Thermal Equilibrium) line radiative transfer code {\sc radex} \citep{VanderTak2007}.

\subsubsection{Description of the procedure adopted for SiO}

This subsection describes the analysis of the SiO lines and intensities in detail. 
The analysis for CO was carried out in a similar manner, thus it is described only briefly in that subsection.

{\sc radex} calculates the molecular line intensities, using the escape probabilities from \citep{Osterbrock1989}, and the uniform sphere method for the gas distribution was chosen for this calculation.
The code involves calculations of level populations, using the Einstein coefficients and collisional cross sections of molecular lines assembled by the {\sc LAMBDA} database \citep{Schoier2005}, and we use H$_2$--SiO collisional cross sections based on the calculations by \citet{Dayou2006}.
In the ISM, H$_2$ is widely assumed to be the dominant collisional partner in molecular clouds; however, that is probably not the case for the ejecta of SNe.
As a consequence of a series of nuclear burning processes, the progenitor star's core will have built up layers of different newly synthesised elements, with hydrogen being depleted.
In the two layers containing abundant Si, the major elements are O and S \citep[e.g.,][]{Woosley1988}, and the collisional partner of SiO is likely to be O$_2$ and SiS.
This would potentially change the collisional cross section by a factor of 1--10, depending on the transitions \citep{Matsuura2017}.

One of the {\sc radex} input parameters is the FWHM of the line width of the Gaussian ($\Delta v$). 
We adopted a $\Delta v$ of 400\,\kms, whose integrated area over the Gaussian profile would be equivalent to that of a box-shaped 300\,\kms\ line profile.
If the line is optically thin, the assumed line profile is not a major issue.
However, as we will see later in this analysis, the lines are mildly optically thick at the line center, but not at the side of the line profile, so we therefore make the assumption that the line profile only moderately affects the line ratios and the line intensities in this ``mildly'' optically thick regime.

The main parameters involved in {\sc radex} calculations are the kinetic temperature  ($T_{\rm kin}$), the density of the collisional partner ($n_{\rm coll}$) and the column density ($N_{\rm SiO}$).
In the optically thin regime, as found in the calculated parameter range where solutions are found, the SiO line intensities are determined by $N_{\rm SiO}$ together with the area filling factor and the expansion velocity $v_{\rm exp}$, while the SiO line \textit{ratios} are determined by $T_{\rm kin}$ and $n_{\rm coll}$, independent of $N_{\rm SiO}$.
The filling factor is defined as the area of the line emitting fraction within the beam/pixel, following \citet{Goldsmith1999}.
Previous analyses of SN\,1987A with lower angular resolution suggested the range of 2.5--45\,\% \citep{Kamenetzky2013, Matsuura2017}; thus, we assume that the filling factor of 1--50\,\% is a reasonable range.
In this analysis, we adopted a column density grid of $N_{\rm SiO}$=$10^{12}$--$10^{18}$\,cm$^{-2}$ in factor of 10 increments, and searched for the predicted line intensities that can match the measured ones within the assumed range of the filling factor.
The adopted ranges of the parameters are $n_{\rm coll}$=$10^{3}$--$10^{10}$\,cm$^{-3}$ and $T_{\rm kin}$=10--200\,K.
\cite{Matsuura2017} suggested the temperature range is below 190\,K for SiO, with the CO kinetic temperature between 30--50\,K, so we restricted the analysis to below 200\,K.
Although we include temperatures up to 200\,K in the {\sc radex} calculations, it is very unlikely that majority of SiO gas has such a high temperature, and most likely, the overall SiO gas should have a temperature close to the CO temperature.
We searched for matching SiO line ratios within these parameter ranges.

The ALMA data of the three SiO transitions have different beam sizes and orientations, so \siofivefour\ and \siosixfive\ were convolved to match the lowest spatial resolution -- that of the \siosevensix\ beam -- with a uniform pixel width of 0\farcs015.
The convolved and regridded line ratio maps, made on a channel-by-channel basis, are displayed in Fig.~\ref{fig:molecule_ratios}.
Intensities were averaged over 5$\times$5 pixels in order to increase the signal to noise ratio in the \siosevensix\ image.
As the minor-axis FWHM of the beam is 0\farcs17 for \siosevensix, there is a small loss of spatial information by this averaging.

One caveat: the continuum subtraction for \siosixfive\ was performed on the final imaged data cube, whereas for \siofivefour\ and \siosevensix\ the continuum subtraction was done in $uv$ space before imaging (\texttt{imcontsub} vs. \texttt{uvcontsub} in \textsc{Casa}).
Continuum subtraction in $uv$ space is generally considered preferable if the continuum dominates the line emission, and the difference could slightly affect the ratios, but at high S/N as in the case of \siosixfive\ the two methods should give similar results.
As discussed in \S~\ref{sec:analysis}, there was no appreciable difference between the two methods for the \cosixfive\ data. 

The uncertainties in the measured SiO lines and line intensities are dominated by calibration uncertainties, and we adopted 10\,\% of the line intensities for this analysis.
The intensity uncertainties that were measured as a fluctuation of the `blank' sky level are 3\,\% for \siosixfive\ and \siofivefour, and 5\,\% for \siosevensix\ after 5-pixel averaging, so that the uncertainties based on the observations are smaller than the systematic uncertainties from calibration errors.
Because the dominant uncertainties are systematic, these propagate in an asymmetric way: i.e., if the actual line intensity of \siosixfive\ were higher than the measured value, the line ratio of \siosixfive/\siofivefour\ would increase, but the ratio of \siosevensix/\siosixfive\ would decrease.
The Cycle 2 flux densities being systematically lower than Cycle 0 as discussed in \S~\ref{sec:photometry} is not an issue for this analysis, as the important result here is that the relative change of temperature and density within the ejecta can explain the difference in molecular line ratios.

\subsubsection{Results of SiO analysis}
\label{sec:radexsioresults}

\begin{figure}
\centering
\includegraphics[width=1.0\linewidth]{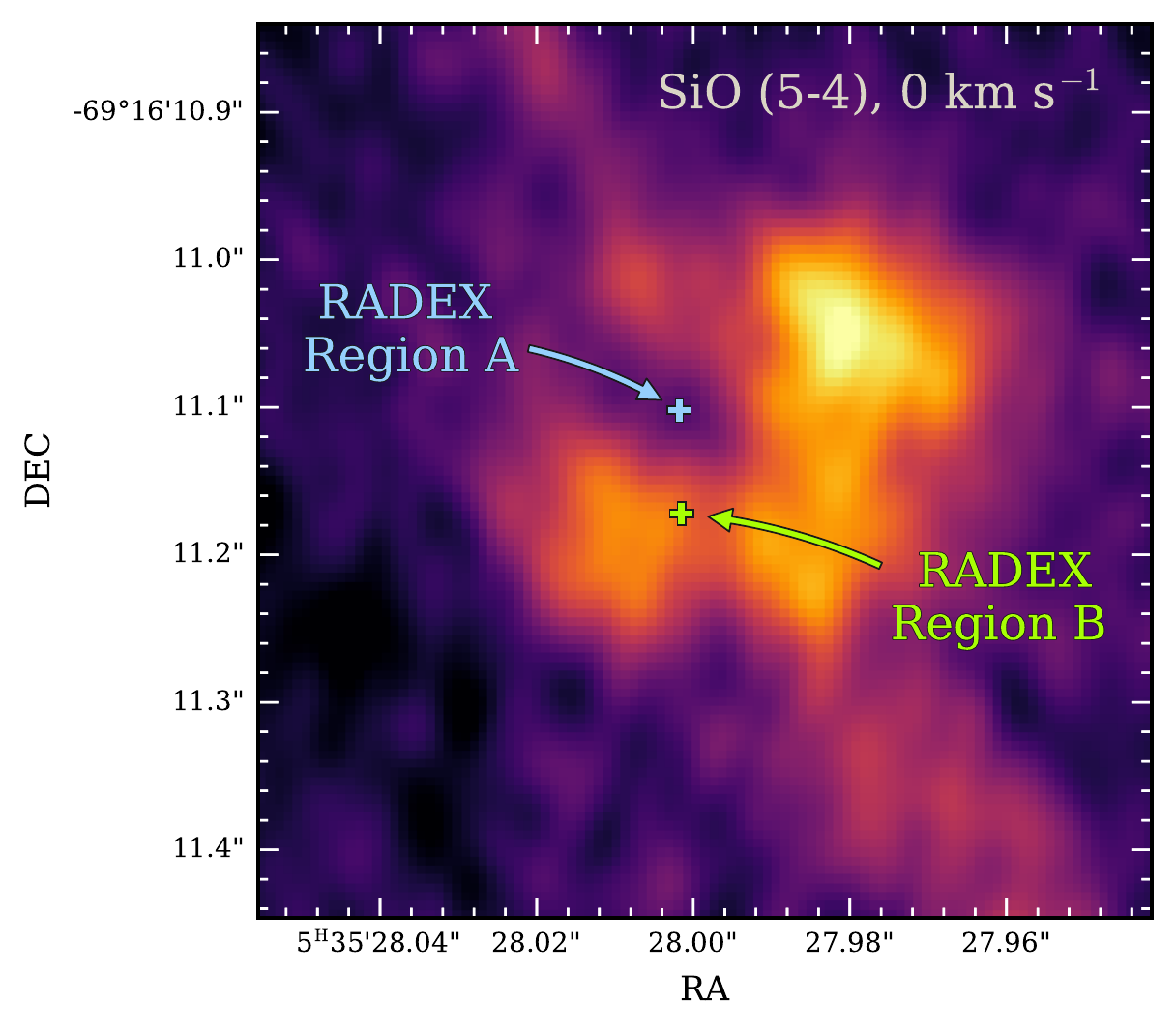}
\caption{
Locations of regions A and B used for the {\sc radex} analysis of the SiO transitions.
The background image is the \siofivefour\ line at 0 \kms\ (LSRK) channel, which clearly shows the molecular hole at region A.
Region B is a representative region of the general SiO emitting ejecta.
}
\label{fig:radexregions}
\end{figure}

%====================
\begin{figure*}[t!]  %figure fig-ratio-1
\centering
\resizebox{0.4\hsize}{!}{\includegraphics[trim=1.5cm 12cm 5.8cm 4cm]{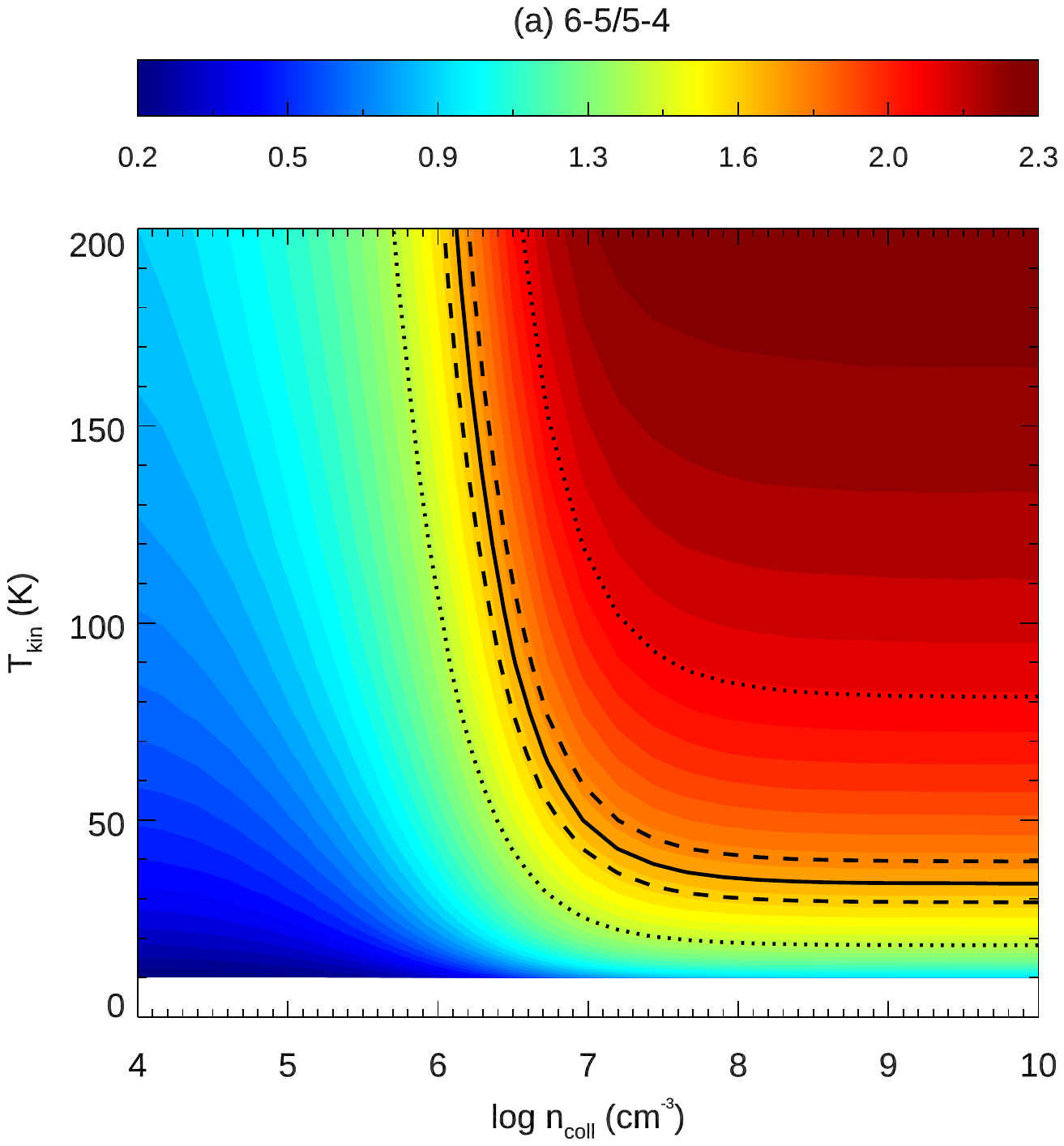}}
\resizebox{0.4\hsize}{!}{\includegraphics[trim=1.5cm 12cm 5.8cm 4cm]{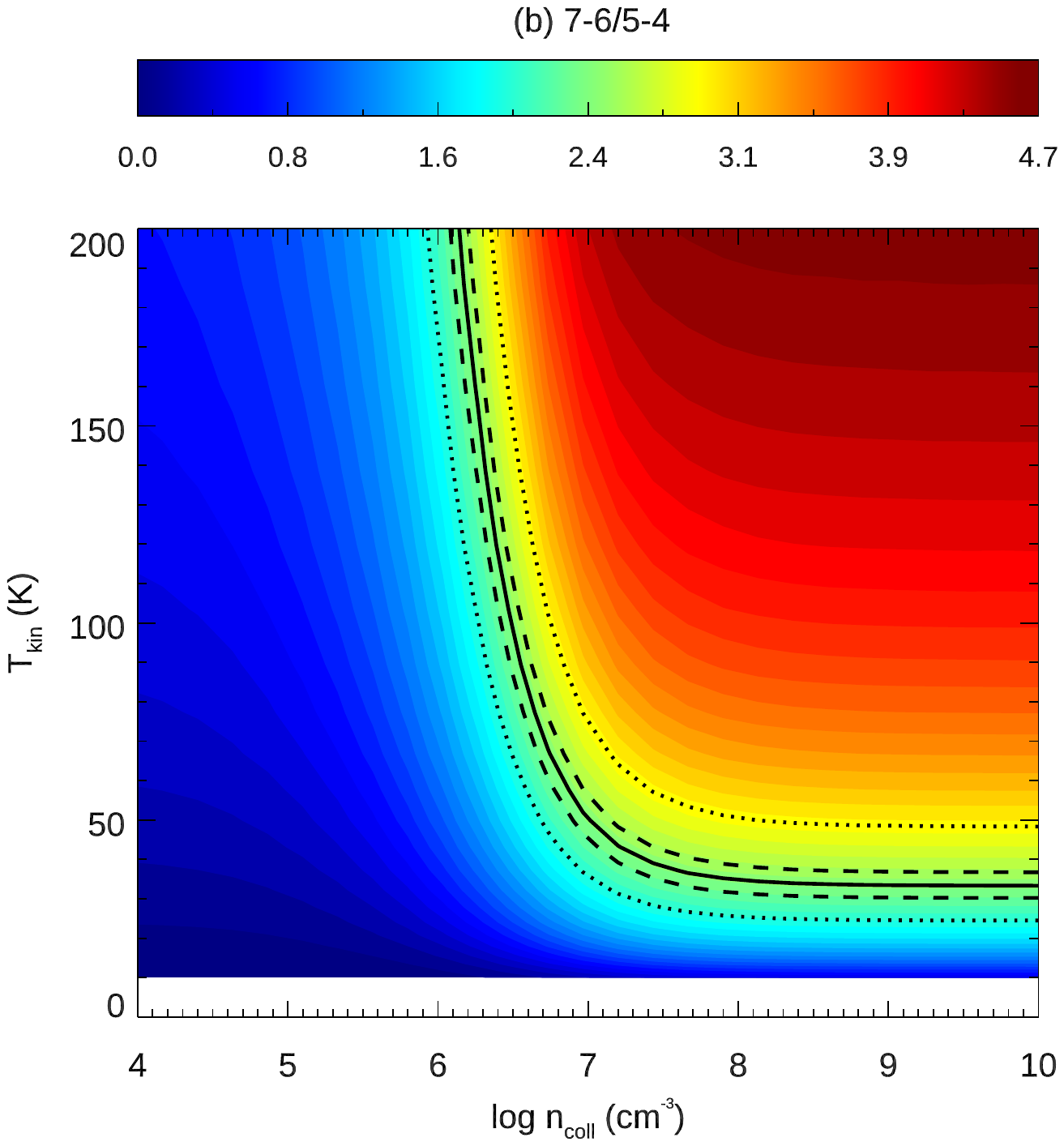}}
\resizebox{0.4\hsize}{!}{\includegraphics[trim=1.5cm 11cm 5.8cm 2cm]{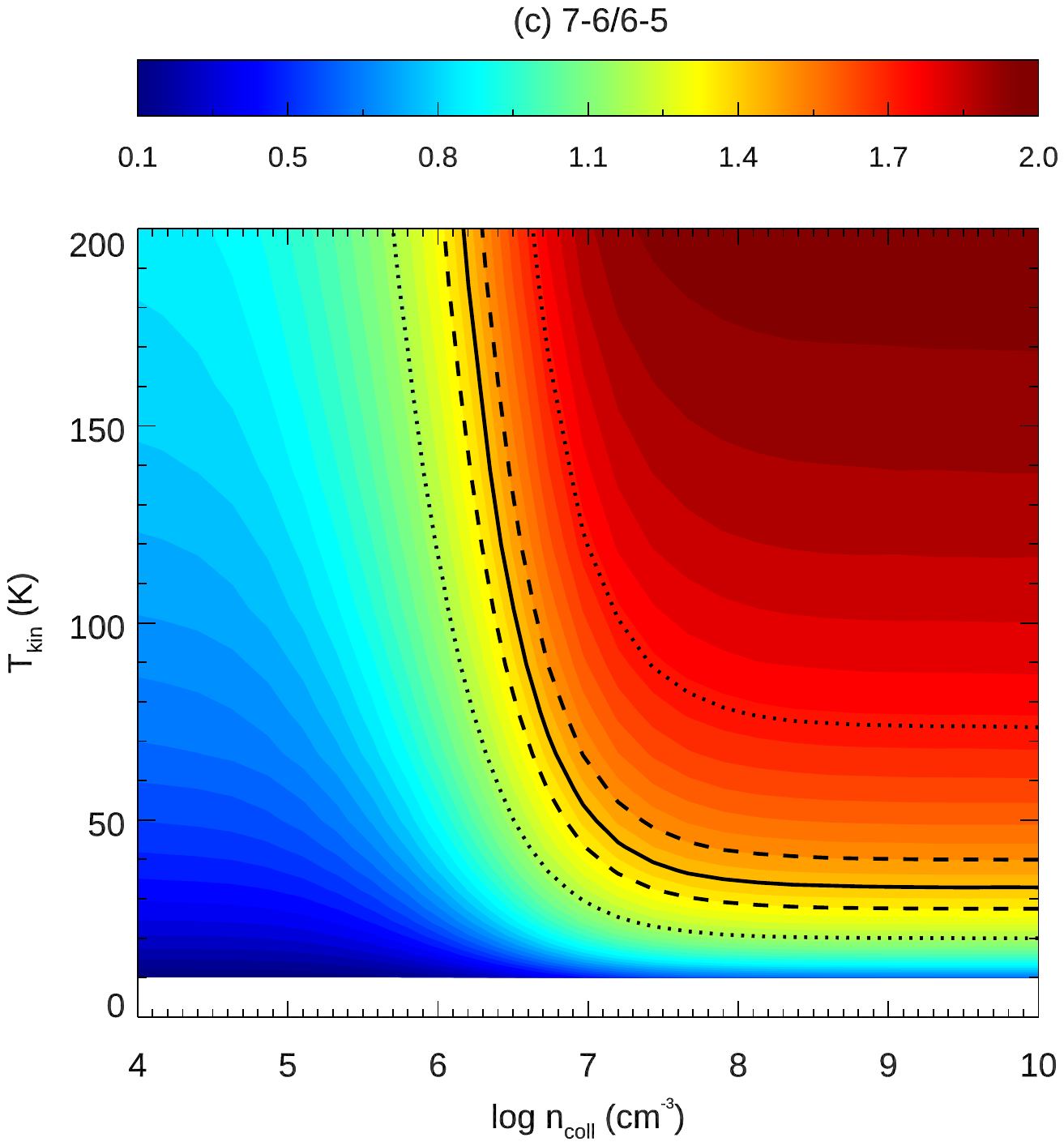}}
\resizebox{0.4\hsize}{!}{\includegraphics[trim=1.5cm 11cm 5.8cm 2cm]{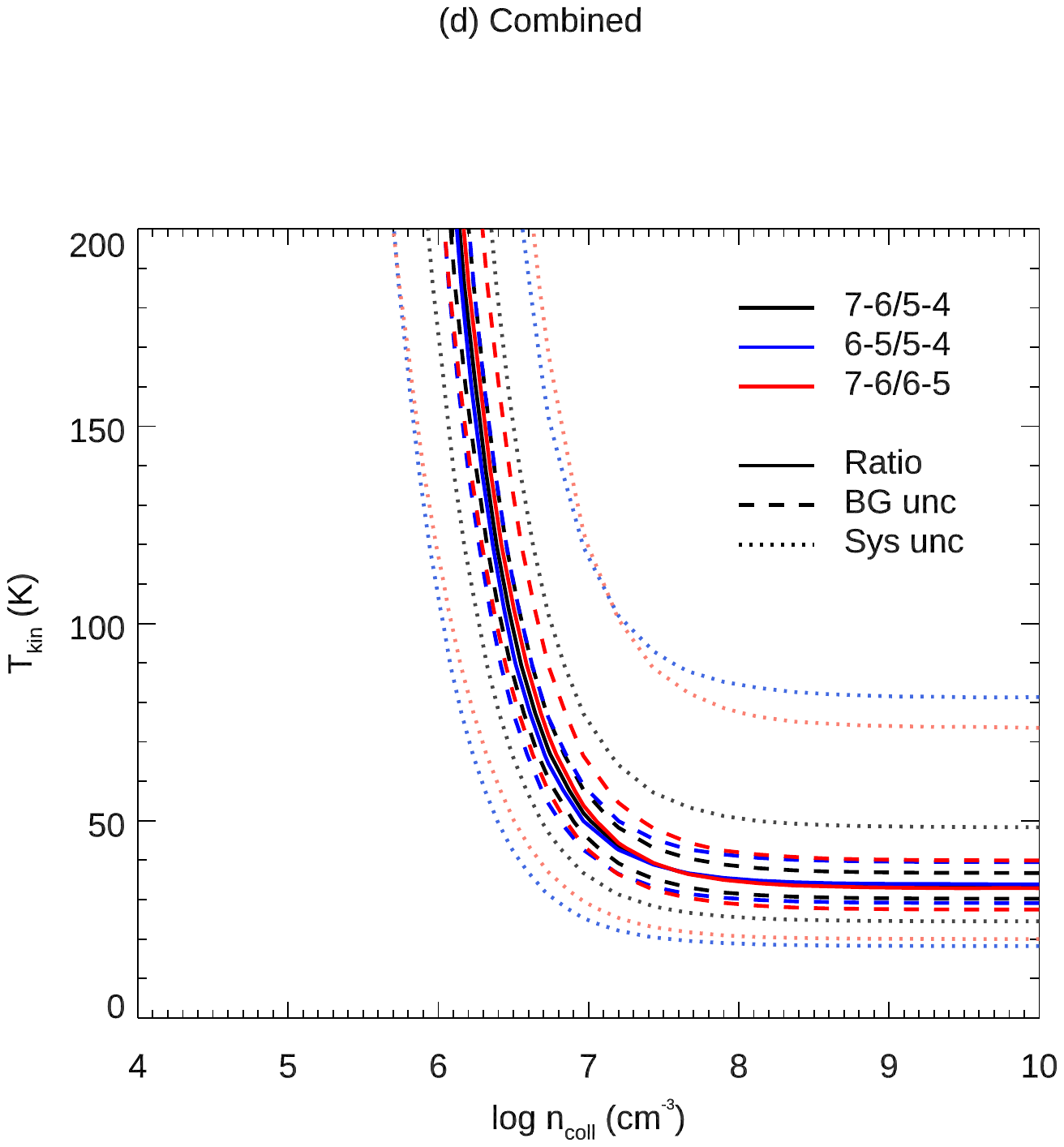}}
\caption{
(a)--(c) Results of {\sc radex} modeling of the SiO line ratios at region A, representing the molecular ``hole''.
These calculations assume a column density of $N_{\rm SiO}=10^{16}$\,cm$^{-2}$.
(d) The combination of all three best-fit curves.
For each line, the solid curve denotes the best fit locus of the  grid of kinetic temperature and \htwo\ density values to the observed line ratios, while the dashed and dotted curves represent the 1\,$\sigma$ background uncertainties and 1\,$\sigma$ systematic uncertainties, respectively.
\label{fig:radxfer1}
}
\end{figure*}
%===============

%====================. intensity rules out this posibility
\begin{figure}  %figure fig-ratio-1-17
\centering
\resizebox{\hsize}{!}{\includegraphics{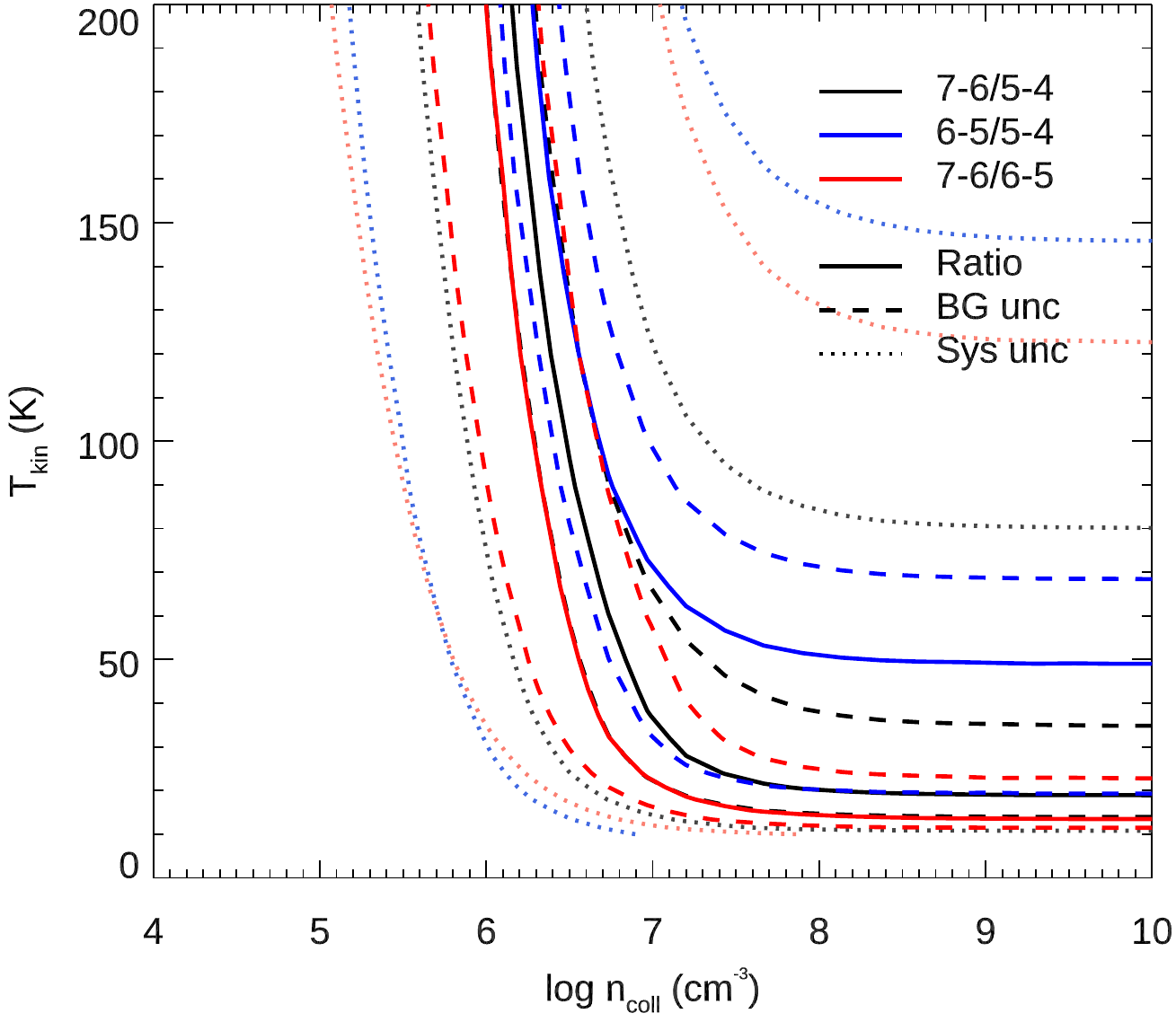}}
\caption{ 
The SiO line ratios that match the measured ALMA values at the SiO hole. 
The same as Fig.\,\ref{fig:radxfer1} (d) but for a column density $N_{\rm SiO}=10^{17}$\,cm$^{-2}$. 
\label{fig:radxfer1_17} 
} 
\end{figure} 
%=============== 
%==================== 
\begin{figure}  %figure fig-ratio-2 
\centering 
\resizebox{\hsize}{!}{\includegraphics{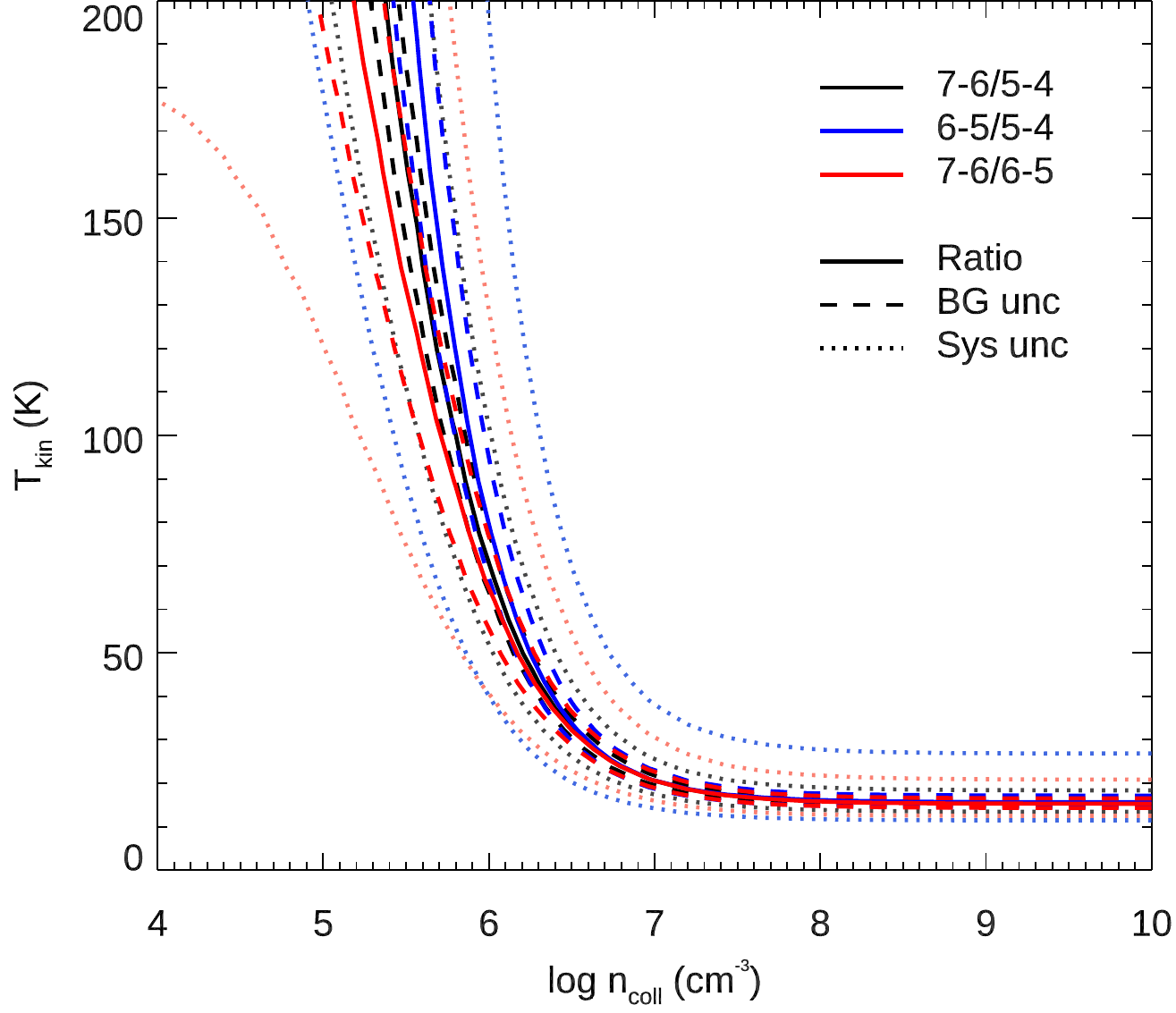}} 
\caption{ 
The same SiO calculations as in Fig.\ref{fig:radxfer1} (d) but for region B with column density $N_{\rm SiO}=10^{16}$\,cm$^{-2}$. 
\label{fig:radxfer2}
}
\end{figure}
%===============

We selected two representative spatial regions, shown in Fig.~\ref{fig:radexregions}, in order to understand the change in physical parameters ($T_{\rm kin}$, $n_{\rm coll}$) within the ejecta, which in turn affect the SiO emission.
One is close to the molecular hole seen in \siofivefour\ and \cotwoone\, which we call region A, and the other is representative of the neighboring `typical' SiO ejecta, named region B.
The actual hole itself has almost negligible \siofivefour\ line intensity, so we chose the nearest possible point for region A.
All ratios in this analysis were determined from the 0~\kms\,LSRK channel, which has good S/N for all three SiO lines, and the hole is clearly identified.

In region A, which has a hole in \siofivefour, we found that a column density of $N_{\rm SiO}=10^{16}$\,cm$^{-2}$ is reasonable.  
For $N_{\rm SiO}=10^{15}$\,cm$^{-2}$ or below, the predicted line intensities do not reach the measured levels, assuming a filling factor of 50\%. 
Slightly larger values of $N_{\rm SiO}=10^{17}$\,cm$^{-2}$ can match the measurements, but beyond $10^{18}$\, cm$^{-2}$ the predicted and measured line intensities do not match.

Fig.~\ref{fig:radxfer1} demonstrates the plausible ranges for the kinetic temperature ($T_{\rm kin}$) and collisional partner's density ($n_{\rm coll}$) at region A, for $N_{\rm SiO}=10^{16}$\,cm$^{-2}$.
The collisional partner for the {\sc radex} calculations is H$_2$; see the last paragraph of this section for discussion of this point.
In Fig.~\ref{fig:radxfer1} (a)--(c), the colored contours show the calculated ratios, the black lines show the measured SiO line ratios, and the 1\,$\sigma$ background uncertainties and 1\,$\sigma$ systematic uncertainties are indicated with dashed and dotted lines, respectively.
Fig.~\ref{fig:radxfer1} (d) compares the possible ranges of the kinetic temperature ($T_{\rm kin}$) and collision partner density ($n_{\rm coll}$) from the three SiO line ratios.
The uncertainty of the \siosevensix/\siofivefour\ ratio has the tightest constraint, so only the uncertainty of this ratio is plotted in Fig\,\ref{fig:radxfer1} (d).
Fig\,\ref{fig:radxfer1} (d) shows that the ratios of all three SiO lines can be matched with a very similar set of $T_{\rm kin}$ and $n_{\rm coll}>10^7$\,cm$^{-3}$, as indicated by the solid lines of the three different ratios almost overlapping each other.
There are multiple scenarios that match the measured SiO line ratios: the first possibility is $T_{\rm kin}$=34\,K with LTE conditions at $n_{\rm coll}>10^7$\,cm$^{-3}$, the second possibility is a non-LTE condition with a range of ($10^6<n_{\rm coll}<10^7$\,cm$^{-3}$) and $50<T_{\rm kin}<$200\,K, and finally the curves connecting these two conditions.
The line center turns optically thick at the column density of $N_{\rm SiO}=10^{16}$\,cm$^{-2}$, but the majority of the lines at off-center velocities are optically thin, so the overall analysis is not strongly affected by optical thickness at this column density. 
The filling factor at this column density is 9--14\,\%.

By increasing the column density from $N_{\rm SiO}=10^{16}$\,cm$^{-2}$ to $N_{\rm SiO}=10^{17}$\,cm$^{-2}$, the SiO ratios change much more gradually as a function of the kinetic temperature ($T_{\rm kin}$) and collision partner density ($n_{\rm coll}$) (Fig.~\ref{fig:radxfer1_17}), expanding the feasible parameter space.
This is the effect of higher optical depth.
The sets of $T_{\rm kin}$ and $n_{\rm coll}$ from the three different calculated SiO ratios do not overlap in Fig.\,\ref{fig:radxfer1_17}, but it is still possible to consider a wide range of $T_{\rm kin}$ at $n_{\rm coll}>10^7$\,cm$^{-3}$, from 19--22\,K within the 1\,$\sigma$ uncertainty.
In the non-LTE range, the required $n_{\rm coll}>10^7$\,cm$^{-3}$ is more or less comparable to that for $N_{\rm SiO}=10^{16}$\,cm$^{-2}$.
In order to reproduce the line intensities with this higher column density, the filling factor of the beam area is 0.8--2\,\%, much lower than for the $N_{\rm SiO}=10^{16}$\,cm$^{-2}$ case.

At region B, the physical parameters are slightly different from those at the hole (region A).
Fig.\,\ref{fig:radxfer2} shows the possible parameter space which fits the SiO ratios for region B; only $N_{\rm SiO}=10^{16}$\,cm$^{-2}$ gives a feasible range.
The most plausible temperature is $T_{\rm kin}$=18\,K for LTE conditions ($n_{\rm coll}>10^7$\,cm$^{-3}$), and an alternative possibility is non-LTE with $3\times10^5<n_{\rm coll}<10^7$\,cm$^{-3}$ and $18<T_{\rm kin}<$200\,K.
In the optically thin regime, the filling factor and the column density are inversely correlated, thus the accuracy of the filling factor is limited by our column density grid.

In summary, the difference in the modelled line ratios and intensities near the \siofivefour\ and $J$=6~$\!\rightarrow\!$~5 hole (region A) with respect to the `representative' ejecta region B can be explained in the following three ways.
The first possibility is that the hole region has a higher temperature (35\,K) than the surrounding locations (19--22\,K) with both having LTE conditions -- i.e., high density of the collisional partner.
The second possibility also requires LTE conditions, but instead of high temperature, the hole region has a higher column density in a much smaller area, represented by a small beam filling factor.
The third possibility is that the entire area is non-LTE, i.e., a lower density of the collisional partner, but with the hole having a factor of a few to a few tens higher density of the collisional partner than the surrounding region. 
These three explanations are not exclusive to each other; a mixture of these three scenarios is possible.

The uncertainties involved with this analysis arise from uncertainties in the collisional cross section.
The collisional partner is most likely not H$_2$, but rather other molecules such as O$_2$ or SiS, according to chemical models \citep{Sarangi2013}.  
Therefore, instead of higher H$_2$ density at the hole the collisional partner may be different in the hole and in region B -- for example, region B may be dominated by collisions with O$_2$, whereas in the hole the dominant partner could be SiS. 
However, as will be discussed in \S~\ref{sec:blob_models}, hydrodynamic simulations predict that such spatial distributions for the Si and O atoms are unlikely, therefore, our conclusion that there is a higher temperature, column density or density at region A is still valid even with this uncertainty.

\subsubsection{CO analysis and results}
\label{sec:radexcoresults}

%====================
\begin{figure}
\centering
\includegraphics[width=0.9\linewidth]{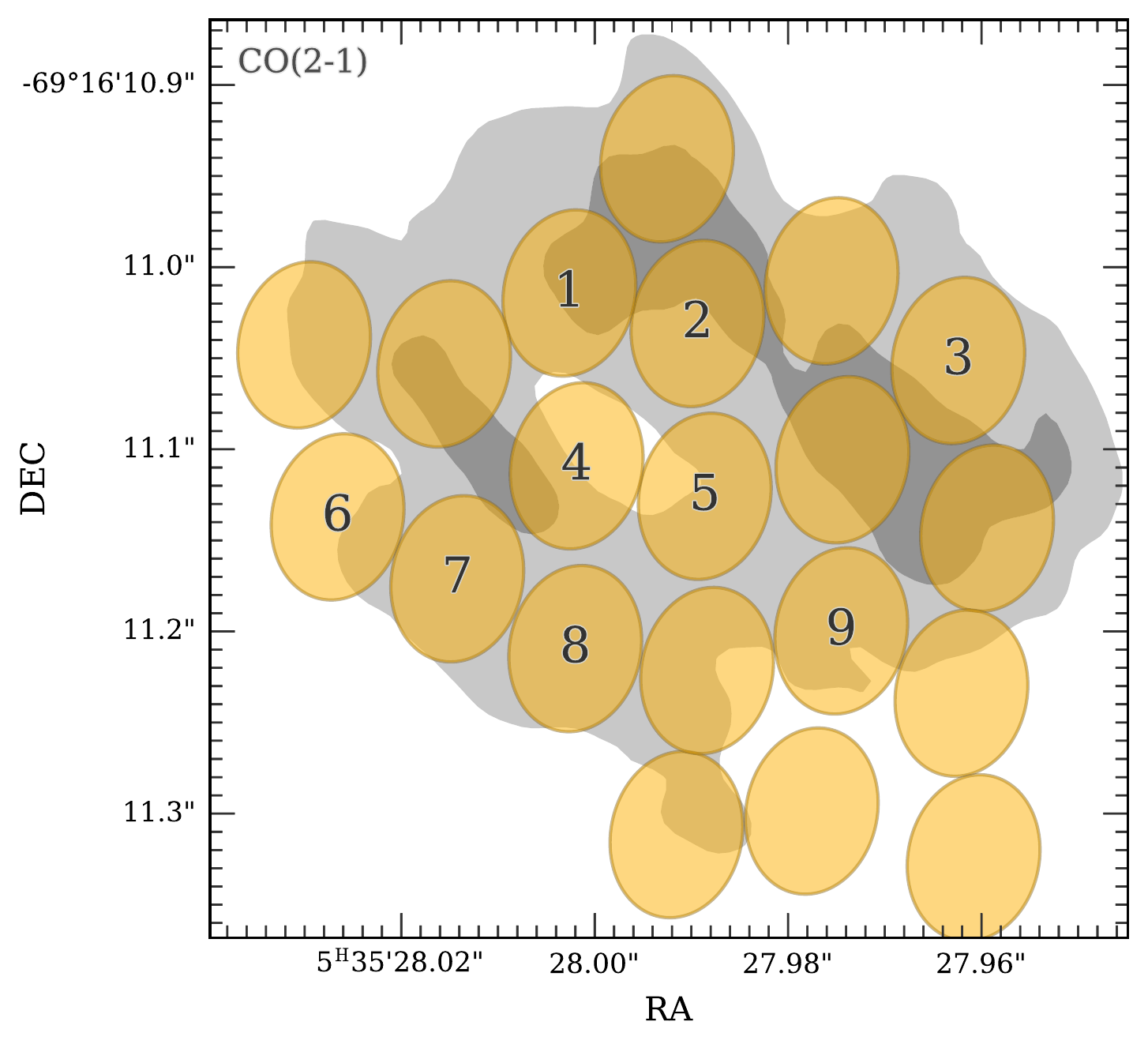}
\includegraphics[width=0.9\linewidth]{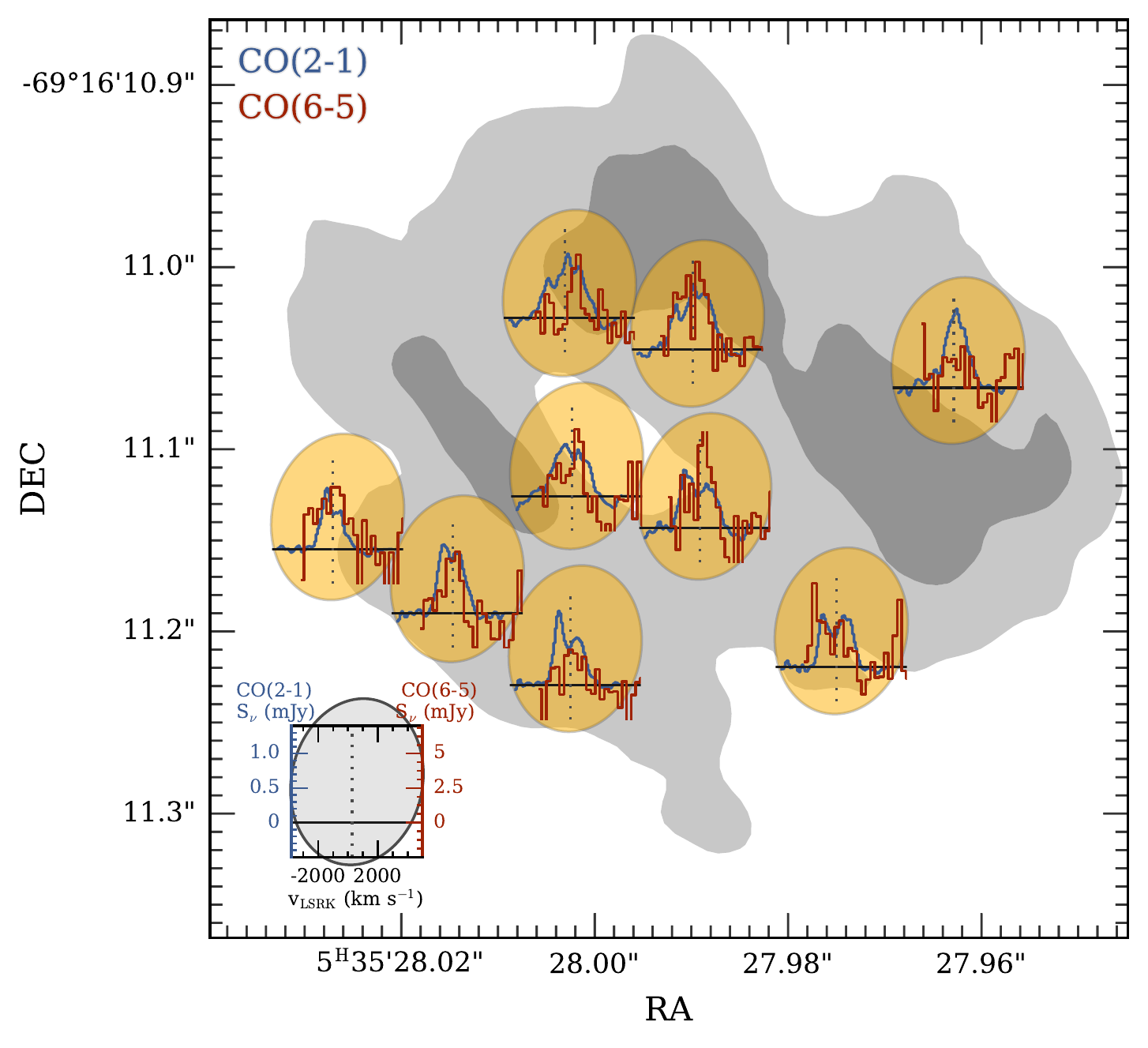}
\includegraphics[width=0.9\linewidth]{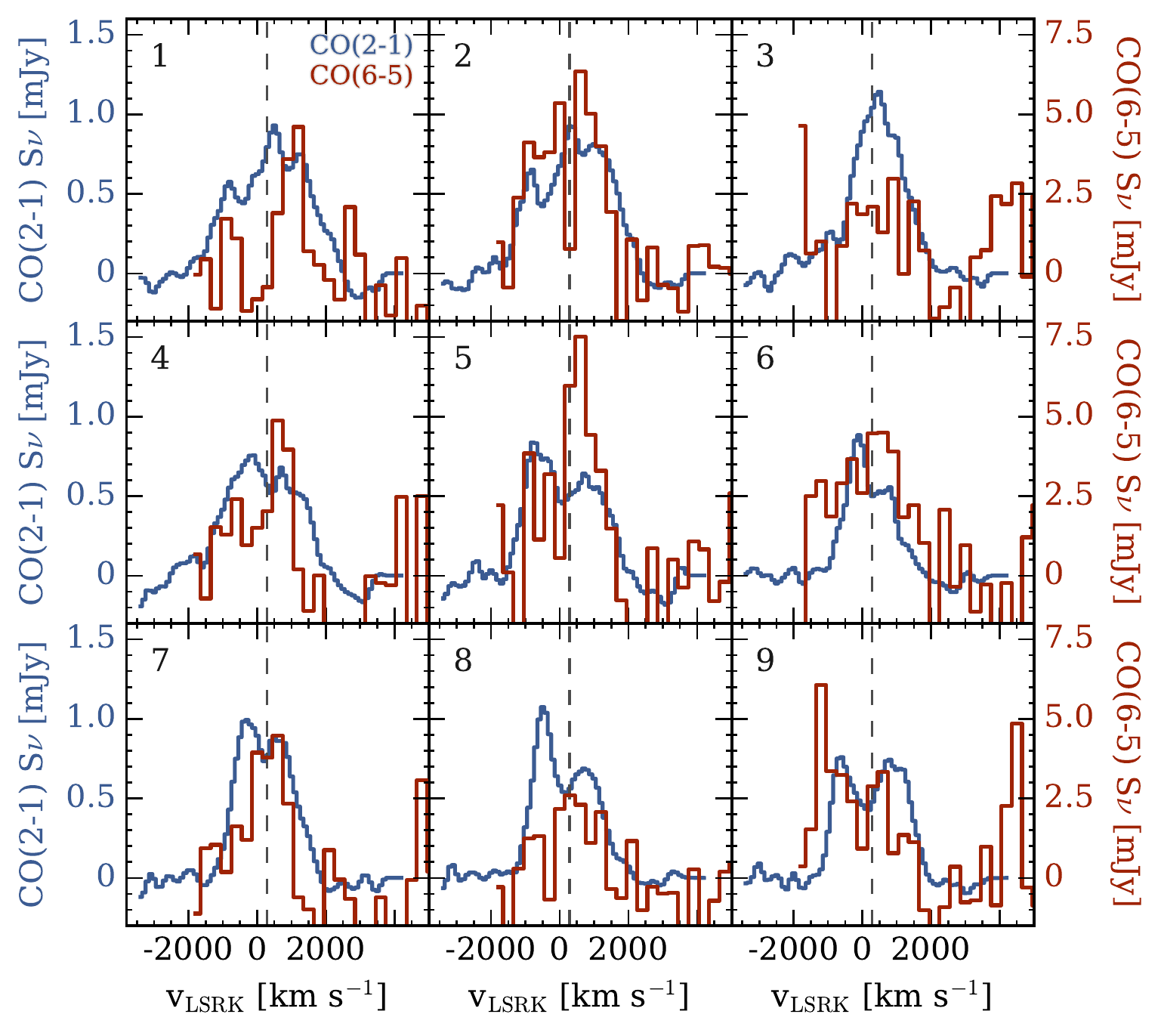}
\caption{
\textit{Top:} Regions chosen for comparing the \cotwoone\ and \cosixfive\ fluxes and line profiles; each region spans one \cosixfive\ beam. Regions of particular interest are labeled 1--9.
The shaded greyscale contours illustrate the \cotwoone\ emission (lower left panel of Fig.~\ref{fig:multiwav}).
\textit{Middle:} A qualitative comparison of the stacked \cotwoone\ (blue) and \cosixfive\ (red) spectra in each region with a representative scale shown in the bottom left corner.
\textit{Bottom:} A zoomed in view of the spectra for regions 1--9 shown in the middle panel. The vertical dashed line is the systemic velocity of SN~1987A, 287 \kms\ (LSRK).
\label{fig:COregionratios}
}
\end{figure}
%===============

Although the \cosixfive\ line does not have enough S/N for a quantitative analysis on a pixel-by-pixel basis, we can sum the spectra in independent (single beam-width) regions to aid our analysis.
The top panel of Fig.~\ref{fig:COregionratios} shows the location of 20 regions selected across the CO-emitting ejecta, overlaid on top of the \cotwoone\ emission.
Nine regions, each the size of the \cosixfive\ beam, are highlighted as areas of interest, potentially probing different conditions (numbered 1--9 in rows from left to right).
The middle panel of Fig.~\ref{fig:COregionratios} compares the summed spectra of \cosixfive\ and \cotwoone\ (with the latter convolved to the \cosixfive\ beam before integration) for these 9 regions showing their location with respect to the \cotwoone\ emission.
The spectra are in units of mJy, having been spatially integrated over each region.

The bottom panel of Fig.~\ref{fig:COregionratios} provides a zoomed in view of these spectra for a more detailed comparison.
Across the CO ejecta, we see that the majority of the line profiles in the \cotwoone\ and \cosixfive\ transitions are similar to each other in the scaled spectra (see for example regions 6, 7, and 9 in Fig.~\ref{fig:COregionratios}).
However, in regions 1 and 8, the \cosixfive\ profile is suppressed at negative velocities with respect to the \cotwoone\ line.
The \cosixfive\ emission is also suppressed with respect to \cotwoone\ in region 3, though this is across the entire velocity profile. 
At the location of the CO molecular hole (regions 4 and 5), we see strong \cosixfive\ emission at velocities of --1000--+1000 \kms\ with respect to \cotwoone\ (bottom panel Fig.~\ref{fig:COregionratios}); this continues to neighboring region 2. 
We note that regions A and B from the SiO analysis fall within the CO map regions 4 and 8, respectively, but they are not interchangeable.
They were selected independently, and are centered at different locations and serve different purposes (region B is representative of the general SiO emitting properties across the molecular ejecta based on the SiO line ratios, whereas Fig.~\ref{fig:COregionratios} demonstrates that region 8 has very different CO gas properties to most of the other regions).

\begin{figure}  %figure fig-ratio-3
\centering
\resizebox{\hsize}{!}{\includegraphics[trim=2.0cm 11cm 5.8cm 4.5cm]{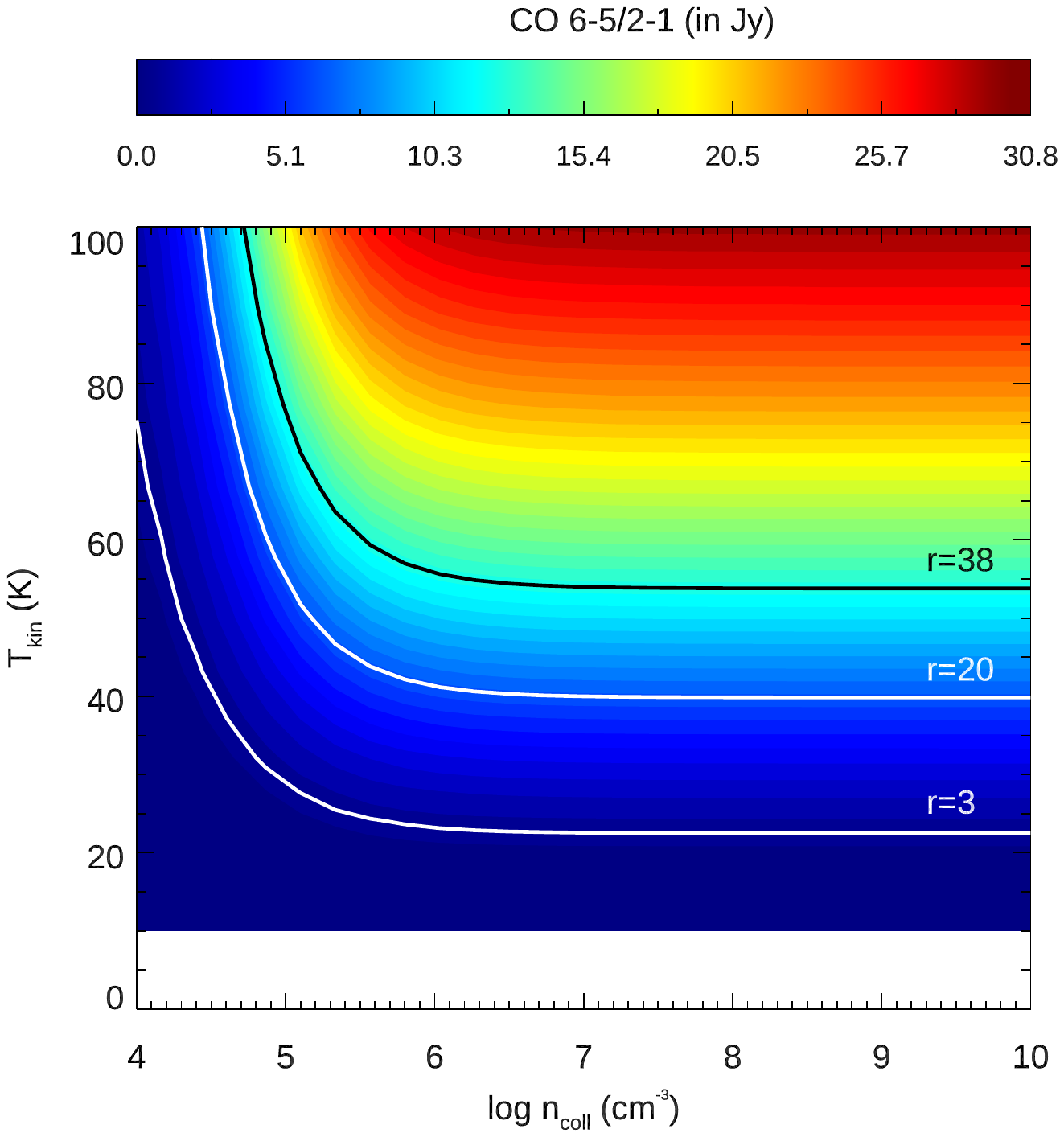}}
\caption{
Ratios of \cosixfive\ and \cotwoone\ flux densities in Jy units derived using {\sc radex} for the optically thin case.
The black and two white lines show ratio values of 38, 20 and 3, respectively indicating high, intermediate and low CO 6--5/2--1 end ratios observed in the 20 regions in Fig.~\ref{fig:COregionratios}.
}
\label{fig:radxfer3}
\end{figure}

Using this information, we carry out an analysis with {\sc radex} similar to that for SiO.
Fig.\ref{fig:radxfer3} shows the resulting {\sc radex} calculations of the \cosixfive/2~$\!\rightarrow\!$~1 ratios.
The three black and white lines show the curves for flux ratio values of 38, 20 and 3, respectively, corresponding to the higher, intermediate and lower ends of the line ratios observed across the 20 regions.
We note that the units of the flux densities used to derive the line ratios in this (CO) Section are in Jy, and are therefore different to the W\,m$^{-2}$ used in the SiO ratio calculations. 
This is because the CO spectra are compared in velocity space, whereas we use integrated line intensities for the SiO analysis\footnote{The spectra and channel maps in Figs.~\ref{fig:COregionratios} and \ref{fig:chanmaps_sio} are spectral density units (mJy and mJy per beam). 
In order to compare the line ratios in the integrated fluxes in $\rm W\,m^{-2}$, the flux densities in mJy per velocity channel units need to be multiplied by a factor of $\partial f$/$\partial$v=--$f_0$/c to account for the change from integrating in velocity v to frequency $f$.  For the CO 6~$\!\rightarrow\!$~5/2~$\!\rightarrow\!$~1 ratio, multiply by $f_{0,\mathrm{CO65}}$/$f_{0,\mathrm{CO21}}$ $\sim$ 3.}.

As demonstrated originally in Fig.\,3 in \citet{Kamenetzky2013}, the \cosixfive\ line is sensitive to temperature change.
We propose here that the \cosixfive\ suppression with respect to \cotwoone\ indicates the gas is at a lower temperature in regions 1 and 8 (where we also see a peak in the dust emission, \S~\ref{sec:analysis}, Fig.~\ref{fig:3color}) compared to the surrounding regions. 
In these regions, the blue wing of the \cosixfive\ emission is lower, thus if the dust and \cosixfive\ are spatially coincident in regions 1 and 8, this could imply that the CO and dust originate from a discrete region on the near side of the ejecta, though this is speculative as we have no velocity information on the dust. 
Due to the low S/N of the \cosixfive\, line, we cannot specify the exact excitation temperatures in these regions.
However, the CO excitation temperature is higher near the \cotwoone\ hole (regions 4 and 5) at velocities of $-$1000--1000\,km\,s$^{-1}$.

%====================
%====================

%%%%% ------------------- Discussion ------------------- %%%%%

\section{Discussion}
\label{sec:discussion}

\subsection{Dust Properties from the Integrated SED}
\label{sec:SEDdiscussion}

Any dust model that produces a mass higher than the total abundance of metals formed in the SN ejecta is clearly unphysical. 
Here we discuss whether the observed, or predicted, metal yields formed in the ejecta of SN~1987A can be used to rule out some of the dust varieties and compositions that produce good fits to the SED (Table~\ref{table:modBBfits_kappacurves}). 
Inferred dust masses of several tens of solar masses, as in the case of TiO$_2$ \citep{Posch2003} which requires 82\,\msun\ of dust, or MgAl$_2$O$_4$ \cite{Fabian2001} which requires 122\,\msun\ for example, are clearly untenable as they are larger than the progenitor star mass (18--20\,\msun, \citealp{Woosley1988}). 
This rules out a further three dust varieties in Table~\ref{table:modBBfits_kappacurves} (all of which satisfy our $\chi_\nu^2$ criteria for a good fit). 

Simple upper limits to the dust masses of various dust species can be estimated by calculating the resulting mass for 100\% of the elements in the nucleosynthesis models being locked into dust, ignoring all chemistry, mixing, and other physical limitations.
Such a highly unrealistic scenario is useful for winnowing out dust models that yield dust masses that are too large.
Considering the 18\,\msun\ $Z=0.008Z_\odot$ progenitor model from \cite{Nomoto2013}, the total mass of the key limiting metals C, N, Ne, Mg, Si, S, and Fe is 0.77\,\msun, with 1.21\msun\ of oxygen.
Focusing on individual dust varieties, the limits can be further differentiated.
The \cite{Nomoto2013} 18\,\msun\ model predicts 0.149\msun\ of carbon, putting a limit on the mass of graphite and amorphous carbon grains, as well as a rough limit for PAH varieties since their masses are dominated by carbon.  
The carbonaceous grain model that gives both a good fit to the SED and produces the nearest fitted mass to the predicted carbon yield is the amC (AC1 sample) from \cite{Rouleau1991}, with a dust mass of 0.43\msun. 
For silicate dust, good fits to the SED produce masses of $0.3-0.7$\,\msun, though the yields from \citet{Nomoto2013} would result in a maximum combined silicate metal mass (and therefore dust mass, limited by the available Mg) of $<0.4$\,\msun.

The iron yield from the \cite{Nomoto2013} 18\,\msun\ nucleosynthesis model is 0.079\,\msun\ ($^{56}$Fe only) or 0.085\,\msun\ (including all isotopes); this is orders of magnitude less than the inferred pure iron dust model \citep{Henning1996}, which requires 3.97\,\msun\ of dust to fit the SED.
The \cite{Woosley1995} 15\,\msun\ and 18\,\msun\ $Z=0.1Z_\odot$ progenitor models predict iron masses ranging from 0.14--0.20\,\msun, where roughly half of the iron originates from $^{56}$Ni decaying to Fe. 
This is an order of magnitude less mass than the fit requires. 
We can therefore rule out a scenario where iron-rich grains in SN~1987A are producing the bulk of the thermal emission due to either not fitting the SED (in the case of FeO and FeS, \citealt{Henning1996,Henning1995}) or resulting in unrealistically high dust masses for the pure iron model. 

Limits from the \cite{Nomoto2013} 18\,\msun\ explosive synthesis model for some other common dust varieties include 0.337\,\msun\ for forsterite (Mg$_2$SiO$_4$), 0.481\,\msun\ for enstatite (MgSiO$_3$), 0.014\,\msun\ for alumina (Al$_2$O$_3$), and 0.373\,\msun\ for silica (SiO$_2$) -- assuming 100\% of all isotopes are locked in dust grains.
These are all notably lower than the dust masses resulting from the modBB fits -- e.g., 4.0\,\msun\ for forsterite, 4.1\,\msun\ for enstatite and 0.9\,\msun\ for alumina (Table~\ref{table:modBBfits_kappacurves}).
Given the discussion above, one can place a rough upper limit on the total mass of dust that could form in the ejecta of SN~1987A of $< 1.5$\,\msun\ given the available metal budget. 
From this it is possible to immediately rule out a further seven dust varieties listed in Table~\ref{table:modBBfits_kappacurves} as producing unphysical dust masses.

Many of the fits to the SED of SN~1987A require more mass in dust grains than the mass of available metals for the corresponding species. 
This can be explained if the SED is made up of a mixture of several species each contributing to the overall dust budget, as originally proposed in \citet{Matsuura2015}.
For example, locking all available mass into a combination of C+MgSiO$_3$+FeS would give a total metal mass, and an upper limit on the total dust mass, of 0.72\,\msun. 
However, taking the yields of the \cite{Woosley1988} 18\,\msun\ 0.1$Z_\odot$ progenitor model, for example, would give 0.55\,\msun\ of total metals available for dust formation for this combination of species, or $\sim$30\% less mass than the simple sum of the total species indicates.

Using the predicted metal yields as an upper limit to rule out dust varieties and determine the mass of the ejecta dust also has its own challenges in that model abundances for core-collapse supernovae vary with different models, uncertainties in the nuclear process assumed, rotation, and the implementation of the artificially induced shock explosion model. 
We therefore caution that the model abundance yields can only provide loose upper limits.  
Considering the various limitations and caveats for the different dust models considered in this study, a likely overall dust composition, based on the measured SED and nucleosynthesis limits, is a combination of amorphous silicates (especially those of reasonably high emissivity, such as the \citealp{Demyk2017} model) and amorphous carbons as also concluded by \citet{Matsuura2015}, limiting the total dust mass to potentially $<$0.7\msun.

\subsection{The Spatial Distributions of Dust, CO and SiO, and Chemistry Leading to Dust Formation}
\label{sec:chemistry}

Our spatially resolved images (Fig.~\ref{fig:multiwav}) show that the dust distribution is clumpy and asymmetric.
Comparing the spatial distribution of the dust with the \cosixfive\ reveals a weak anti-correlation with the integrated \cosixfive\ and dust distributions, while there is little spatial correlation between the dust and SiO images.
We suggested earlier that this anti-correlation occurs because \cosixfive\ is suppressed compared with \cotwoone\ in the dust bright regions, indicating that the excitation temperature of CO is lower than in other neighboring regions.
This provides new information about the chemistry and physics involved in the formation of dust.

Note that we see an exception to this in one region (region 4 in our CO analysis and roughly corresponding to region A in our SiO {\sc radex} analysis).
Here, the hole in SiO and \cotwoone\ coincides with the dust peak, with relatively strong \cosixfive\ emission observed at $-$1000--1000\,km\,s$^{-1}$ with respect to \cotwoone, and this strong \cosixfive\ continuing to its neighboring region (region 5 in Fig.~\ref{fig:COregionratios}).
We will discuss this region separately in \S~\ref{sec:blob_models}.

The SN chemistry after the explosion inherits a series of nuclear synthesis processes at the stellar core prior to and during the SN explosion \citep[e.g.,][]{Sarangi2013}.
The outermost region is the H-envelope, followed by the He shell, which can contain more carbon atoms than oxygen atoms \citep{Woosley1995, Rauscher2002}.
The He shell can also form CO, as it enriches with C and O \citep[e.g.,][]{Sarangi2013}.
The inner region is roughly represented by an O+Ne zone, which can also contain C, followed by an O+Mg+S+Si zone, and finally with a $^{56}$Ni core, that also contains Si, but very low C or O.
Here we interpret an apparent anti-correlation between the dust and \cosixfive\ spatial distributions as the result of both CO and dust components having originated from the He-envelope or O+Ne nuclear burning zone containing both C and O prior to the explosions.
Thus we propose that the dust grains formed in this region could be carbonaceous.

Our suggestion of carbonaceous dust contradicts predictions of supernova dust formed in chemical models.
\citet{Sarangi2015} predicts that SiO molecules formed in the O+Mg+S+Si zone, and SiO molecules directly condense into silicate dust.
The majority of nuclear zones have more O atoms than C atoms, and the formation of CO blocks the formation of graphite or amorphous carbon.
\citet{Deneault2006} and \citet{Clayton2011} previously discussed the formation of carbonaceous grains by dissociating CO via highly energetic electrons, making unbound carbon available for carbonaceous dust formation.
However, the \citet{Clayton2011} calculations involved only a few reaction rates involving C and CO.
In contrast, \citet{Sarangi2013} and \citet{Sarangi2015} included more extensive chemical networks, as well as dissociation of molecules by energetic electrons, and found this resulted in very few carbonaceous dust grains ($6\times10^{-3}$\,$\rm M_{\odot}$ of carbonaceous dust out of total of 0.04\,$\rm M_{\odot}$ dust mass for a 19\,$\rm M_{\odot}$ star).
One simple explanation is that \citet{Sarangi2013} and \citet{Sarangi2015} modelled the SN chemistry only up to day 1500, and the dust composition might have changed since then.
However, it still requires CO to be dissociated in order to remove the blockage of carbonaceous dust formation, suggesting that any dissociation process must be more efficient on longer time scales.
Alternatively, the chemical reaction rate used in the models might have large uncertainties, particularly involving the highly energetic electrons. Together with the recent detection of HCO$^+$ \citep{Matsuura2017}, which was largely under-predicted in the chemical model of \citet{Sarangi2013}, this suggests some tensions exist in molecular chemistry models of SNe. 
However, this tension between our proposed C-rich grain formation and the lack of C-rich grains ``grown'' in the SN dust models could be alleviated if macroscopic mixing is efficient enough to allow Si and C dust to form in the same regions.
Indeed, some 3D hydrodynamical models \citep[e.g., B15, N20; see][]{Utrobin2019} suggest that $\sim$30--70\% of the Si can be mixed out of the central regions into the C shell (A. Wongwathanarat, private communication).

An alternative explanation for the anti-correlation between dust and CO suggested in this work is that carbonaceous dust could originate from the He layer while the CO gas is restricted to the C+O core, possibly resulting in a projected anti-correlation between dust and CO. 
However, this does not explain why the gas temperature is lower at the dust emitting region (though, the presence of dust can lower the gas temperature, see below).  
In this scenario, the CO might have originated from two different layers of nuclear burning zones. 

Our ALMA observations have also shown that regions of bright dust emission have lower CO excitation temperature than other regions.
There are two possibilities to explain this.
First, more dust may have led to cooler gas temperatures due to radiative cooling via dust emission, i.e., the cooler gas temperature is the consequence of dust formation.
Second, the temperature of these regions may have already been lower in the early days when the dust grains were formed, and the temperature reached the dust sublimation temperature while the gas density was reasonably higher, driving more efficient dust condensation.
In this case, dust formation is the consequence of cooler gas temperature.
Currently, the data do not provide a way to distinguish between these two cases.

We note that we see faint diffuse dust emission that might be more extended beyond the Band--9 dust structure in Fig.~\ref{fig:multiwav}, for example, in comparison with the 315\,GHz image, but the surface brightness is lower in those locations.
The CO-dust anti-correlation is not as obvious in these fainter, extended regions, therefore we do not claim that the entire dust content of the SN\,1987A ejecta is carbonaceous in nature; some of the dust components within the system could be associated with silicates as well. 

The high resolution ALMA observations in this work show that the dust distribution in the ejecta of SN~1987A is clumpy even at small scales (as also seen in the CO and SiO ejecta shown originally by \citealp{Abellan2017}).
Aside from SN~1987A, we know from the knots and filaments observed across the Galactic SNR Cassiopeia\,A \cite[e.g.,][]{Milisavljevic2015} that the gas ejected in SN explosions can be clumpy.
Interestingly, \citet{Sarangi2015} included clumpiness in their chemical models and showed that clumpy gas, compared with smoothly distributed gas, provides density enhancements in the ejecta resulting in larger grain sizes for SN-formed dust.

\subsection{Interpreting the Bright Point Source (the Blob)}
\label{sec:blob}

\subsubsection{The blob: predictions from hydrodynamical models}
\label{sec:blob_models}

The bright dust peak (the blob) is observed at the location of the hole in the \cotwoone\ and \siofivefour\ line-emitting ejecta, and we also see emission located in the hole in the \siosevensix\ transition. 
To explain this additional SiO emission, our analysis of the SiO line ratios gives three possibilities: a higher SiO temperature, a high column density with high area filling factor, and high density of the collisional partner.
The higher intensity of the emission of the dust blob compared to its surroundings could be explained by higher dust (`column') density or by higher dust temperatures inside the blob.  
For both dust and SiO molecules, the two key physical parameters are temperature and density, where the latter is also associated with the column density.

We first discuss the possibility of enhanced density (of dust, as well as SiO and CO) in the blob based on model predictions.  
To form the SiO and dust, a significant amount of Si must be present in the blob region. 
The dust could also be carbon based, such that instead of Si, C could also be enhanced. 
However, recent hydrodynamical simulations (Gabler et al. \textit{in preparation}) compared with ALMA observations (Fig. 4 in \citealt{Abellan2017}) show, that in the explosions of three out of the four simulated progenitor models, the final density distributions of Si and C (each multiplied with the oxygen density) have a void in the center. 
This void is caused by a strong reverse shock from the He/H interface within the ejecta immediately after the explosion. 
This shock first slows down the expansion speed of the inner ejecta (also containing Si and C) when passing them, but then it compresses the material such that the entropy increases significantly and an outwards moving shock forms. 
This feature is termed the `self-reflected' shock, first discussed in \cite{Ertl2016} (with forthcoming details in Gabler et al. \textit{in preparation}). 
The self-reflected shock accelerates the innermost ejecta compared to homologous expansion and leaves a region with lower density in the center. 
When passing more and more of the ejecta, the shock loses energy. 
It finally dissipates and cannot accelerate the outermost ejecta. 
This acceleration of the low-velocity ejecta leads to a formation of a higher density shell-like configuration as observed in the emission of the transition lines \siofivefour\ or \cotwoone\ (Fig. 4 in \citealt{Abellan2017}). 
However, in the hydrodynamic simulations, one model based on the progenitor model from \citet{Shigeyama1990} (model N20 in \citealt{Wongwathanarat2015}, \citealt{Abellan2017}) leads to a weaker reverse shock and, hence, a weaker self-reflected shock. 
The latter then is not able to significantly accelerate the central ejecta. 
Therefore, the central region of this explosion model still has similar densities of Si and C compared to that of the fastest moving Si- and C-rich ejecta. 
While the N20 model may offer the possibility of a high central density, the light curve \citep{Utrobin2015,Utrobin2019} disfavors that model.

\subsubsection{The blob: gas and dust heated by the compact source}
\label{sec:compactsource}

To explain the higher SiO and CO gas temperature and increased brightness for the observed properties of the blob, we suggest two possibilities:
(i) gas heated by a compact object or (ii) a clump heated by $^{44}$Ti decay. 
Here we argue that the most probable explanation for the detected dust blob is that the innermost part of dust and gas is heated by radiation from the compact object, with an early development of a pulsar wind nebula.

$^{44}$Ti was synthesized at the time of the SN explosion, and its decay energy is the main source of the heating of the inner ejecta \citep{Jerkstrand2011,Larsson2016}.
SN explosion models show that $^{44}$Ti is located more or less at a similar radial extent as $^{56}$Ni and $^{28}$Si \citep{Woosley1995,Wongwathanarat2015,Wongwathanarat2017} with almost identical bulk velocities, though the modeled distribution of $^{44}$Ti could be more uncertain than the other elements due to its greater sensitivity to the explosion physics (Jerkstrand et al., \textit{in preparation}).
Observations show that the $^{44}$Ti ejecta is  redshifted, suggesting that the bulk is moving away from the observer \citep{Boggs2015}.  
Since the predicted distribution of $^{44}$Ti shows qualitatively similar properties as Fig. 4 in \citet{Abellan2017}, because it is subject to effectively the same hydrodynamical history (Jerkstrand et al., \textit{in preparation)}, it is unlikely that gas heated by $^{44}$Ti decay would be more centrally distributed and also be co-located with a small blob of gas and dust at one small region in the very inner ejecta.

Therefore, we suggest that the central blob is due to warmer ejecta, and the most likely source of the heating energy is from the compact object (i.e., possibility (i)).
The dust and gas in the blob could be directly heated by X-rays from the surface of the compact object \citep{Alp2018a}, or the dust could be heated by synchrotron radiation generated by the compact object. 

In the latter case, previously it was shown that synchrotron radiation from neutron stars can heat up the dust grains in pulsar wind nebulae, as seen in Galactic SNRs including the Crab Nebula \citep{Temim2006, Gomez2012b, Owen2015}, G54.1$+$0.3 \citep{Temim2017, Rho2018}, G11.2$-$0.3, G21.5$-$0.9 and G29.7$-$0.3 \citep{Chawner2018}.
In the Crab Nebula, the radiation from the pulsar wind can partially dissociate gas, contributing to the formation of the molecules OH$^+$ and $^{36}$ArH$^+$ \citep{Barlow2013}.
SN\,1987A may currently be undergoing this phase, i.e., just beginning to develop a pulsar wind nebula in the innermost region.
If the compact source is a black hole \citep{Brown1992,Blum2016}, instead of a neutron star, jets from the black hole can also heat up gas and dust locally \citep[e.g.,][]{Russell2006}. 

Instead of synchrotron, direct thermal radiation from the neutron star can also heat the gas and dust locally \citep{Alp2018a}.
The SN 1987A compact object is still surrounded by a dense metal-rich gas. 
Metals absorb and scatter (soft) X-ray, UV and optical light efficiently, so that it would be challenging to detect light from the compact source directly \citep{McCray2016,Alp2018b}. 
The absorbed energy is re-processed into longer wavelengths, and eventually ends up heating dust and molecules. 
However, this picture does not consider any radio emission produced directly by the compact object.

From the neutron star kick inferred by the distribution of intermediate-mass elements \citep[e.g.,][]{Katsuda2018} and the redshifted $^{44}$Ti spectrum \citep{Boggs2015}, the direction of the compact object is predicted to be moving toward us and extending along the north-east direction in the sky projection \citep[e.g.,][]{Janka2017}. 
The location of this blob is consistent with that prediction. 
There is an offset between the dust blob and the estimated location of the progenitor star as derived by
\citet{Alp2018a}: 
$\alpha = 05^\mathrm{h} 35^\mathrm{m} 27^\mathrm{m}.9875$, $\delta = -69^\circ 16^\prime 11^{\prime\prime}.107$ (ICRS). 
The brightest pixel in the 679~GHz emission is offset 72 mas to the east and 44 mas to the north of the position of the progenitor star from \citet{Alp2018a}. 
This offset is $\sim$3--5 times the total alignment uncertainty, however, it is still within the range of the neutron star kick.
\citet{Zanardo2014} proposed that the compact object may have traveled 20-80 mas from the site of the SN, towards the west, in comparison to the circular radius of 100 mas used in \citet{Alp2018a}.  
The peak of the dust blob from our work falls within this range (though it is at the upper end, at approximately 700 \kms\ in our data, assuming the SN~1987A position of \citet{Alp2018a}).

Moreover, the location of the compact object and the brightest part of the pulsar wind in the Crab Nebula are also not coincident \citep{Weisskopf2000, Gomez2012b}. 
Although the Crab's pulsar is powerful enough to affect its environment dynamically whereas the compact object in SN~1987A would be at an earlier evolutionary stage, nevertheless, a misalignment between the estimated location of the progenitor star and the location of the dust blob is not unprecedented.

The approximate temperature of the dust of the blob can be estimated from the 679\,GHz flux density. 
The peak is a factor of $\sim$2 brighter than the surrounding ejecta continuum.
A factor of two increase in flux density at the dust peak can be explained if the dust is at a higher temperature compared to the global dust ejecta (an increase to 33\,K from 22\,K for \citet{Zubko1996} or to 26\,K from 18\,K for \citet{Jaeger2003} dust grains would be required to explain the peak).

\begin{figure}[]  %Compact Object Limits
\centering
\includegraphics[width=1.0\linewidth]{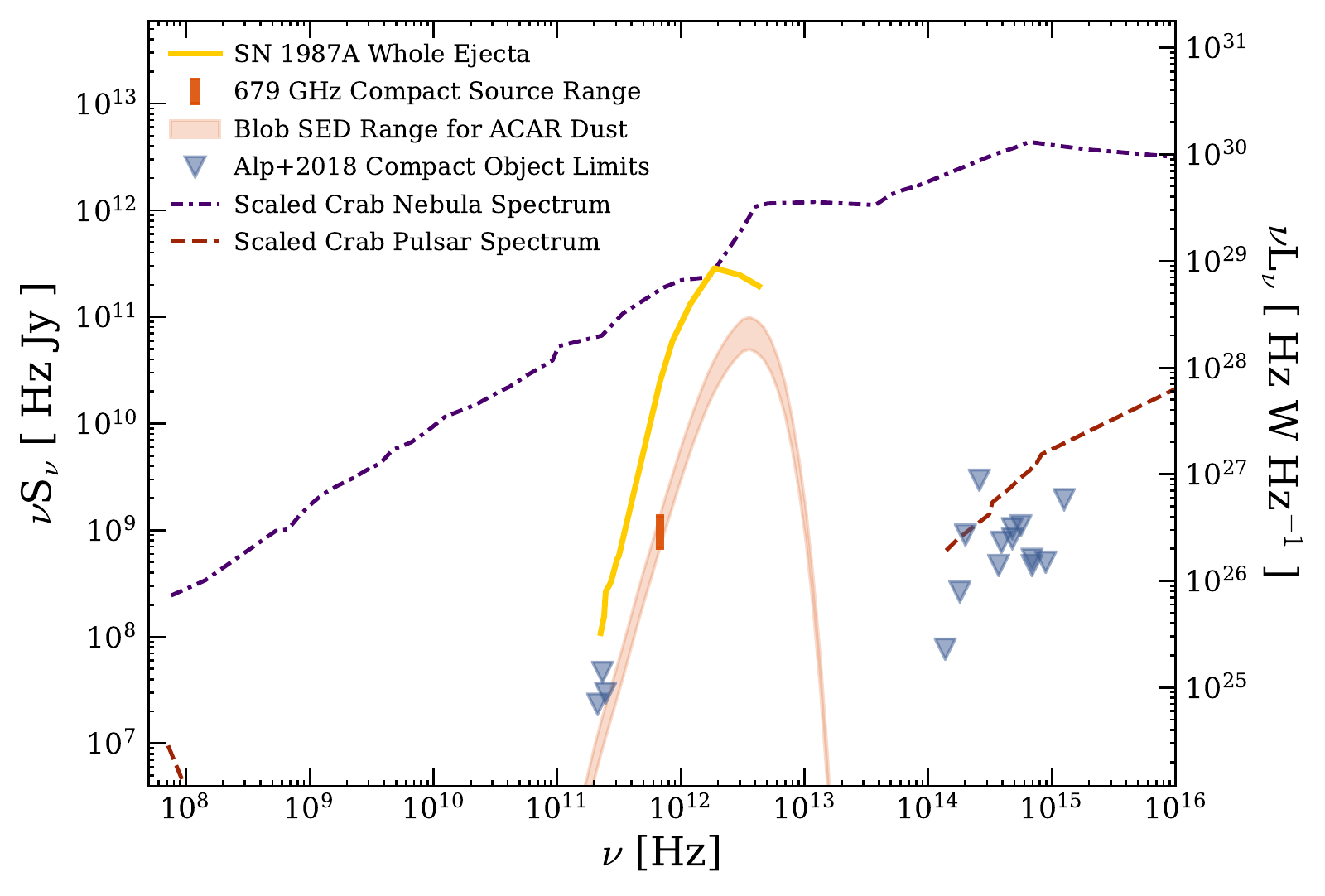}
\caption{
Limits on the luminosity of the compact object in SN~1987A. 
The flux densities for the entire ejecta are shown by the yellow curve. 
The estimated 679\,GHz compact object flux density range (1--2\,mJy) is denoted by the vertical orange bar. 
The light orange shaded region shows the \cite{Zubko1996} ACAR model for this flux density range, and assumes it corresponds to a temperature of 33K. 
The flux densities assuming this ACAR model and 1--2\,mJy at 679\,GHz are consistent with the limits from \cite{Alp2018a} (blue-grey triangles).  
For comparison, the spectra of the Crab Nebula and the Crab Pulsar \citep[purple and red curves, respectively;][and references therein]{Buhler2014} are shown scaled to the distance of the LMC. 
\label{fig:CompactObjectLims}
}
\end{figure}

We estimate the dust peak flux density as S$_\mathrm{679GHz}$ = 3--5 mJy, depending on how a 2D gaussian is placed on the 679 GHz map; 
the peak pixel flux densities are at S/N$\sim$7 (above the RMS level), and the uncertainty is caused by the source being blended with other nearby features. 
This flux density contains contamination from the underlying continuum emission, which can contribute about 2 mJy, thus, the estimated flux density of the compact source is of order 1--2\,mJy. 
Figure~\ref{fig:CompactObjectLims} shows our estimated 679~GHz flux density range of 1--2 mJy (orange bar), along with spectra for the Crab Nebula and its central pulsar \citep{Buhler2014}, scaled to the distance of SN~1987A, for comparison. 
The estimated blob flux density falls between the total spectrum of the pulsar wind nebula and the sole pulsar spectrum, using the Crab as a template. 
Using our 1--2 mJy range and assuming a dust model, one can estimate the millimeter spectrum and the luminosity of the compact source. 
For the \cite{Zubko1996} ACAR model and a temperature of 33\,K, we obtain the light orange shaded region in Fig~\ref{fig:CompactObjectLims}.  
\cite{Alp2018a} derived upper limits to the flux densities of the compact object, shown here as blue triangles. 
At 213\,GHz, our ACAR modBB curve gives S$_\nu$ of order 0.1\,mJy, which is consistent with their flux density limits around that frequency. 
An alternative limit on the compact object emission can be estimated from the bolometric luminosities of different components of the ejecta.    
Integrating the blob modBB curve assuming ACAR dust results in a localized bolometric luminosity L$_\mathrm{bol,dust}$=40--90~L$_\odot$ for 679\,GHz flux densities of 1--2 mJy. 
This is an order of magnitude estimate, incorporating uncertainties in the flux density measurements and temperature estimate.
This luminosity range is an upper limit assuming that the compact source heats dust from 0\,K to 33\,K. 
However, the compact source is most likely additional heating, on top of $^{44}$Ti-heated 22\,K dust, which might be partly subtracted as the underlying 2 mJy continuum, but some of this contribution might not be subtracted, so that the power coming from the compact source could be lower than 40--90~L$_\odot$.

In summary, we suggest that the dust blob seen in the ALMA Cycle 2 Band 9 images could be due to dust heated by the compact object and potentially an emergence of a pulsar wind nebula based on the following arguments: (i) we expect a compact source to be present; (ii) we see one and only one blob which is difficult to reconcile with the expected geometry of $^{44}$Ti; (iii) this scenario would produce a temperature increase at the location of the blob as proposed in this work; and (iv) the position of the dust blob is within the predicted SN kick, though towards the high end. 
However we caution that we only have one frequency band and as such, the nature of the dust peak is not clear: this argument is only valid if the blob is thermally or non-thermally heated emission from dust. 
Alternative possibilities, such as the direct detection of the compact object spectrum cannot be ruled out in this work.

%====================
%====================

\subsection{Dust as a Source of Extinction}
\label{sec:dustextinction}

The precise origin of the H$\alpha$ hole, whether from a physical lack of material (due to the reverse shock or a simple void due to the expansion of ejecta), or illumination of the ring X-rays brightening the outer rim of the ejecta, or from dust extinction, has been discussed for many years \citep[e.g.,][]{McCray2003,Larsson2011,Fransson2013,Larsson2016,Larsson2019}.
A simple estimate of the extinction can be made by assuming the dust fills a spherical shell with uniform density, and calculating the optical depth using the mass extinction coefficient $\kappa_\mathrm{ext}$ (related to, but different from the $\kappa_\mathrm{abs}$ used in the modBB fits) according to $\tau_\lambda = \int \kappa_\mathrm{ext,\lambda} \, \rho \, ds$.
For shell edges corresponding to the features in the images (0\farcs05--0\farcs2, or $\sim$0.4--1.5$\times 10^{15}$m), the \cite{Zubko1996} ACAR model and its corresponding dust mass fit give an opacity at 6563\AA\ of $\tau_{\mathrm{H}\alpha} \sim$ 560.
Silicate dust varieties tend to have lower $\kappa_\mathrm{ext,\lambda}$ at a given wavelength -- the amorphous forsterite model of \cite{Jaeger2003}, for example, gives $\tau_{\mathrm{H}\alpha} \sim$ 400.
$\kappa_\mathrm{ext,\lambda}$ at optical wavelengths can depend strongly on the assumed grain size for some models, typically decreasing as grain size increases.
Yet even for very large grains of $a$=5\um, the optical depth of ACAR dust at H$\alpha$ is 10.
Even this level is optically thick, meaning that dust extinction can play a non-negligible role in the observed \halpha\ distribution.

%%%%---------------- Summary --------------------------

\section{Summary}
\label{sec:summary}

We have observed SN~1987A with ALMA in Cycle 2, 10352--10441 days after the explosion, in Bands 7 and 9. 
This follows on from the first ALMA results from Cycle 0 in \cite{Kamenetzky2013,Indebetouw2014,Zanardo2014} and \cite{Matsuura2015}, and complements the molecular line studies of \cite{Matsuura2017} and \cite{Abellan2017} from Cycles 2 and 3.
In this paper we describe the observations, data reduction, calibration, and photometry in the ALMA bands at the highest angular resolution to date for the continuum of SN~1987A.

\begin{itemize}

\item We find that the dust emission in the ejecta is clumpy and asymmetric, fitting within the H$\alpha$ ``key hole'', with a peak of emission that we name ``the blob''.  The dust ejecta region is smaller in scale than the cool, clumpy CO and SiO ejecta regions.  Dust, in the amounts we fit here in a simple uniform spherical shell geometry, is optically thick at optical wavelengths, with $\tau\sim$500.  Dust extinction could be a factor in the appearance of the \halpha\ hole.

\item We see an anti-correlation between the \cosixfive\ emission and dust, and this anti-correlation is not seen when comparing dust and \cotwoone. Our {\sc radex} analysis suggests this is the result of a lower CO gas temperature where the dust emission is stronger, compared to the surrounding ejecta, hinting that the dust may be C-rich and may have formed due to dissociation of CO, contrary to chemical predictions.

\item We observe a dust peak (the blob) at the location of the molecular hole detected at lower CO and SiO transitions, and the higher CO and SiO line transitions are stronger in that location than the emission in the lower transitions.  We suggest that this is the result of warm gas and dust at the location of the blob, and discuss the possibility that this could be due to slow moving reverse shock material originating from the explosion (the self-reflected shock), heating from a high concentration of radioactive decay, or the compact source.  The most likely scenario is an indirect detection of the compact source.

\item  We fit the spectral energy distribution of the dust emission from the ejecta with modified blackbody profiles.  The derived dust masses and temperatures depend on the submm emissivity of the dust, which is not very well determined.  Temperatures from the fits are generally between 18--23K. Amorphous carbon and graphite models have dust with high emissivity and so yield the lowest dust masses, around $0.4-1.6~{\rm M}_\odot$. Silicates return higher dust mass estimates in the range $0.6-4~{\rm M}_\odot$.  ``Typical ISM'' dust varieties from the Milky Way and nearby galaxies give masses between $1-2~{\rm M}_\odot$.  Taking the total mass of available metals excluding oxygen predicted to be ejected by a SN with progenitor mass appropriate for SN~1987A, we rule out several grain models and compositions for the ejecta dust. We revise the possible range of dust masses in the ejecta to $0.2-0.4$\msun\ for carbon or Si grains, or a total of $<0.7$\msun\ for a mixture of grain species.  A mixture of dust species, including silicates and carbonaceous grains, seems necessary to reconcile the continuum SED, nucleosynthesis model yields, and the molecular line analysis.

\end{itemize}

%%%%---------------- Acknowledgements --------------------------
\acknowledgments{{\it Acknowledgments:}
We acknowledge Arka Sarangi, Loretta Dunne, Steve Maddox, and the anonymous reviewer for interesting and useful discussions. 
PJC and HLG acknowledge support from the European Research Council (ERC) in the form of Consolidator Grant {\sc CosmicDust} (ERC-2014-CoG-647939, PI H\,L\,Gomez).
MM acknowledges support from the UK STFC Ernest Rutherford fellowship (ST/L003597/1).
Research at Garching was supported by the ERC through grant ERC-AdG No.\ 341157-COCO2CASA, and by the Deutsche Forschungsgemeinschaft through Sonderforschungsbereich SFB 1258 ``Neutrinos and Dark Matter in Astro- and Particle Physics'' (NDM) and the Excellence Cluster ``ORIGINS: From the Origin of the Universe to the First Building Blocks of Life'' (EXC 2094; \url{http://www.origins-cluster.de}).
MJB acknowledges support from ERC grant SNDUST (ERC-2015-AdG-694520).
JCW is supported in part by NSF AST-1813825.
This paper makes use of the following ALMA data: \newline \mbox{ADS/JAO.ALMA\#2013.1.00063.S}, \newline \mbox{ADS/JAO.ALMA\#2013.1.00280.S}, \newline \mbox{ADS/JAO.ALMA\#2015.1.00631.S}, \newline \mbox{ADS/JAO.ALMA\#2016.1.00077.S}, and \newline \mbox{ADS/JAO.ALMA\#2017.1.00221.S}. %Full listing of ADS/JAO.ALMA\#1..., ADS/JAO.ALMA\#2... as required by ALMA
ALMA is a partnership of ESO (representing its member states), NSF (USA) and NINS (Japan), together with NRC (Canada), MOST and ASIAA (Taiwan), and KASI (Republic of Korea), in cooperation with the Republic of Chile.
The Joint ALMA Observatory is operated by ESO, AUI/NRAO and NAOJ.
The National Radio Astronomy Observatory is a facility of the National Science Foundation operated under cooperative agreement by Associated Universities, Inc.

%Recognition of facilities used. Optional listing of specific instrument in parentheses: \facility{HST(WFPC2)}
\facility{Atacama Large Millimeter Array} 

\software{
  \texttt{astropy} \citep{Astropy2018},
  \textsc{Casa} \citep{CASA},
  \texttt{emcee} \citep{emcee},
  \texttt{ipython} \citep{ipython}, %\citep{http://dx.doi.org/10.1109/MCSE.2007.53},
%%  \texttt{IRAF}/\texttt{PyRAF} \citep{tody86, tody93, stsci12},
  \texttt{lmfit} \citep{lmfit},
  \texttt{matplotlib} \citep{matplotlib},
  %\texttt{miepython} \citep{https://github.com/scottprahl/miepython}, %--> no DOI or paper
  \texttt{multicolorfits} \citep{multicolorfits}, %\citep{http://dx.doi.org/10.5281/zenodo.3256060},
  \texttt{numpy}  \citep{numpy}, %\citep{http://dx.doi.org/10.1109/MCSE.2011.37}
  %\texttt{pywcsgrid2} \citep{https://github.com/leejjoon/pywcsgrid2}, %--> no DOI or paper
  \textsc{radex} \citep{VanderTak2007}, 
  \texttt{SAOImage DS9} \citep{DS9},
  \texttt{scipy} \citep{scipy},
  \texttt{WWCC} \citep{Arshakian2016}.
}

%%%---------------- Bibliography --------------------------

\bibliographystyle{aasjournal}

%%%============== End Bibliography =========================

\newpage

%%%%%---------------- Appendix --------------------------
\appendix

\renewcommand{\thesection}{\Alph{section}}

\section{Molecular Line Channel Maps}
\label{app:chanmaps}
\renewcommand\thefigure{\thesection.\arabic{figure}}
\setcounter{figure}{0}

\begin{figure*}[h!]
\centering
\includegraphics[width=0.49\textwidth]{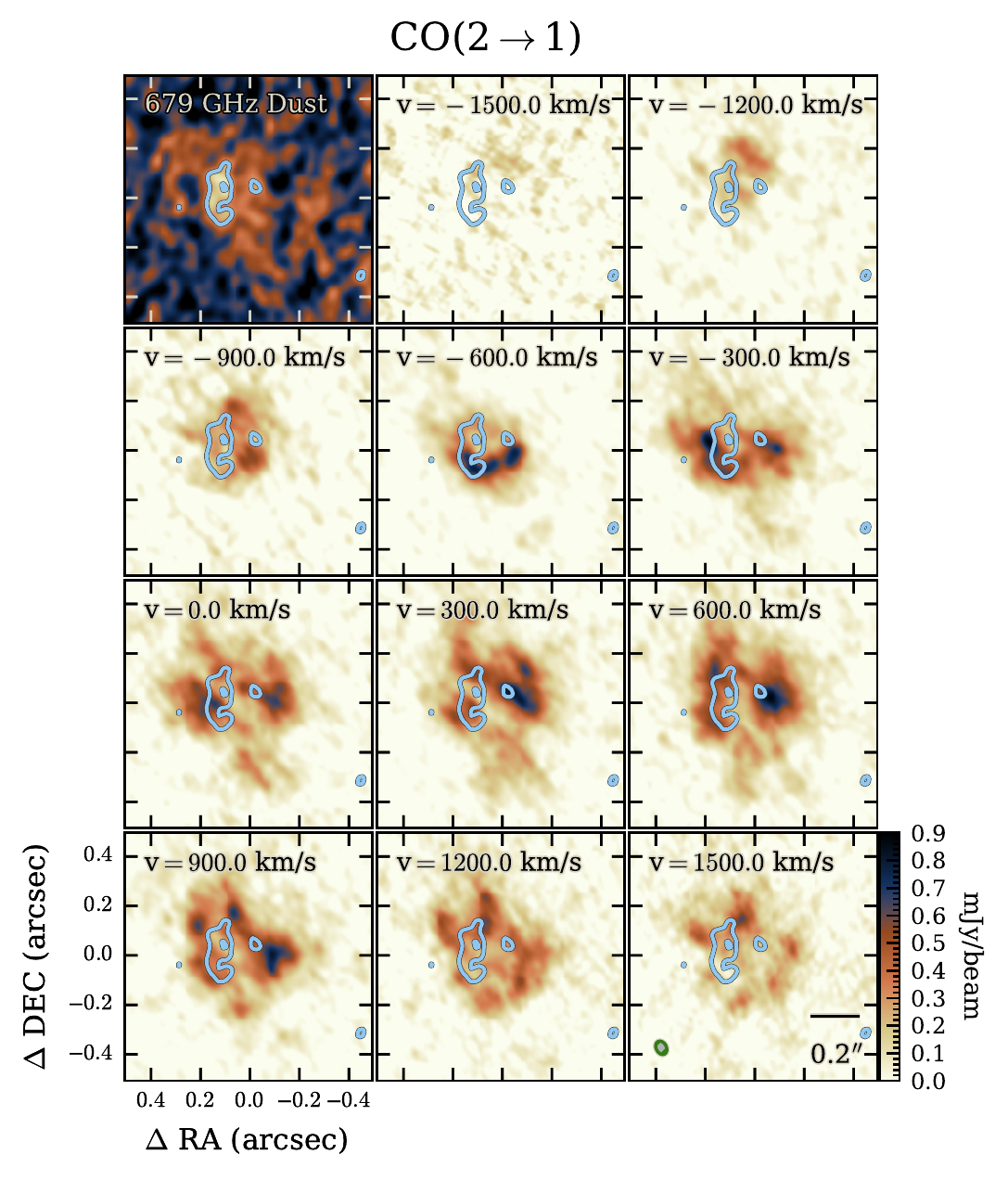}
\includegraphics[width=0.49\textwidth]{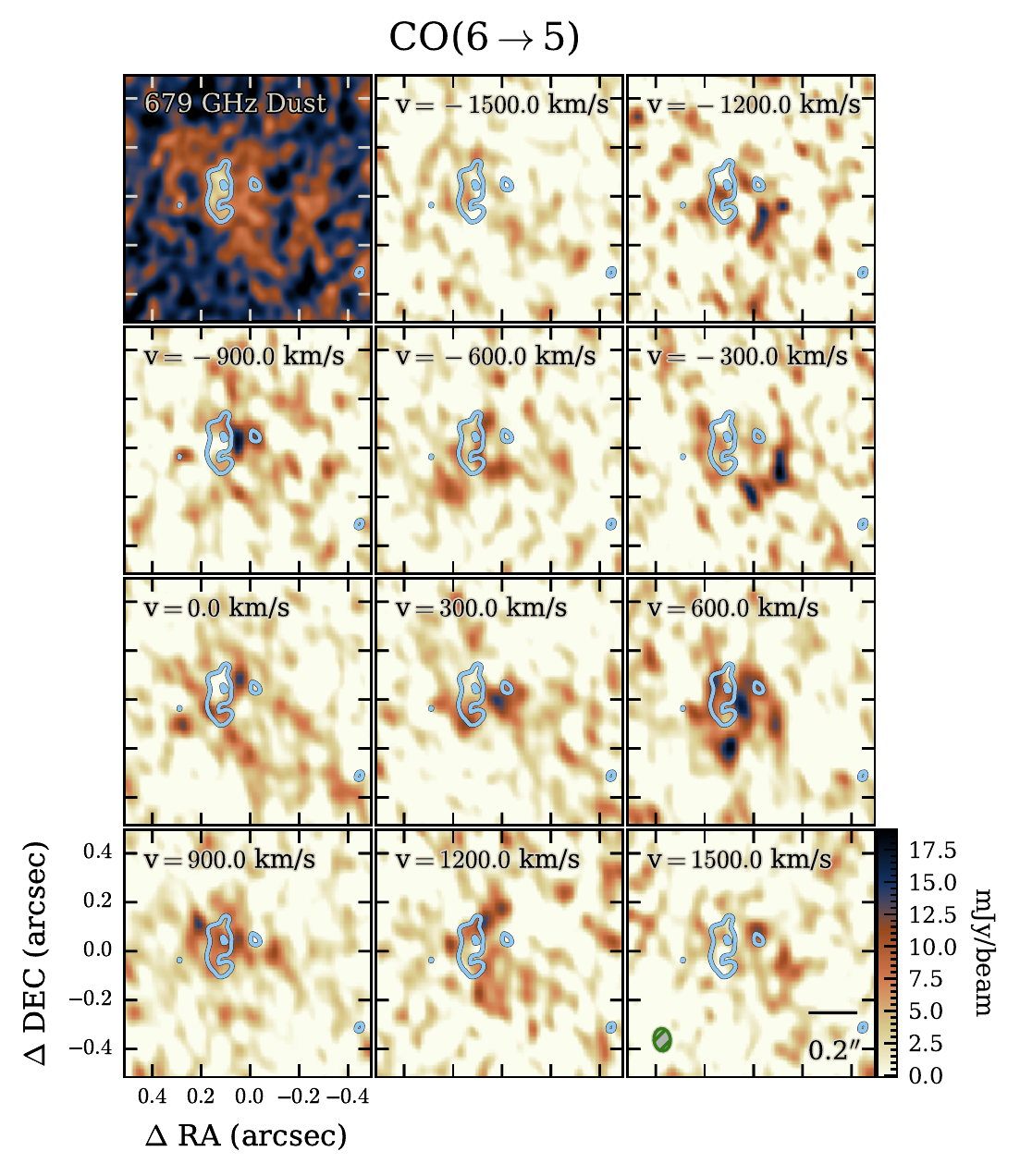}
\caption{CO channel maps in 300 km s$^{-1}$ steps, LSRK; for reference, the SN~1987A systematic velocity is 287 \kms.  \textit{Left}: \cotwoone\ as presented by \cite{Abellan2017}, with the notable central hole in the molecular emission that persists all the way through the line of sight.   \textit{Right}: \cosixfive\ from the present work.  Even with the poorer S/N due to atmospheric transmission in Band--9, the emission profile is noticeably different from that of \cotwoone.  Cyan contours are Band--9 dust at 3$\sigma$ and 5$\sigma$ levels, and the beam size is given by the green ellipse in the lower right panel.
}
\label{fig:chanmaps_co}
\end{figure*}

\begin{figure*}[h!]
\centering
\includegraphics[width=0.49\textwidth]{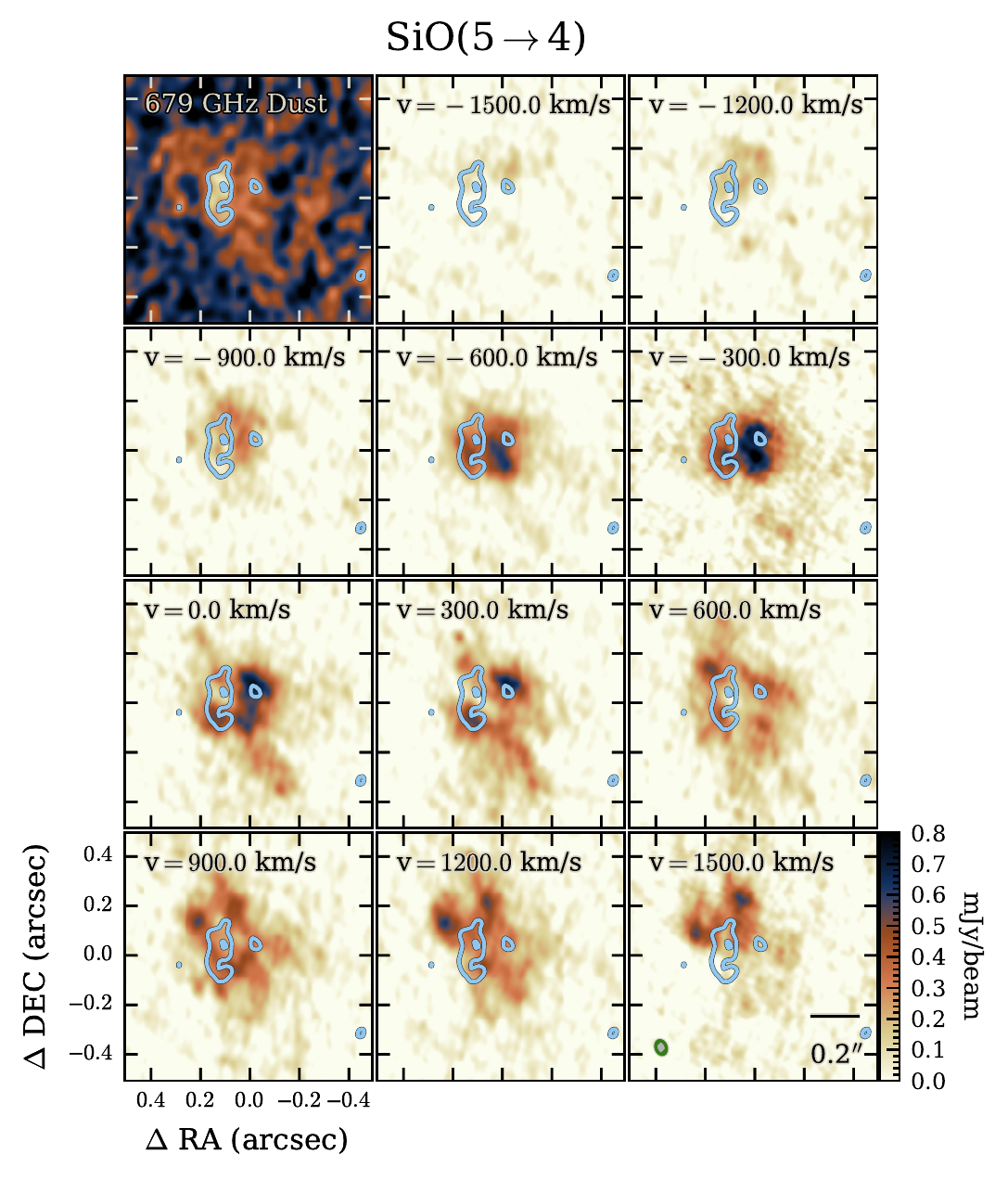}
\includegraphics[width=0.49\textwidth]{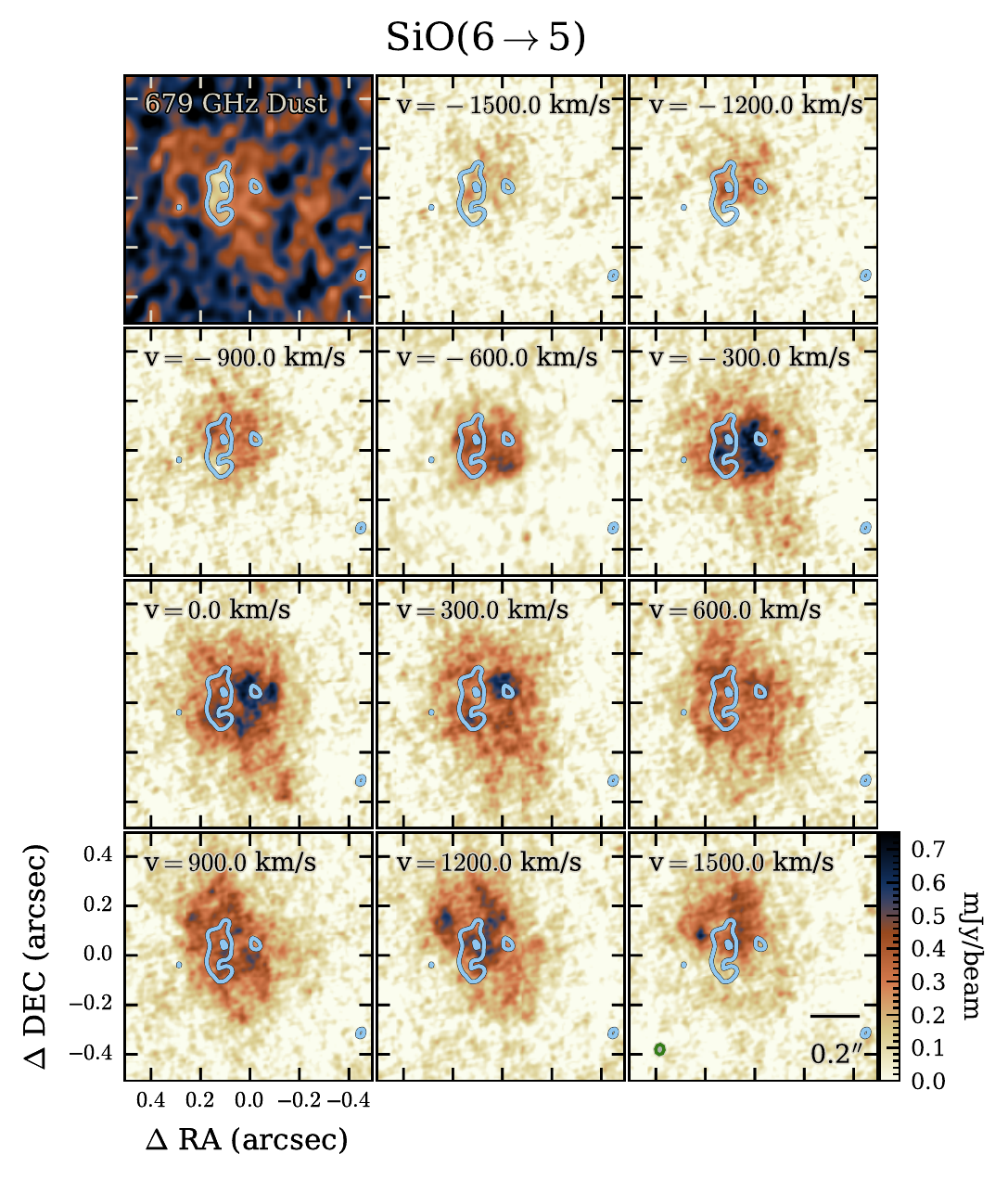} \\
\includegraphics[width=0.49\textwidth]{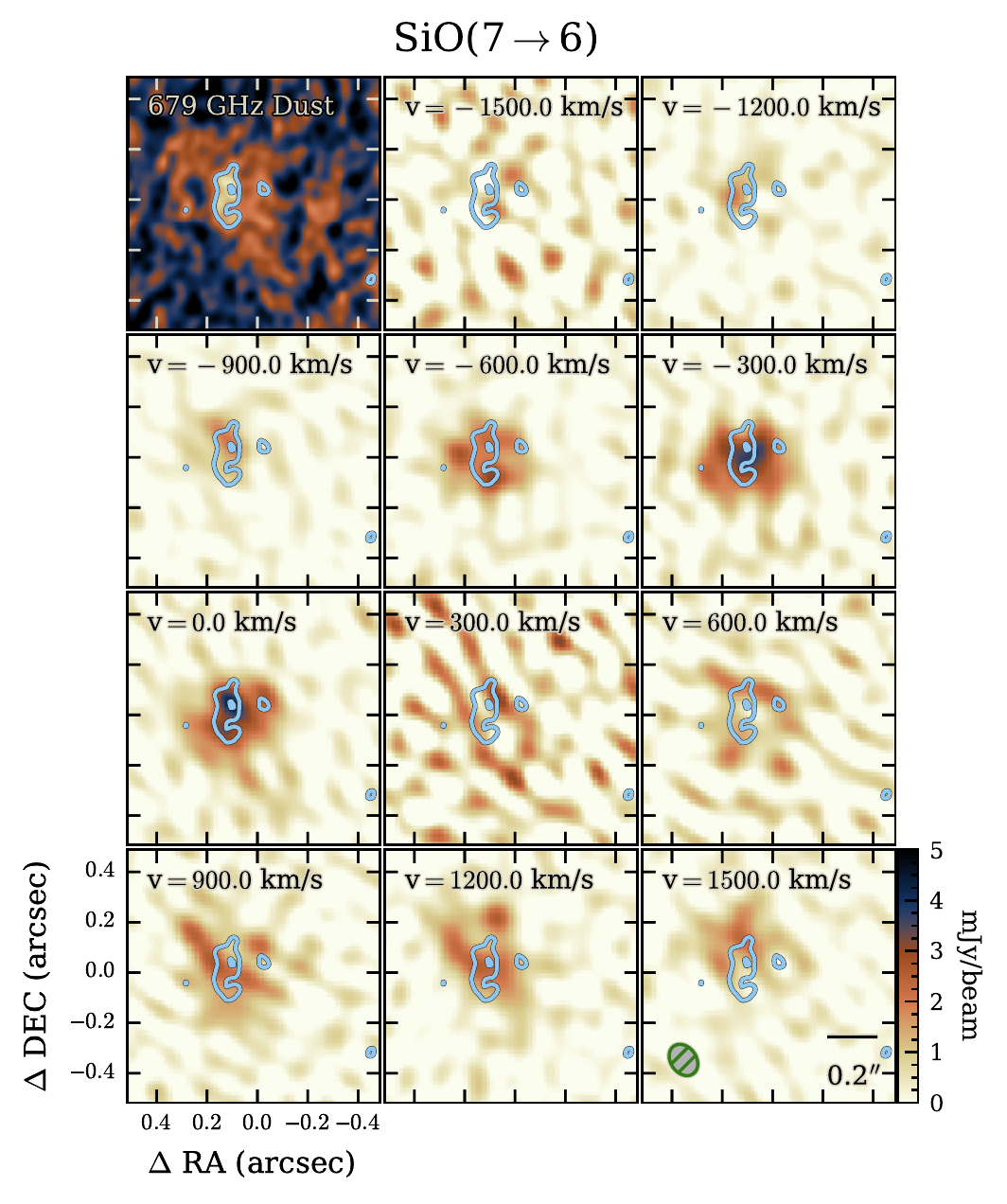}
\caption{SiO channel maps in 300 km s$^{-1}$ steps, LSRK; for reference, the SN~1987A systematic velocity is 287 \kms.  \textit{Top Left}: Continuum + \siofivefour\ as presented by \cite{Abellan2017}, also exhibiting the central molecular hole seen in \cotwoone. \textit{Top Right}: Continuum + \siosixfive\ from \cite{Abellan2017}.  \textit{Bottom}: \siosevensix\ from the present work.  
The emission profiles of the three lines tend to be similar for a given channel, with the conspicuous exception of an excess in \siosevensix\ at the spatial position of the molecular hole but slightly in front (at 0 km s$^{-1}$). Cyan contours are Band--9 dust at 3$\sigma$ and 5$\sigma$ levels, and the beam size is given by the green ellipse in the lower right panel.  
}
\label{fig:chanmaps_sio}
\end{figure*}

\newpage

\section{Defining the Center for Photometric Measurements}
\label{app:center}
\renewcommand\thefigure{\thesection.\arabic{figure}}
\setcounter{figure}{0}

The inferred center of the SN~1987A system appears to vary slightly in different parts of the spectrum, and it is important to determine it carefully for analysis of the expanding ejecta material.   
\cite{Alp2018a} used the hotspots in the ring to determine the center of SN~1987A. 
This was done by fitting 2D Gaussians to the hotspots in the \hst/ACS \textit{R}--band image from 2006 and then fitting 1D Gaussians along the same directions in all the other 32 \textit{R} and \textit{B} images from 2003 to 2016 (Table~8 of \citet{Alp2018a}).
They derive a ring center position of $\alpha$=5$^{\rm h}$35$^{\rm m}$27\fs9875, $\delta$=--69$^{\circ}$16$^{\prime}$11\farcs107 (ICRF J2015.0), with uncertainty from bootstrapping the hotspot locations of (11,4) mas.
Here we explore similar methods where we fit the ring emission to determine the central position of the SN~1987A system, but we use the Cycle 2 ALMA data at 315 GHz because it has the highest spatial resolution and S/N ring emission among the new images presented in this work.  
First, the ejecta and all emission clearly exterior to the ring were masked.  
Then, a ridge of emission peaks around the ring was determined.  
This was carried out by starting from the rough center of the map, and at 50 different angles the pixel with the brightest flux along each ray was located. Finally, an ellipse was fit to the ridge using three methods (Fig.~\ref{fig:centercoordsmethod}) (i) fitting using a quadratic curve method (red); (ii) standard least-squares minimization of the parametric ellipse equation (blue) and (iii) weighted least squares, using the inverse of the squared pixel intensities as the weights (green) so as to favor bright regions over tenuous emission.  
The latter method weights the pixels higher on lines of sight with brighter values than those lines of sight with very faint emission. 
All three methods, though the resulting ellipses have slightly different shapes, returned center positions within about 10 mas of each other -- less than one pixel (Figure~\ref{fig:centercoordsmethod}).  
The final weighted fit gives a ring emission center of $\alpha$=5$^{\rm h}$35$^{\rm m}$27\fs998, $\delta$=--69$^{\circ}$16$^{\prime}$11\farcs107 (ICRS).  
We estimate an uncertainty in these fit coordinates of 5.9 mas by varying the number of search angles and by averaging the results from the three fit methods. 
Combined with the 12 mas astrometric uncertainty for the 315~GHz image (\S~\ref{sec:observations}), the total uncertainty on the inferred center position is 18 mas, slightly larger than one pixel width.
The main (systematic) uncertainty in the position comes from assuming that the progenitor/SN should coincide with the center of the ring: the optical ring probes high-density gas, whereas ALMA probes lower density gas at higher latitudes.
The revised center from the ALMA data used here falls on the southern edge of the central hole of the \halpha\ emission.

\begin{figure}[h]
\centering
\includegraphics[trim=0mm 0mm 0mm 0mm clip=true,width=0.7\textwidth]{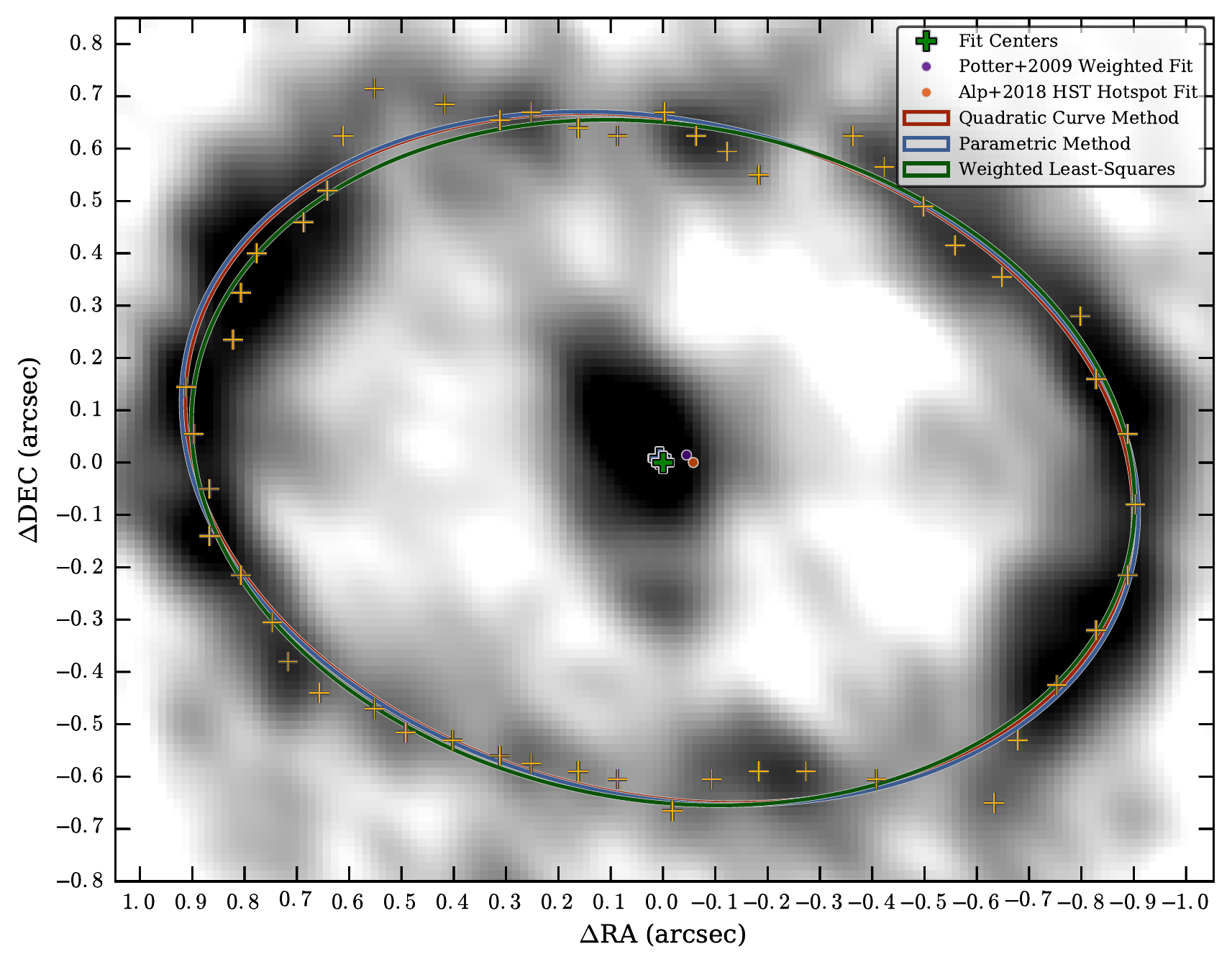}
\caption{Illustrating the three methods used for determining the center of the SN using the ring emission as seen in the ALMA 315 GHz image using the quadratic curve method (red), parametric ellipse fitting (blue) and least squares fitting (green). 
The centers derived by fitting the ATCA radio ring continuum in \citet{Potter2009}, and by fitting to \hst\ image ring hotspots in \citet{Alp2018a} are also shown.}
\label{fig:centercoordsmethod}
\end{figure}

\newpage

\section{Simulating Potential Flux Loss}
\label{app:casasimulations}
\renewcommand\thefigure{\thesection.\arabic{figure}}
\setcounter{figure}{0}

\begin{figure*}[h]
\centering
\includegraphics[trim=0mm 0mm 0mm 0mm clip=true,width=\textwidth]{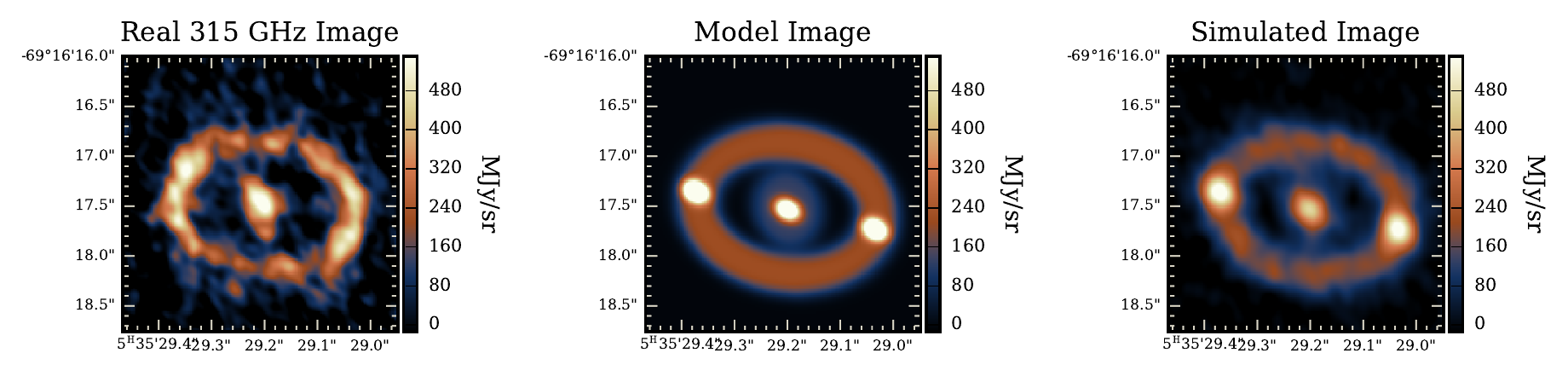}
\caption{ \casa\ simulations of ALMA observations at 315 GHz to test if diffuse flux is lost in the observing process.  
Modelled emission that resembles the true image is recovered qualitatively and quantitatively and suggests we are not missing a significant component of extended diffuse emission.   }
\label{fig:simulations}
\end{figure*}

Simulations exploring the possibility of over-resolving extended dust emission were performed with \texttt{simobserve} and \texttt{simanalyze} in \casa.  
The same antenna locations and integration times from the observations were used, to ensure consistency.   
The most extreme case we considered is a uniform ellipse spanning to the edge of the ring (2\farcs4 across).  
This is similar to the largest angular scales these observations were sensitive to (see Table~\ref{table:Obs}).  
Even in this worst-case scenario, the differences in the integrated ring and ejecta flux densities between the simulated model and true observations were roughly 10\% different when run through the same photometry prescription.  
A more realistic input model is shown in Fig.~\ref{fig:simulations}: a uniform broad annulus for the ring plus a fainter diffuse ellipse spanning the entire ejecta, plus more compact clumps; their integrated flux densities match within a few percent.

Notably, the uniform broad ellipse model does not translate to a simulated model that resembles the true observations, we therefore discount the possibility of the real submillimeter source being more diffuse and uniform as highly unlikely.   
By matching the input model more closely to the emission seen in the ALMA observations we obtain reasonable-looking simulated maps of SN~1987A with a flux density difference of several percent.  
We conclude that there may be some missed extended flux, on the level of a few percent, if the true source has a slightly broader distribution.  
However, this is below the uncertainty level in the ALMA photometry.

An overly conservative approach to this issue would be to include an additional 10\% systematic uncertainty in quadrature to photometry values for this effect.  However, the uncertainty estimation presented in the main text already accounts somewhat for differences in faint features, because it includes the standard deviation of flux densities from same-sized apertures placed at many random locations in each map.

\renewcommand\thefigure{\arabic{figure}}

%%%%%============== End Appendix =========================

\listofchanges

\end{document}